\documentclass[journal,article,submit,pdftex,moreauthors]{Definitions/mdpi} 

\usepackage{pgf}
\usepackage{lineno}

\linenumbers

\firstpage{1} 
\pubvolume{1}
\issuenum{1}
\articlenumber{0}
\pubyear{2025}
\copyrightyear{2025}
\datereceived{ } 
\daterevised{ } 
\dateaccepted{ } 
\datepublished{ } 
\hreflink{https://doi.org/} 

\Title{The history of galaxy mergers in IllustrisTNG}

\TitleCitation{The history of galaxy mergers in IllustrisTNG}

\Author{Bendeguz Koncz $^{1,2}$*\orcidA{},
Istvan Horvath$^{3}$\orcidB{},
Andras Peter Joó$^{1}$\orcidE{}
Andreas Burkert $^{4}$\orcidC{}
and
L. Viktor Tóth$^{1,2}$\orcidD{}}

\AuthorNames{Bendeguz Koncz, Istvan Horvath, András P. Joó, L. Viktor Tóth, Andreas Burkert}

\isAPAStyle{%
       \AuthorCitation{Lastname, F., Lastname, F., \& Lastname, F.}
         }{%
        \isChicagoStyle{%
        \AuthorCitation{Lastname, Firstname, Firstname Lastname, and Firstname Lastname.}
        }{
        \AuthorCitation{Lastname, F.; Lastname, F.; Lastname, F.}
        }
}

\address{$^{1}$ \quad Department of Astronomy, E\"otv\"os University, Budapest, Hungary\\
$^{2}$ \quad University of Debrecen, Faculty of Science and Technology, Debrecen, Hungary\\
$^{3}$ \quad University of Public Service,  Budapest, Hungary\\
$^{4}$ \quad University Observatory, Ludwig Maximilians University Munich, Scheinerstr. 1, 81679 Munich, Germany\\
}

\corres{Correspondence: kbendeg398@mailbox.unideb.hu}

\abstract{
The process of galaxy evolution over cosmic time is not yet fully understood, since there is a debate on the impact of galaxy collisions on the star formation and metallicity. The local environment of the galaxy mergers could also have a large impact on the evolution of the galaxies, but it has not yet been possible to examine it in detail. Modern simulations with larger capacity, including the newest physical knowledge and new observations with JWST, help us to answer these questions. Using the IllustrisTNG cosmological simulation, we processed the catalogue data and the merger tree files of the TNG100-1 and the TNG300-1 simulations. We calculated the galaxies' average star formation rate (SFR) and mass at redshifts between 0 < z < 15. We investigated the environment of galaxy mergers, with the focus on the local density, and also examined how the SFR changes in merging galaxies. We compared our findings with JWST results and highlighted differences in the star formation rate density (SFRD) history between the models and observations.
}


\begin{document}



\section{Introduction}

Star formation in galaxies is a widely studied area, but external processes can strongly influence the further evolution of a galaxy, leaving many questions unanswered. The impact of galaxy mergers on star formation activity is highly debated. On one hand, galaxy interactions can increase star formation (\citet{Hwang_2019}; \citet{1996ApJ...464..641M}), and can also trigger a rise in the star formation rate (SFR) in the merger companions (\citet{Shah_2022}). Recent X-ray discoveries suggest that dwarf galaxies are likely sources of the hot intra-cluster medium (\citet{1995ApJ...454..604N}), and these merger events can lead to energetic phenomena such as multiple core-collapse supernovae. On the other hand, many galaxies are quenched following interactions (\citet{2013ApJ...767...50M}; \citet{Ellison_2022}; \citet{2017MNRAS.465..547P}; \citet{2010MNRAS.407..749G}). Other studies have found that galaxy mergers have only a small effect on the star formation rate (\citet{2019A&A...631A..51P}).

The IllustrisTNG model \citet{Nelson_2017}, \citet{Marinacci_2018}, \citet{Pillepich_2017}, \citet{Springel_2017}, \citet{Naiman_2018} is a series of different-volume cosmological gravo-magnetohydrodynamic simulations, which include comprehensive models for galaxy formation and evolution. The TNG simulations solve for the coupled evolution of dark matter, cosmic gas, luminous stars, and supermassive black holes from redshift z = 127 to 0, generating 100 resulting snapshots from z = 20 to 0. 
For our analysis, we used the TNG300-1 run, which has the volume of (302.6 Mpc)$^3$

\subsection{Galaxy mergers}

Galaxy mergers are intensive events when two or more galaxies interact with each other. In recent decades, several studies have dealt with the evolution of galaxy mergers since the investigation of the mergers mechanism can help us to understand the galaxies' evolution, morphology, and dynamics.

The merger galaxies' evolution is significantly different from those that do not interact with other galaxies. Through the gravitational interaction of the collision, gas and dust transmission can trigger different processes. Galactic inflows can enhance the central nuclear activity \citet{1985ApJ...296...90C},  \citet{2011MNRAS.418.2043E}. The star formation rate is increasing \citet{2019MNRAS.485.1320M}, \cite{1985MNRAS.214...87J}, which is referable with the galactic outflows \citet{2005ApJS..160..115R}, \citet{2009ApJ...697.2030S} and with the nuclear activity \citet{2005ApJ...632..751R}, \citet{2017ApJ...839..120W}. \citet{2014ApJ...789L..16L} and \citet{2019MNRAS.485.5631C} showed that 10-20 $\%$ of the galaxies with high star formation rate are galaxy mergers.

\citet{Ma_2022} studied the AGN feedback on the cold gas in quenching galaxies and compared the results in the IllustrisTNG, Illustris, and SIMBA hydrodynamical simulations. They found that galaxies contain more HI in the TNG than in the observed data, the cumulative AGN feedback is the primary driver of the cold gas depletion, which causes galaxy quenching. The kinetic feedback mechanism of the TNG is insufficiently strong to push enough cold gas away from the central galaxies, leading to significant inconsistencies with cold gas observations. 

The history of the mergers' metallicity is a highly researched area as well. In previous works \citet{1979A&A....80..155L} and \citet{1981ApJ...243..127K} have shown the correlation of mass and metallicity of the galaxies and the relation between the luminosity and metallicity \citet{1984ApJ...281L..21R}. \citet{2006AJ....131.2004K} investigated the luminosity-metallicity relation of galaxy pairs, and they found that the low metallicity galaxy pairs produce more energetic central outbursts. Others, like \citet{Ellison_2008} found that the interactions between galaxy mergers indicate gas inflows into the central region, which leads to the enhanced star formation rate and feeds the central nuclear activity as well.

\citet{Hwang_2019} investigated the relation between the star formation rate and density relation in the IllustrisTNG simulation between 0 < z < 2 with galaxy samples. The galaxies in each sample have the same comoving number density. They found that the SFR decreases with the local density at z = 0, but at larger redshifts z $\geq$ 1, the SFR increases with the density. They also showed that the gas fraction of molecular hydrogen is decreasing with the density, independently of the redshift.

Galaxy mergers have been examined in previous publications using the IllustrisTNG simulations:  \citet{2024ApJ...975..104C} constructed a sample of the orbits of isolated
low-mass (10$^8$ < M$_{*}$ < 5 × 10$^9$ M$_\odot$) and high-mass (5 x 10$^9$ < M$_{*}$ < 10$^{11}$ M$_\odot$) major pairs (stellar mass ratio > 1:4) in the TNG100 simulation. They showed that the merger timescales of major pairs in TNG100 vary with redshift, but that low-mass and high-mass pairs have equal merger timescales at all redshifts from z = 0 to 6 if the correct separation-selection criteria are used to pick equivalent samples of pairs. \citet{2024MNRAS.527.6506B} presented a large public dataset of over 750,000 synthetic galaxy images from TNG50 and TNG100 simulations. They investigated how different types of mergers (major, minor, mini) relate to galaxy asymmetry and star formation activity. In their analysis, they found that mini mergers, despite their small mass ratios, are the primary drivers of enhanced asymmetry and elevated star formation rates in star-forming galaxies, more so than major or minor mergers. These mini mergers leave longer-lasting effects on galaxy structure and SFR and play a dominant role in stellar mass assembly and morphological disturbance across redshifts 0.1–0.7. \citet{2023MNRAS.521..800M} examined 700 galaxy clusters from the TNG300 simulation, focusing on the brightest cluster galaxies (BCGs) with M$_{200}$ $\geq$ 5 × 10$^{13}$ M$_{\odot}$ and their evolution. On average, BCGs undergo $\sim$5 major mergers ($\mu$ $\geq$ 1/4) over their lifetimes, with the combined number of major and minor mergers being roughly 1.5-2 times higher. We define the mass ratio ($\mu$) as the mass of the smaller galaxy divided by the mass of the larger galaxy. At z = 0, ex situ stars contribute about 70\% of the BCG stellar mass, increasing to 90\% in the intracluster light (ICL). BCG star formation rates (SFRs) peak at z $\approx$ 3–4 and drop sharply around z $\approx$ 2–3, with most BCGs becoming quiescent by z $\approx$ 1. At low redshift (z $\approx$ 0.4), median SFRs in BCGs exceed 0.1M$_{\odot}$ and agree well with observations after applying selection thresholds.

Recently, machine learning and deep learning methods have also played an important role in the study of galaxy mergers: \citet{2024MNRAS.533.2547F}, \citet{2024A&A...687A..24M}, \citet{2024ApJ...965..156J}, \citet{2023A&A...679A.142O}.
A comprehensive study of galaxy interaction with the IllustrisTNG simulations has been made in a series of papers, in which several aspects ( e.g.: star formation or quenching of the mergers ) of the interactions were examined: 
\citet{2020MNRAS.494.4969P} explored the link between increased star formation and the presence of nearby companion galaxies, comparing their findings with SDSS observations. In the TNG100-1 simulation, they observed that the average specific star formation rate (sSFR) is boosted by a factor of 2.0 $\pm$ 0.1 at separations smaller than < 16 kpc. Similarly, in the TNG300-1 simulation, they found a consistent enhancement in mean sSFR across the redshift range 0 < z < 1. They estimated that close companions increase the average sSFR of massive galaxies in TNG100-1 by about 14.5\%.
\citet{2020MNRAS.493.3716H} analyzed a sample of post-merger galaxies (PMs) within the redshift range 0 < z < 1 and found that star-forming PMs generally exhibit a twofold increase in star formation rate (SFR), while passive PMs show no significant change. Their results indicate that SFR enhancement: it does not vary with redshift, is inversely related to the stellar mass of the PM, and correlates with the gas content of the progenitor galaxies. Mergers originating from gas-rich progenitors show the strongest SFR boosts (1.75–2.5 times), whereas those from gas-poor progenitors can even exhibit SFR suppression compared to matched controls.
In a follow-up study, \citet{2021MNRAS.504.1888Q} investigated the quenching of star formation in post-merger galaxies. They found that quenching is relatively uncommon among star-forming PMs in the TNG simulations: only about 5\% become quenched within 500 Myr of the merger event. Notably, quenching is more frequent in systems with low gas fractions (f$_{gas}$ $\leq$ 0.1). In the same redshift interval (0 < z < 1), considering merger mass ratios greater than 1:10 and post-merger stellar masses above 10$^{10}$ M$_\odot$, \citet{2023MNRAS.519.4966B} reported elevated supermassive black hole (SMBH) accretion rates in post-merger galaxies. On average, these rates are 1.7 times higher than in their control sample. Additionally, post-mergers are three to four times more likely to undergo a luminous active galactic nucleus (AGN) phase compared to isolated galaxies. The enhanced SMBH accretion can persist for up to $\sim$2 Gyr after the merger, significantly outlasting the $\sim$500 Myr duration of elevated star formation. The simultaneous enhancement of both SFR and SMBH accretion is found to depend on both the merger mass ratio and the post-merger galaxy’s gas content. \citet{2023MNRAS.522.5107B} examined the sSFR of massive galaxies (10$^{10}$ M$_{\odot}$ < M$_{*}$ < 10$^{12}$ M$_\odot$) at z $\leq$0.2 as a function of distance from the nearest companion, using the TNG100-1 and TNG300-1 simulations. They discovered that galaxies with nearby star-forming companions exhibit significantly higher sSFR by a factor of 2.9 $\pm$0.3 in TNG100-1 and 2.27 $\pm$0.06 in TNG300-1 compared to isolated counterparts. These enhancements are evident out to separations of $\sim$300kpc. Conversely, galaxies with passive companions in TNG300-1 show modest sSFR suppression ($\sim$12\%) at separations of 100-300 kpc, and slight increases at distances below 50kpc.

Although these earlier studies examine galaxy interactions in great detail, they still have limitations. Firstly, the redshift interval examined is limited only to z < 1 or z < 2, while the peak of the star formation rate is observed at a higher redshift \citet{Hopkins_2006}. Furthermore, they examine the closest member to the galaxies, while in overdensities more galaxies could interact with each other. Based on this, the following questions remain open: How did the impact of galaxy collisions change in the early universe, and how did this affect the evolution of star formation rate? How are these compared to the recent JWST results? How dense is the environment in which colliding galaxies are located? What effect do large numbers of dwarf galaxy collisions have on massive systems?

The structure of this article is as follows: it begins with a theoretical introduction, providing a brief overview of the evolution of galaxy mergers. The second chapter presents the IllustrisTNG simulation, with a focus on the TNG300-1 model and the Merger Tree data structure used for analysis. In the results section, we first examine the spatial distribution of galaxies and investigate the evolution of galaxy mergers, including key properties such as star formation rate, stellar mass, and gas mass. The latter part of the results focuses specifically on the merger histories of the most massive subhalos. Finally, in the discussion, we compare our findings with recent observations from the James Webb Space Telescope (JWST), leading into our conclusions. 


\section{Methods}

\subsection{IllustrisTNG}

The IllustrisTNG project is the modern version of the original Illustris project, which includes the newest physical knowledge about the evolution of the galaxies. The TNG simulation uses the AREPO code \citet{2010MNRAS.401..791S} and takes the magneto-hydrodynamical and gravitational equations into account \citet{2018MNRAS.473.4077P}, \citet{2017MNRAS.465.3291W}. To investigate the evolution of the galaxies, the TNG simulation uses the following cosmological parameters and calculates for a Newtonian gravitational field and expanding Universe: $\Omega_{\Lambda,0} = 0,6911$, $\Omega_{m,0} = 0,3089$, $\Omega_{b,0} = 0,0486$, $\sigma_{8} = 0,8159$, $n_{s} = 0,9667$, h = 0,6774. Furthermore, the TNG model includes the following physical processes to study the formation and evolution of galaxies: (1) Microphysical gas radiative mechanisms, including primordial and metal-line cooling and heating with an evolving background radiation field. (2) Star formation in the dense interstellar medium. (3) Stellar population evolution and chemical enrichment following supernovae Ia, II, and AGB stars, individually tracking elements: H, He, C, N, O, Ne, Mg, Si, and Fe. (4) Stellar feedback-driven galactic-scale outflows. (5) The formation, merging, and accretion of nearby gas by supermassive black holes. (6) Multi-mode blackhole feedback operating in a thermal 'quasar' mode at high accretion states, and a kinetic 'wind' mode at low accretion states. (7) The amplification of cosmic magnetic fields from a minute primordial seed field at early times. The project includes three different box-size simulations, the TNG50, TNG100 and TNG300. 
The TNG100 and TNG300 were first published, followed by the TNG50 \citet{2019MNRAS.490.3196P}, \citet{2019MNRAS.490.3234N}. The parameters of the three models are summarized in Table~\ref{tab:example_table}., where V is the box volume, L$_{box}$ is the simulated box side length, N$_{GAS}$ is the number of the gas particles, N$_{DM}$ is the number of the dark matter particles, N$_{TR}$ is the tracing particles number, m$_{baryon}$ is the particle mass of the baryonic matter and m$_{DM}$ is the particle mass of the dark matter. 
Each TNG simulation is divided into a different number of snapshots, which are fixed points in the Universe's history. Using these snapshots, we can track the evolution of the galaxies. 
In our work, we are using the TNG300-1 simulation to investigate the evolution of the spatial distribution of the galaxies. This simulation incorporates the cosmological parameters described earlier (Planck2015), the physical models from the TNG simulation, whose parameters are summarised in Table~\ref{tab:example_table2}. The TNG300-1 is divided into 100 snapshots, which are taken between z=20.05 (13.624 Gyr Lookback time) and z=0.00 (present time). The time distribution of the snapshots is approximately uniform, about 150$\pm$50 million years. Using the IllustrisTNG simulation, we show that through the merger events of two galaxies, the average star formation rates in the descendant galaxies are higher than before the interaction. We investigated the cosmic evolution of the properties of progenitor and descendant galaxies and analyzed the spatial distribution of galaxy mergers and their neighborhood. We have to note that for the galaxy mergers only those galaxies have been investigated, which have a cosmological origin, and their stellar mass is larger than 10$^9$M$_{\odot}$. However, for calculating the density parameter, all halos from the FoF group catalogs were used. Since the numerical resolution limit of the galaxies may impact the spatial distribution of subhalos, we examined galaxy distances under different mass cuts, as detailed in Appendix~\ref{appendix B}.

\subsection{Galaxies and the Merger Tree} \label{Merger Tree}

The merger tree in the IllustrisTNG simulation concludes the data of galaxy merger history. Using this data file, we can study the colliding galaxies, before and after the merging event. The merger tree is present in the different IllustrisTNG simulations, therefore, the TNG300-1 has its own merger tree. The merger tree uses the Sublink \citet{2015MNRAS.449...49R} and LHaloTree \citet{2005Natur.435..629S} algorithm, the latter is almost identical to the method used in the Millennium and Aquarius simulations, but in HDF5 format. In the following, we will describe the principles of the Sublink merger tree according to \citet{2015MNRAS.449...49R} and \citet{Pillepich_2017}. 
First, the IllustrisTNG simulations use the Friends-of-Friends (FoF) algorithm to group nearby dark matter particles into halos. Within each FoF group, the SUBFIND algorithm identifies gravitationally bound substructures called subhalos. These subhalos contain not only dark matter but also baryonic components like stars, gas, and black holes. Subhalos with sufficient stellar mass (>10$^9$M$_{\odot}$) are classified as galaxies, with the most massive one in each FoF group typically designated the central galaxy. The merger tree used in the TNG simulation includes the group catalog files, in which the Subfind subhalos are present. These data files are identical to the galaxies.

A subhalo is considered a progenitor of another if and only if the second is its descendant. Each subhalo can have multiple progenitors but only a single descendant, reflecting the hierarchical nature of structure formation in the Universe. To identify the first progenitor of a subhalo, the one with the most massive history is selected. This method is more reliable than simply choosing the most massive subhalo at a given time, as it avoids errors when progenitors have similar masses. To efficiently organize this information, a linked-list structure is created for the simulation. Each subhalo is assigned five key pointers: the first progenitor (with the most massive history), the next progenitor (sharing the same descendant with slightly less mass history), the descendant, the first subhalo in the same FoF group (typically the most massive one), and the next subhalo in the FoF group (ordered by decreasing mass history). This structure is stored using a depth-first method into files, where each merger tree consists of subhaloes connected by progenitor/descendant relationships or FoF group membership. These trees are independent of each other, enabling parallel processing of computationally intensive tasks, such as constructing halo merger trees.

For galaxy mergers, a unique descendant is connected to each subhalo in three steps. The first step is the identification of descendant candidates in the following snapshot for each subhalo. Those candidates that have common particles with the subhalo in question will be the unique descendants. In the next step, the descendant candidates will be scored based on a merit function that takes into account the binding energy rank of each particle. In the last step, the unique descendant is identified with the highest score.

It can happen that a small subhalo is passing through a larger structure, and the algorithm can not detect it because of the low-density contrast. To deal with this problem, it is allowed some of the subhalos to skip a snapshot to find a descendant. Once all descendant connections have been made, the main progenitor of each subhalo is defined as the one with the "most massive history" behind it. \\
Galaxies in different merger trees are independent of each other, therefore, they may have undergone a different evolution. We investigated this question in our previous paper and found that the average values of the merger trees are similar to each other \citet{2023CoSka..53d.153K}. As a consequence, it is not necessary to examine several trees to draw general conclusions about the galaxies' evolution, especially at high redshifts. 

To analyze the time development, we investigated the galaxy mergers at different redshifts and the local environment of those galaxies. We did not take into account redshifts above 15, because the simulation contains only a few galaxies at high redshifts. Using the merger tree, we selected interacting pairs with cosmological origin at each snapshot, which share the same Descendant in the next or following snapshot. In order to show the distribution of galaxy mergers, our analysis includes the most massive subhalos, which are considered galaxy clusters. 
Using the SubhaloID of each galaxy, we selected key parameters: position, star formation rate (SFR), stellar mass, and gas mass. The SFR was computed as the sum of the star-forming gas cells bound to the corresponding Subhalo. The total mass includes all bound cells, regardless of type. Stellar and gas masses were determined within the comoving radius that encloses half of the Subhalo’s total mass. The gas fraction was then derived from these values. Specific star formation rate was derived from the quotient of SFR and stellar mass (M$_{*}$) bound to the corresponding Subhalo. Finally, the star formation rate density (SFRD) was calculated as the amount of stellar mass formed per cubic megaparsec, expressed in units of M$_\odot$ yr$^{-1}$ Mpc$^{-3}$.

\section{Results}

\subsection{Position of the galaxies in the TNG300-1}

We investigated the change of positions and the different properties of the galaxies in the whole volume and, in a single merger tree in the TNG300-1 simulation. To map the three-dimensional environment of the galaxies in a merger tree, we used the group catalog files as well, which contain all of the galaxies in the simulation. 
We followed the method of \citet{Das_2023} in investigating the galaxies' spatial distribution. To measure the density of the three-dimensional local environment, they use the 5th closest neighbor's distance. They define the local density with the $\eta_{k}$ parameter \citet{1985ApJ...298...80C} in the following way:

\begin{equation}
    \eta_{k} = \frac{k-1}{V(r_{k})}
\end{equation}

where $r_{k}$ is the k-th closest galaxy's distance and V(r$_{k}$) = (4/3)$\pi r_{k}^3$ the volume, which belongs to the sphere with the radius r$_{k}$ [ckpc/h].

\begin{figure}[h!]
    \includegraphics[width=\columnwidth]{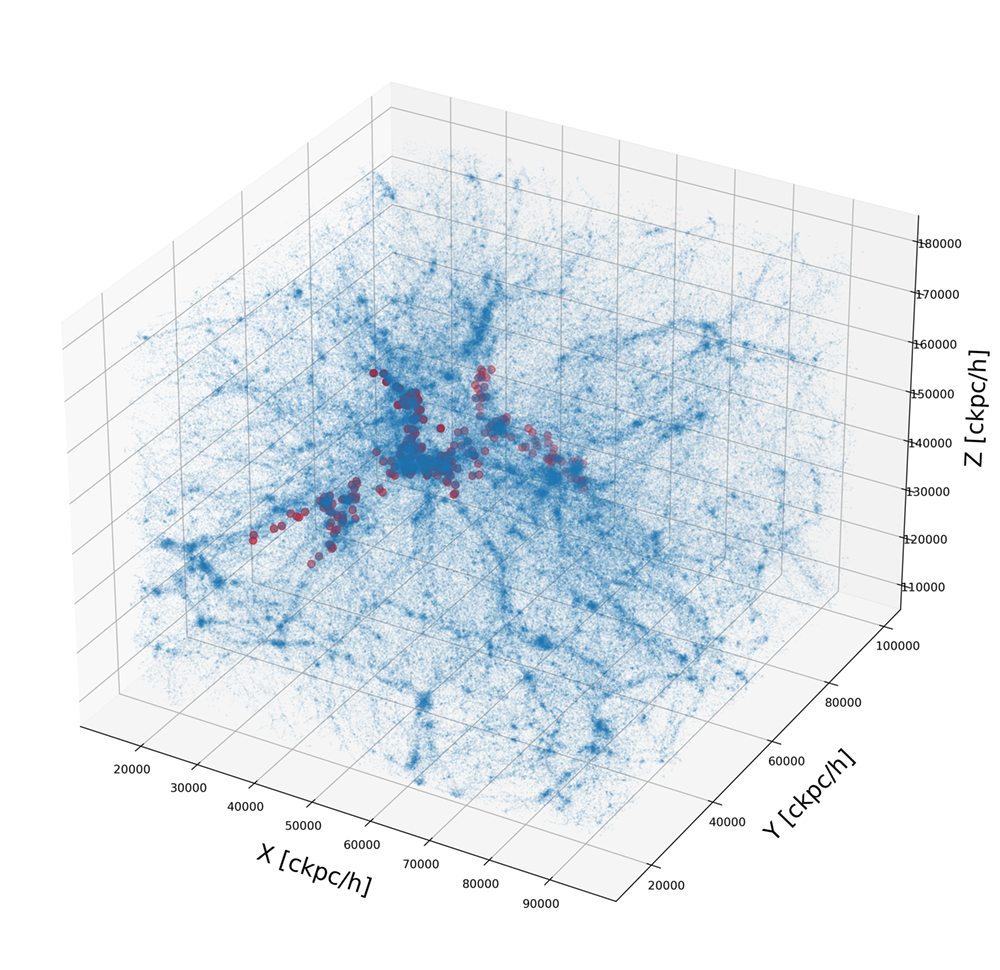}
    \caption{ Spatial distribution of the galaxy mergers in the 0th merger tree (red) and their neighborhood (blue points) at z = 0 redshift in the IllustrisTNG 300-1 simulation.  Spatial coordinates of galaxies (SubhaloPos) are represented by red/blue points. Galaxy collisions occur only in the densest regions. Only those galaxy mergers are shown which are connected to the chosen merger tree, therefore, no collisions can be seen in the surrounding knots.}
    \label{Figure1}
\end{figure}

\begin{figure}[h!]

	\includegraphics[width=\columnwidth]{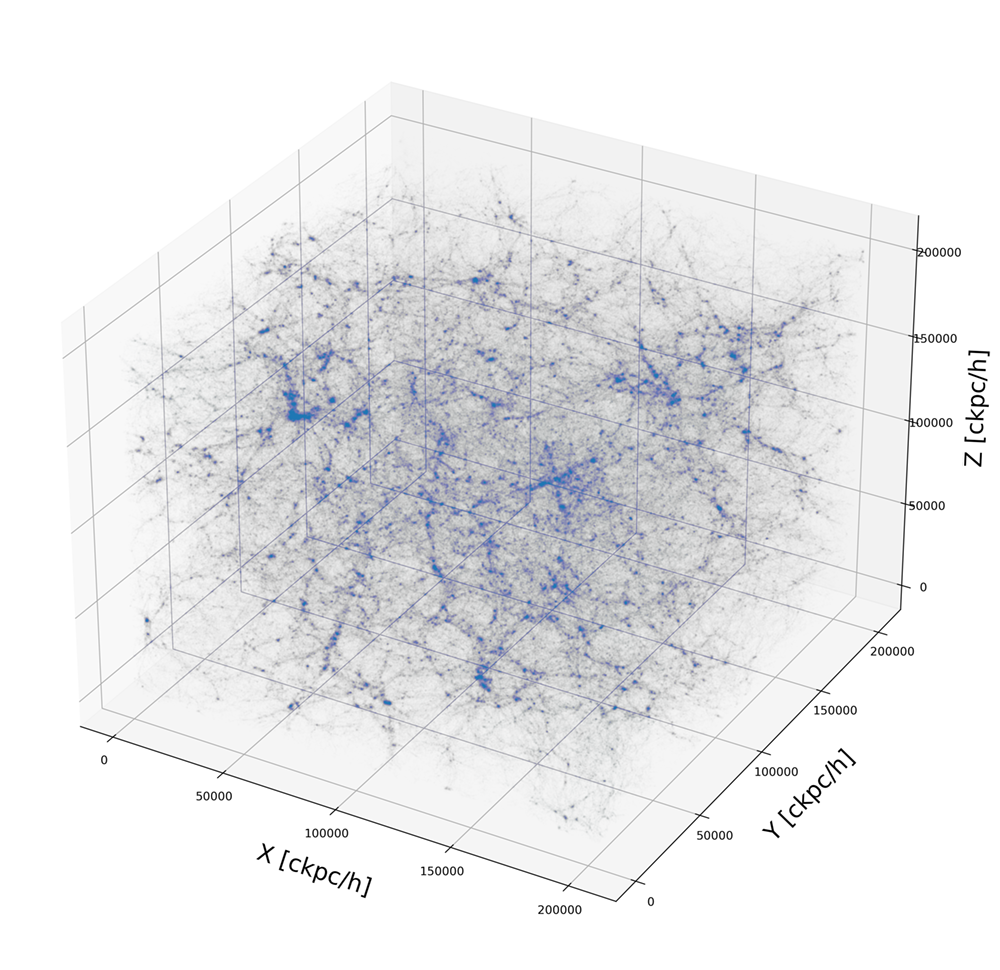}
    \caption{ Spatial distribution of the galaxies in the TNG300-1 simulation at z = 0 redshift. The spatial coordinates of galaxies (SubhaloPos) are represented by blue points. Galaxies form filamentary structures where they build clusters, between these filaments are lower-density regions and voids.}
    \label{Figure_2}
\end{figure}

\begin{figure}[h!]
    \centering
    \includegraphics[width=\columnwidth]{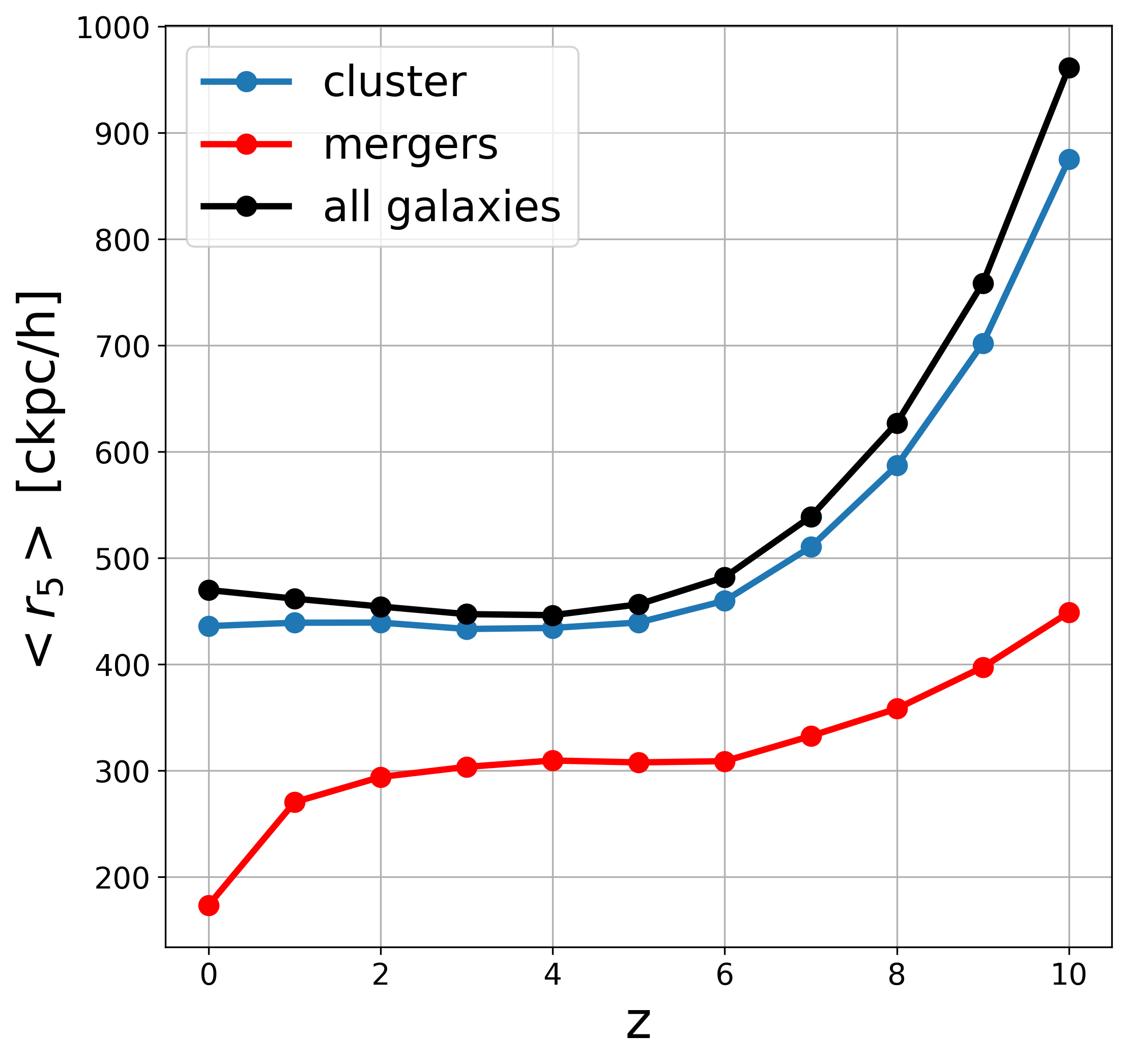}
    \caption{Average comoving distance of the 5th closest neighbor galaxy r$_5$ for all galaxies (black), in a galaxy cluster (blue) and for merger galaxies in a merger tree (red) at different redshifts. Galaxy mergers appear in much denser regions at all redshifts compared to the selected galaxy cluster.}
    \label{fig:r5-tng300}
\end{figure}

\begin{figure}[h!]
    \centering
    \includegraphics[width=\columnwidth]{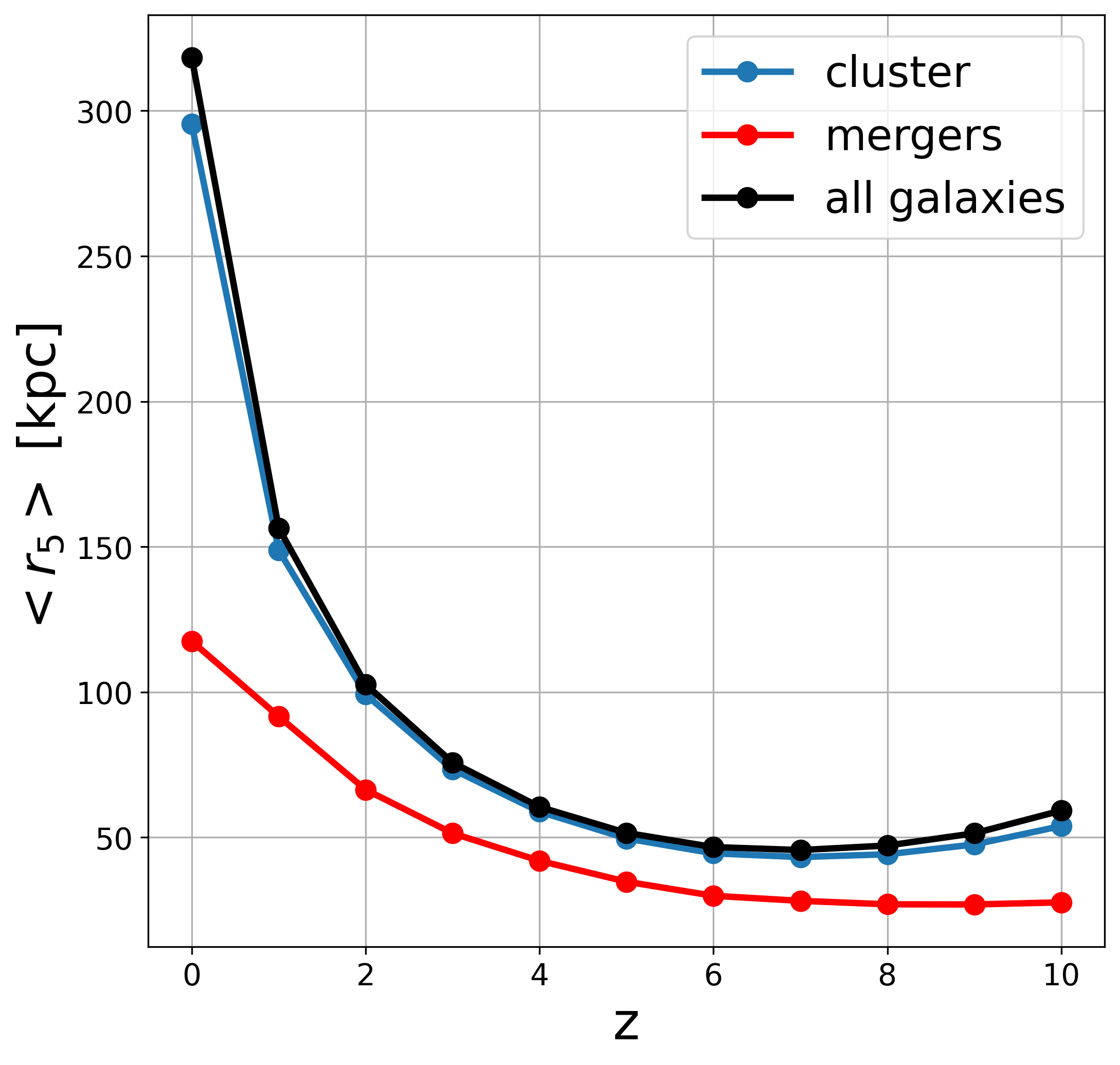}
    \caption{Average physical distance of the 5th closest neighbor galaxy r$_5$ for all galaxies (black), in a galaxy cluster (blue) and for merger galaxies in a merger tree (red) at different redshifts. Galaxy mergers appear in much denser regions at all redshifts compared to the selected galaxy cluster.}
    \label{Figure_phys}
\end{figure}

\begin{figure}[h!]

	\includegraphics[width=\columnwidth]{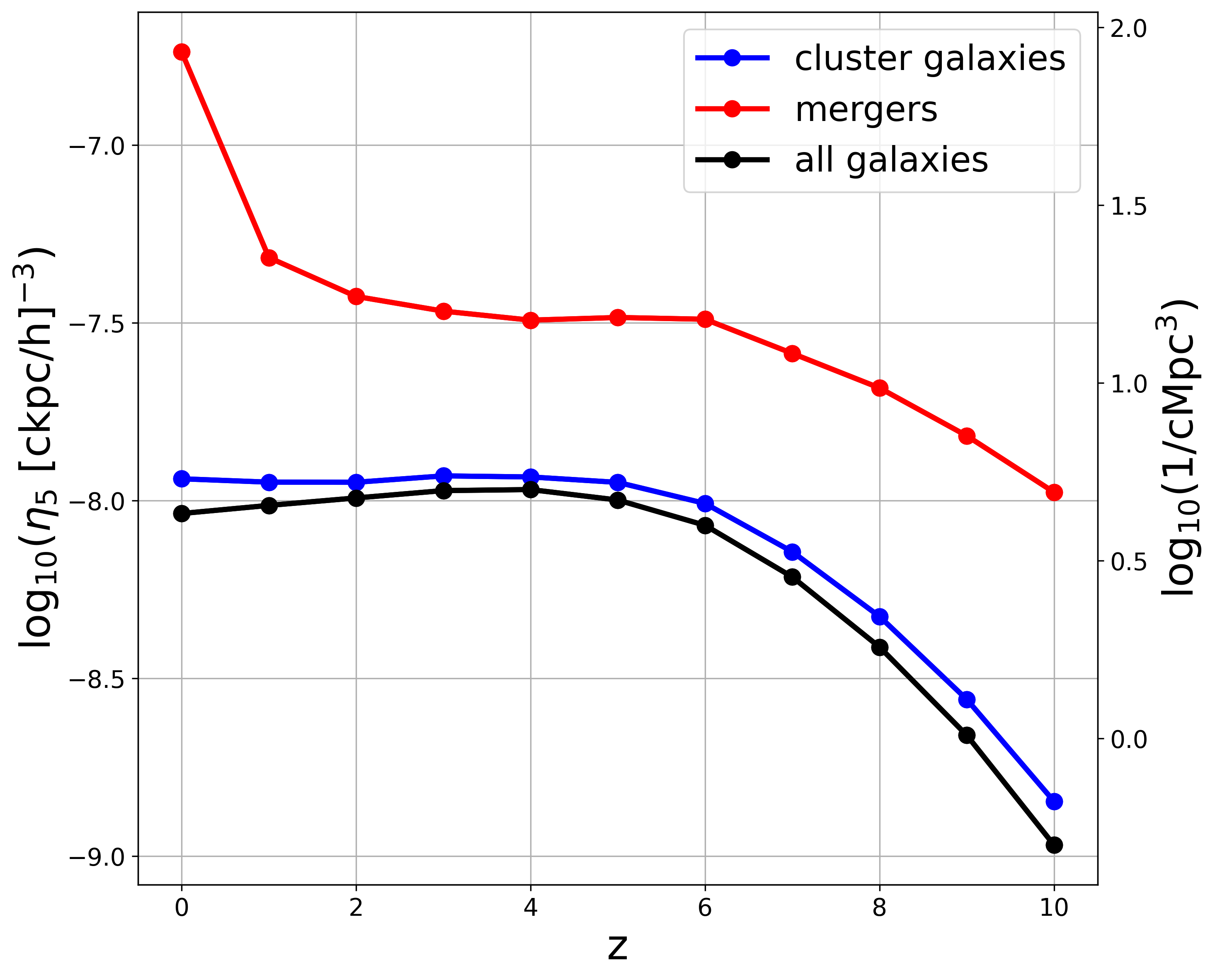}
    \caption{Comparison of the $\eta_5$ density parameters (left y axis) and galaxy number density (right y axis) average at different redshifts (z) for all galaxies in the TNG300-1 simulation (black), for a galaxy cluster (blue) and the galaxy mergers (red). The $\eta_5$ density parameter counts with the distance of the 5th closest neighbor for each galaxy, if its value is larger, the galaxies are closer to each other. Galaxy mergers appear in much denser regions at all redshifts compared to all galaxies. At z = 0, mergers occur only in the densest galaxy filaments.}
    \label{Figure_4}
\end{figure}

Using the spatial positions of the galaxies and the $\eta$ density parameter, we can analyze the evolution of the mergers' local environment. In Figure~\ref{Figure1}, the spatial distribution of the galaxy mergers of a merger tree and their local environment is shown. In Figure~\ref{Figure_2}, all of the galaxies' three-dimensional positions in the TNG300-1 simulation at the redshift z = 0 are shown. 

To show the evolution of the local environment, we calculated the comoving  (Figure~\ref{fig:r5-tng300}) and physical (Figure~\ref{Figure_phys}) distances  of the 5th closest neighbor r$_5$  and density parameter for all of the galaxies in the simulation and galaxy mergers Figure~\ref{Figure_4}.
On these figures, three different groups of galaxies are shown: "All galaxies": The density and r$_5$ parameter were calculated for all galaxies in the IllustisTNG simulation at the redshift in question. "Mergers": We selected the galaxy mergers at the given redshifts (See ~\ref{Merger Tree}). "Cluster": All galaxies in the allocated volume defined by the merger tree in question are counted.
Between redshifts z $\approx$10 and z $\approx$6, the Universe undergoes the period of cosmic dawn, marked by intense star formation and rapid galaxy buildup. From z $\approx$10 to z$\approx$6, comoving distances between galaxies decrease, indicating the early stages of clustering. Physical distances, however, remain nearly constant during this period, around 50 kpc for all galaxies and 30 kpc for merging systems. During this era, numerous small dark matter halos collapse, triggering the formation of the first galaxies and driving a sharp increase in the number of observable systems. This rapid growth leads to a peak in galaxy number density around z $\approx$6. According to the $\Lambda$CDM cosmological model, structure formation proceeds hierarchically: small systems form first and later merge to build larger galaxies. By z $\approx$6, a large population of small galaxies populates the Universe, but as time progresses, many of these begin to merge, reducing the total number of distinct galaxies. After z < 6, comoving distances for all galaxies stabilize, while those for merging galaxies continue to decline beyond z $\approx$3. In physical coordinates, galaxy separations begin to grow after z < 5, reaching up to $\sim$300 kpc for the general galaxy population and $\sim$130 kpc for mergers. Throughout this time, merging galaxies remain systematically closer to one another than the overall population, highlighting their presence in densely populated areas where interactions are more frequent. 
Although the density parameter includes the volume around galaxies, the average values are extremely sensitive to the number density of galaxies. Since we consider the close neighbors around each galaxy, if the majority of galaxies appearing in the total volume are arranged in clusters, their high number density significantly influences the overall average. Galaxies located in voids or near voids contribute negligibly to the average due to their low number density. As a result, when comparing the average density of a galaxy cluster with the total volume, this value will be lower, but only a small difference can be observed.

\subsection{Star formation rate and Gas}

\begin{figure}

	\includegraphics[width=\columnwidth]{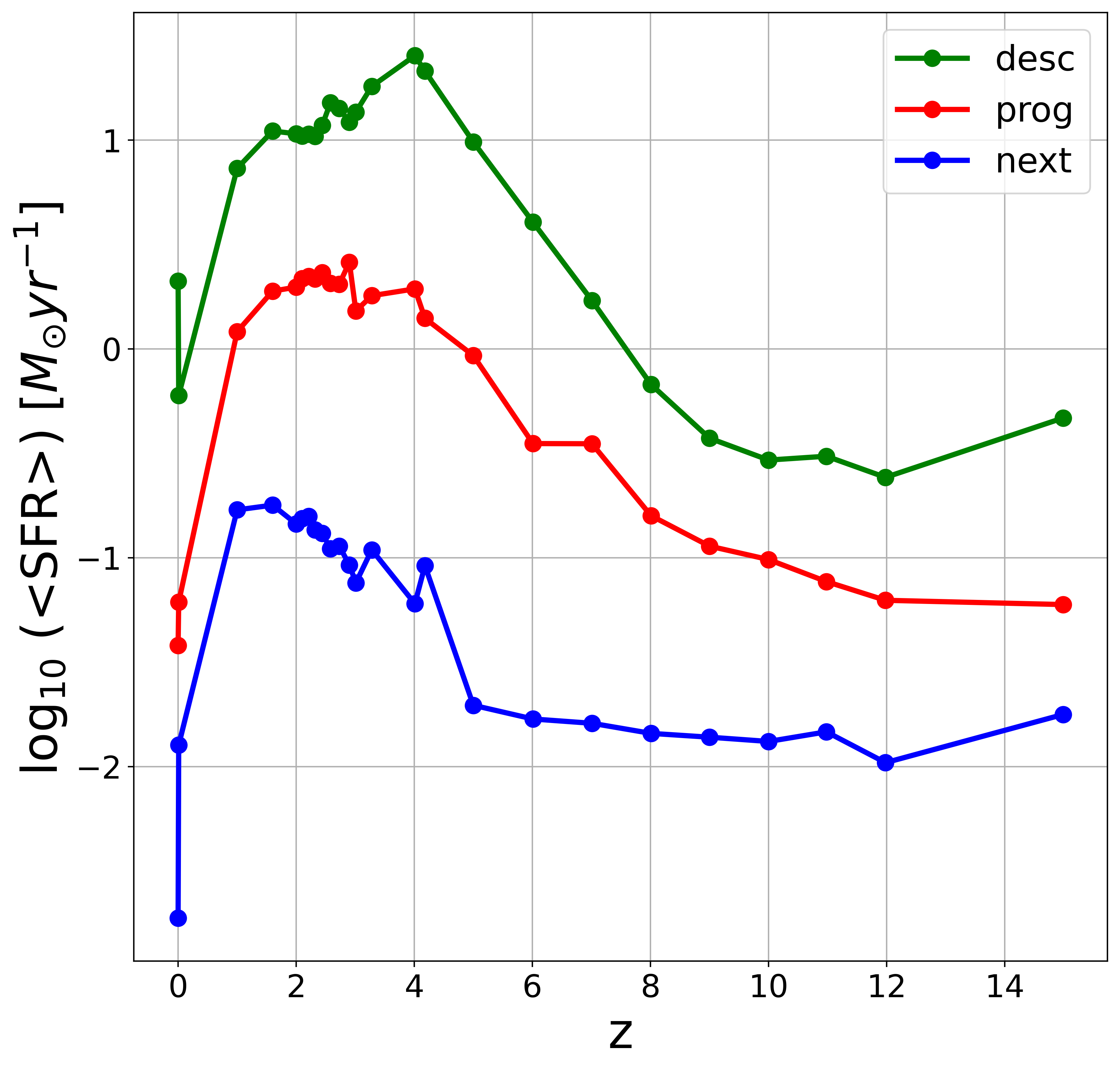}
    \caption{ Merger galaxies average star formation rate (<SFR>) on logarithmic scale versus redshift (z). Each point represents the average SFR at a given redshift. Progenitor galaxies (the most massive ones) are shown with red, NextProgenitors of these galaxies with blue, and their Descendant galaxies marked with green color.
    The star formation rates of the descendant galaxies have higher values at all redshifts than the progenitors. The peak of the curve is at around z = 4, where it is one order of magnitude higher than the Progenitors' value. }
    \label{Figure_5}
\end{figure}

From the different parameters first, we analyzed the star formation rates of the galaxies, comparing the progenitors and descendants values at different redshifts, Figure~\ref{Figure_5}. This figure shows the galaxy mergers mean star formation rate on a logarithmic scale versus the redshift. Galaxies after the merger event are represented by green color, which are the descendant ('desc') galaxies. Galaxies colored red are the progenitor galaxies, which selection method was described earlier.

\begin{figure}
	\includegraphics[width=\columnwidth]{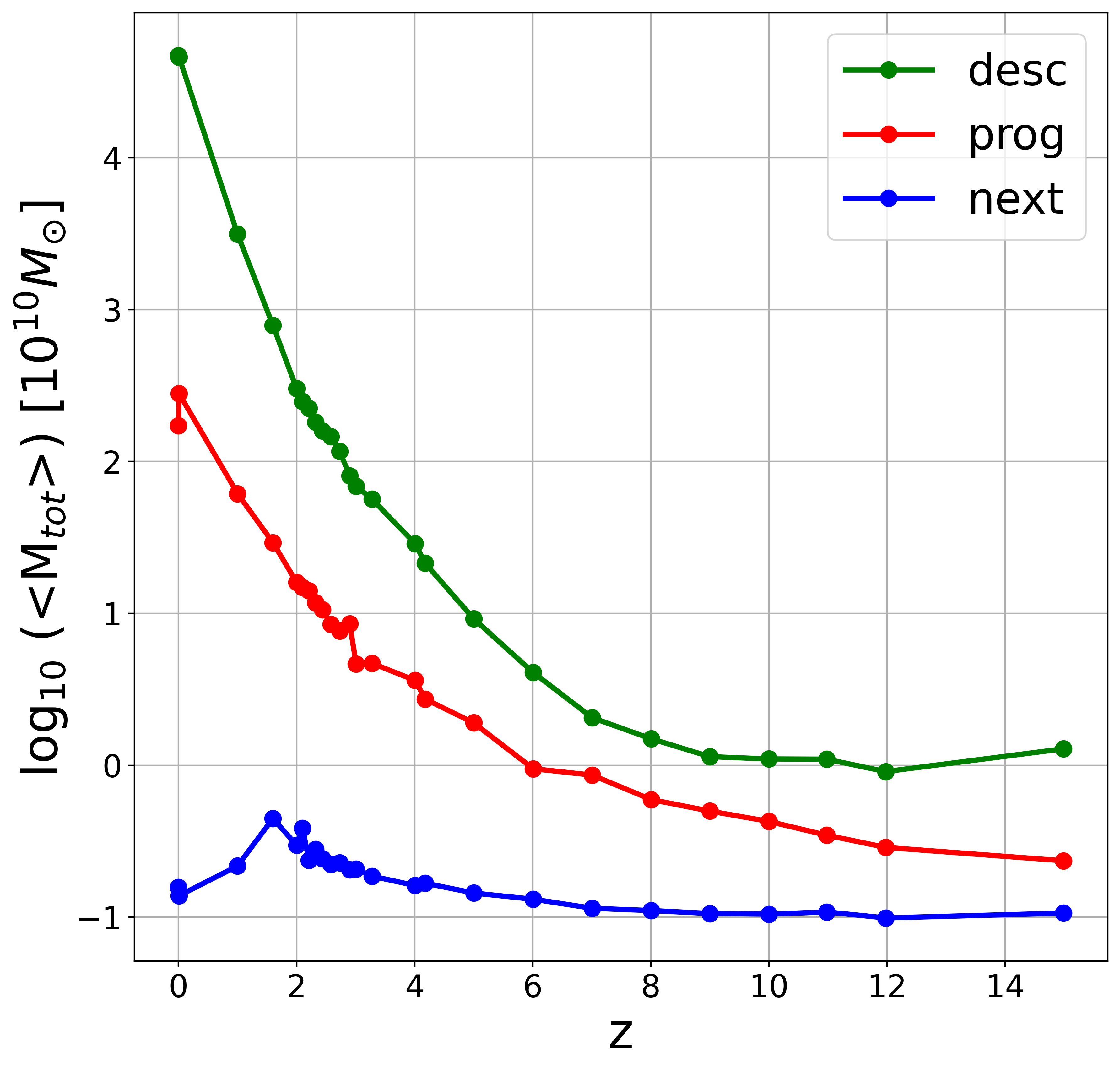}
    \caption{Merger galaxies average total mass (<M$_{tot}$>) on logarithmic scale versus redshift (z). Each point represents the average mass at a given redshift. Progenitor galaxies (the most massive ones) are shown with red, NextProgenitors of these galaxies with blue, and their Descendant galaxies marked with green color.
    The masses of the descendant galaxies have higher values at all redshifts than the progenitors.}
    \label{Figure_6}
\end{figure}

\begin{figure}
	\includegraphics[width=\columnwidth]{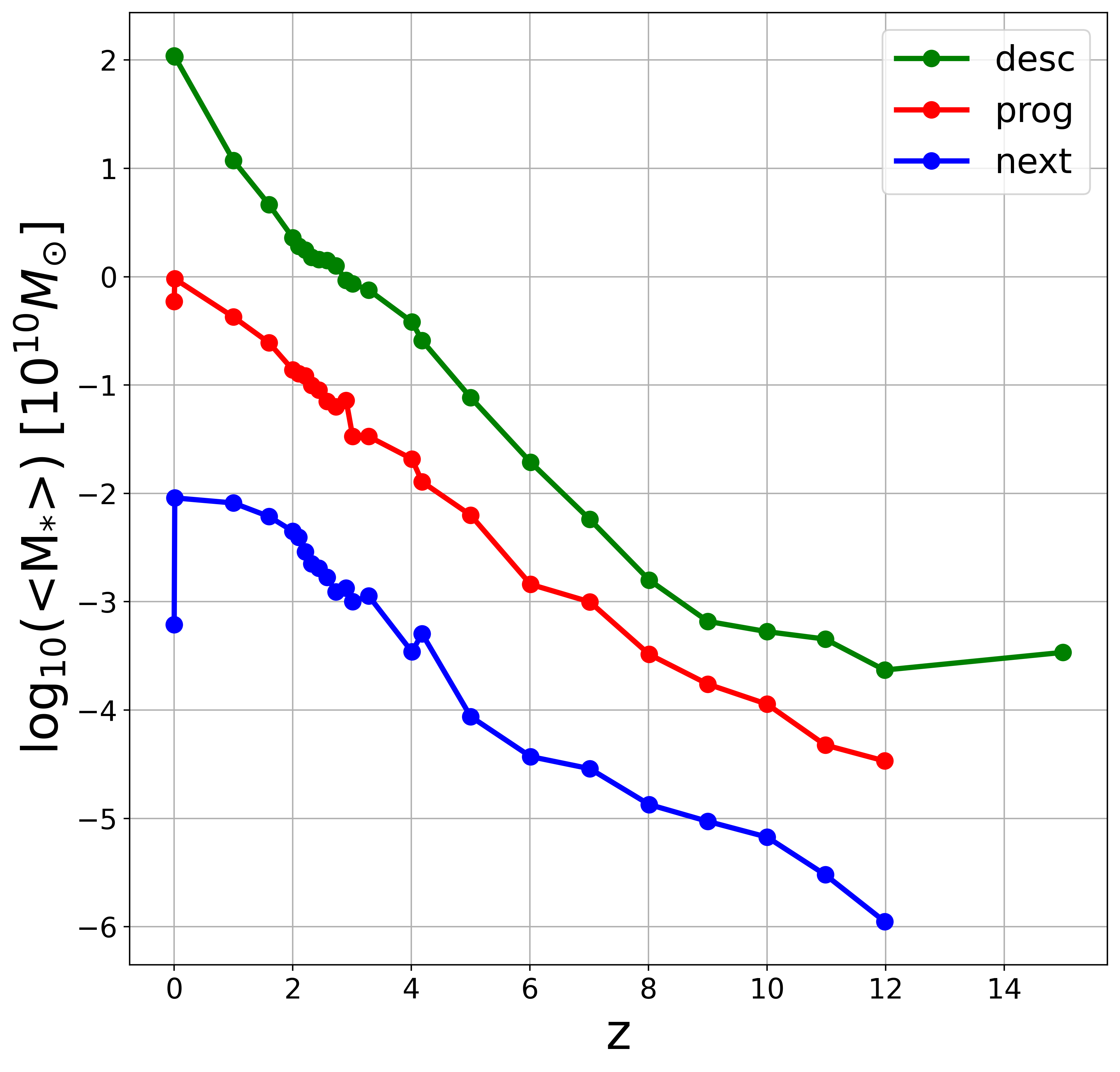}
    \caption{Merger galaxies average stellar mass (<M$_{*}$>) on logarithmic scale versus redshift (z). Each point represents the average stellar mass at a given redshift. Progenitor galaxies (the most massive ones) are shown with red, NextProgenitors of these galaxies with blue, and their Descendant galaxies marked with green color.}
    \label{Figure_7}
\end{figure}

\begin{figure}
	\includegraphics[width=\columnwidth]{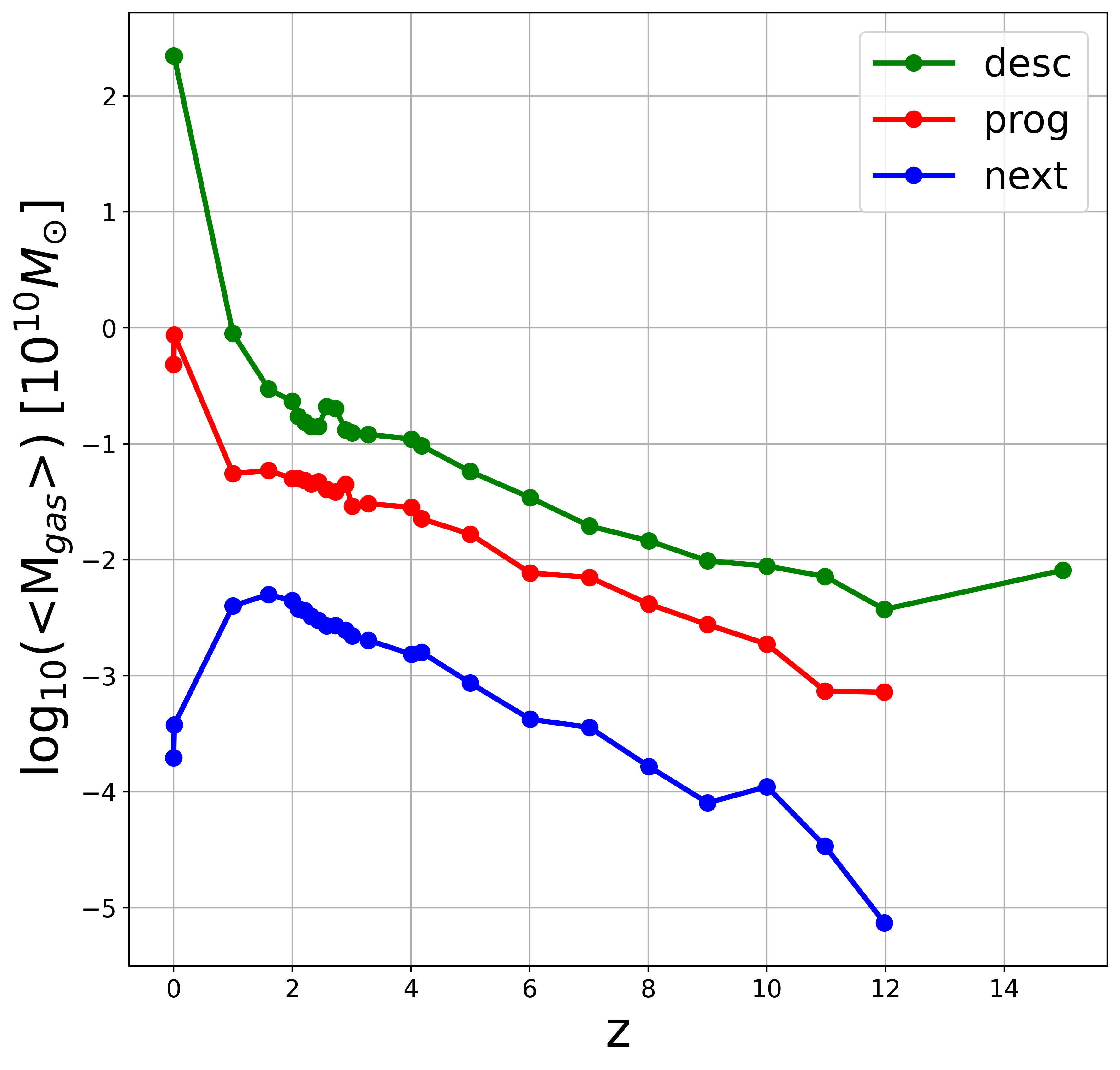}
    \caption{Merger galaxies average gas mass (<M$_{gas}$>) on logarithmic scale versus redshift (z). Each point represents the average gas mass at a given redshift. Progenitor galaxies (the most massive ones) are shown with red, NextProgenitors of these galaxies with blue, and their Descendant galaxies marked with green color.}
    \label{Figure_8}
\end{figure}

\begin{figure}
	\includegraphics[width=\columnwidth]{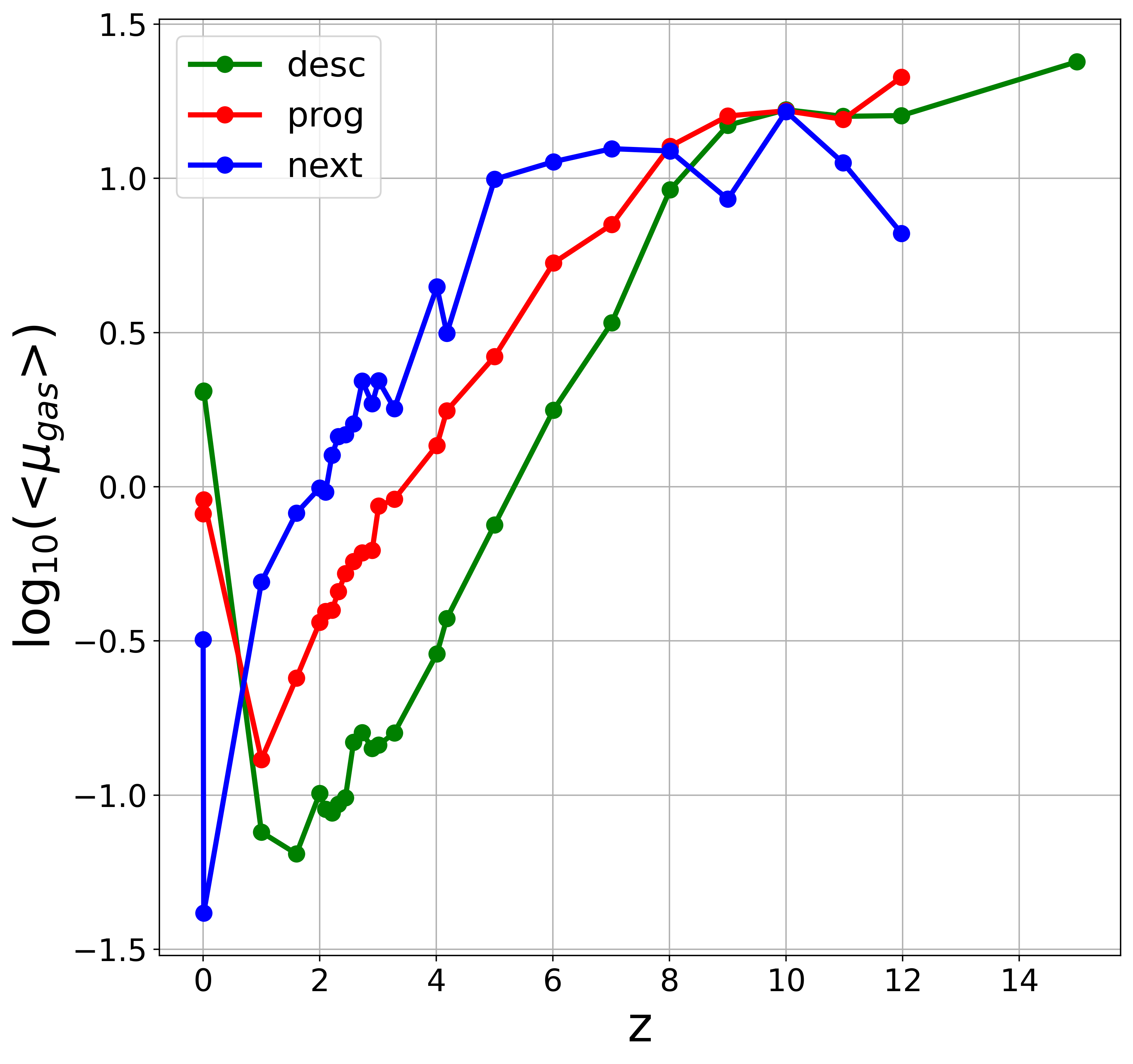}
    \caption{Merger galaxies gas fraction ($\mu_{gas}$) versus redshift (z), where $\mu_{gas}$ = M$_{gas}$/M$_{star}$. Each point represents the average gas fraction at a given redshift. Progenitor galaxies (the most massive ones) are shown with red, NextProgenitors of these galaxies with blue, and their Descendant galaxies marked with green color. At high redshifts, progenitors and descendant galaxies had more and similar amounts of gas fraction, at low redshifts, descendants have less gas compared to the stellar mass.}
    \label{Figure_9}
\end{figure}

\begin{figure}
	\includegraphics[width=\columnwidth]{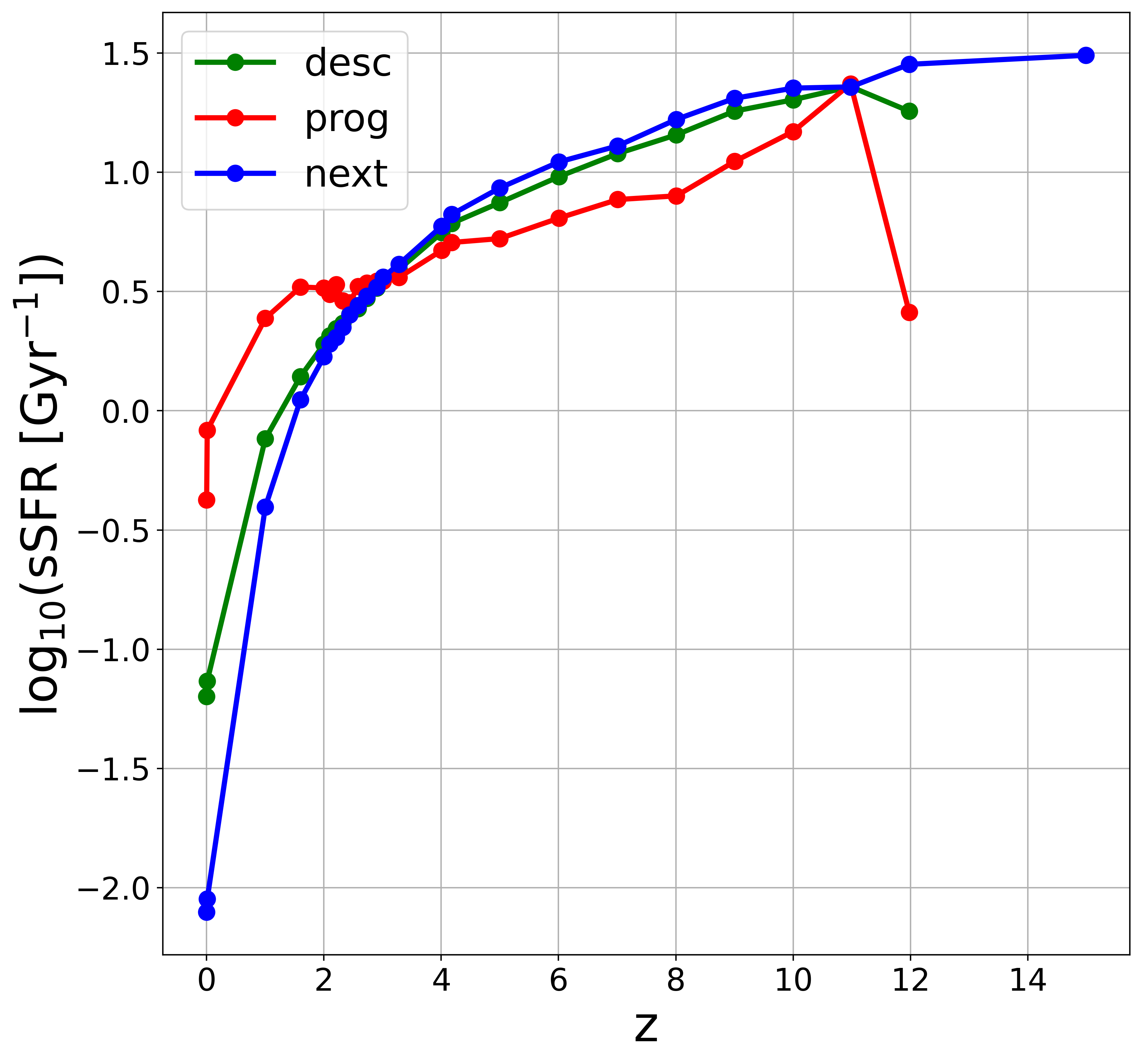}
    \caption{Merger galaxies specific star formation rate (<sSFR>) on logarithmic scale in Gyr versus redshift (z). Each point represents the average specific star formation rates at a given redshift. Progenitor galaxies (the most massive ones) are shown with red, NextProgenitors of these galaxies with blue, and their Descendant galaxies marked with green color.}
    \label{Figure_85}
\end{figure}

\begin{figure}
	\includegraphics[width=\columnwidth]{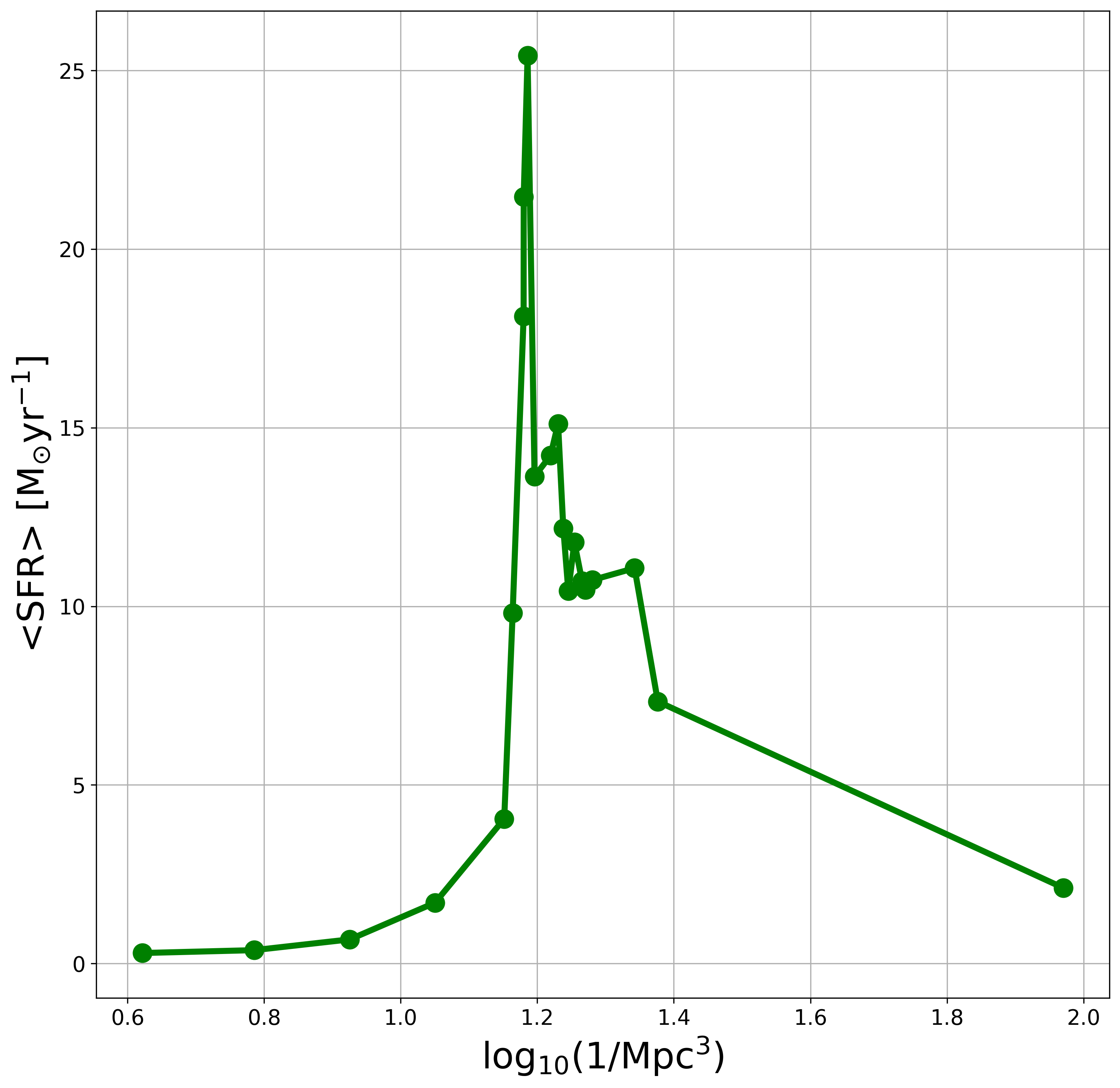}
    \caption{Galaxy mergers average star formation rate versus the galaxy density. Galaxies star formation is highly increasing at the density $\sim$16 galaxy/Mpc$^3$, but at more dense environments the SFR is decreasing. Extremely dense regions could cause the quenching of galaxies.}
    \label{Figure_10}
\end{figure}

NextProgenitor ('next') galaxies are shown in blue, which are the galaxies with the second largest mass history. The Progenitor galaxies were sorted by redshift, therefore, their value is consistent, but the Descendant galaxies can sometimes skip a Snapshot, and the NextProgenitors can occur in the previous or the following Snapshot as well. This means that to a selected redshift, not all of the Descendants are in the following Snapshot (they skip a Snapshot), and not all of the NextProgenitors are in the Snapshot in question, but in the previous one, or some cases in the following one. Despite these, the descendant galaxies are at lower redshifts than the Progenitors and NextProgenitors.    

Since we focus on the average values of the galaxy parameters, not the individual differences, we have used the following approximation: the average values of the NextProgenitor galaxies are shown at the same redshift as the Progenitors, and the Descendant galaxies are in the following Snapshot. It can be seen in Figure~\ref{Figure_5} that the galaxies' average star formation rates are higher than before the merger event at all redshifts. The Descendant galaxies' SFR are higher than the Progenitors at all redshifts. The Descendants SFR has its peak at around z $\approx$ 4, where there is a difference of an order of magnitude compared to the Progenitors, which supports the theoretical and observed hypothesis that through the merger of galaxies, the star formation rate is increasing. The Progenitor galaxies' SFR peak appears between z = 2-4, the Nextprogenitors are, by definition, smaller and younger galaxies than the Progenitors, therefore, their SFR is one order of magnitude smaller in most cases. The evolution of the star formation rate of different types of galaxies is similar, but the galaxy mergers have reached their peak at a higher redshift, while the lower mass galaxies have their peak later.

To answer the question of why there is a difference between the star formation rate history of the different types of galaxies, we investigated the galaxies' mass history. In Figure~\ref{Figure_6} the galaxies' average total mass is shown. The progenitors and descendants galaxies' average mass increases with time after z < 8, while the nextprogenitors' mass remains approximately the same. There is only a slight increase at 0 < z < 7, with a peak at z = 2. Compared to the SFRH, there is no correlation between the average total mass and the star formation rates. Descendant galaxies have higher mass than other galaxies. This is because massive galaxies are involved in several galaxy collisions, therefore their average mass is also higher, especially at low redshifts, where most of the galaxy mergers take place at the center of galaxy clusters.  

In Figure~\ref{Figure_7} and Figure~\ref{Figure_8}, the average stellar mass and the average gas mass are shown, respectively. The stellar mass of galaxies is increasing with time monotonic, like the gas mass. Only the NextProgenitor galaxies have one order of magnitude lower gas mass at redshift 0 than at z = 1 or 2. Smaller progenitor galaxies have less gas content than the trend would indicate. As we look at the gas fraction of these galaxies  Figure~\ref{Figure_9}, it can be seen that until z = 6, Progenitors and Descendants had the same fraction, after which the descendant galaxies' gas fraction became the lowest. In general, all of the galaxies' gas fraction is decreasing, which explains the higher star formation rate.

In Figure~\ref{Figure_85}, we present the time evolution of the specific star formation rate (sSFR = SFR/M${*}$) to explore how star formation efficiency changes over cosmic time. Across all galaxy types, the sSFR exhibits a similar evolutionary trend. At early times, galaxies display relatively high sSFR values, with log${10}$(sSFR [Gyr$^{-1}$]) exceeding 1. However, as the universe evolves, the sSFR steadily declines. By redshift z = 6, the sSFR has already dropped below 1, indicating a reduction in star formation activity per unit stellar mass. Throughout this evolution, the descendant and next-progenitor galaxies maintain nearly identical sSFR values across all redshifts. Progenitor galaxies, however, tend to show slightly lower sSFR values at high redshifts, while at lower redshifts, their sSFR becomes somewhat higher. In the redshift range 0 < z < 1, the log sSFR values lie between –2.0 and 0, marking a significant decline in star formation efficiency. This drop in sSFR at z < 1 can be attributed to multiple factors. One major contributor may be the increase in stellar mass of the most massive galaxies, particularly among the descendants. However, the decline in star formation rate (SFR) itself appears to play an even more critical role during this period. Interestingly, not only does the gas fraction rise, but the absolute gas mass also increases significantly. This accumulation of gas is likely linked to the high frequency of galaxy mergers occurring in this epoch, a topic that will be discussed in more detail in the following section.

In Figure~\ref{Figure_10}, we analyze the relationship between the SFR of descendant galaxies and their surrounding local galaxy density. At relatively low local densities, specifically below  $\approx$12 (galaxy/Mpc$^3$), we observe minimal star formation activity associated with galaxy mergers. However, as the local density increases into the intermediate range of roughly 14 to 25 galaxy/Mpc$^3$, there is a marked and rapid increase in star formation, suggesting that moderate-density environments may be more conducive to triggering star-forming events, possibly due to increased interactions and gas inflow. Interestingly, beyond a threshold density of about 25 galaxy/Mpc$^3$, the star formation rate begins to decline. This suppression of star formation at higher densities may indicate the onset of environmental quenching mechanisms, such as ram-pressure stripping, or feedback from active galactic nuclei (AGN), which inhibit the gas supply necessary for continued star formation. These trends highlight the complex interplay between environment and star formation in galaxy evolution.\\

\subsection{Massive subhalos}

The most massive central galaxies are key players in the evolution of galaxy clusters. Due to their dominant positions and large gravitational influence, these galaxies tend to be involved in a significantly higher number of interactions and collisions compared to less massive systems. This high frequency of mergers and dynamical activity makes them particularly valuable for studying the effects of galactic interactions on galaxy evolution. Unlike smaller galaxies that may experience only a handful of minor encounters, these massive systems provide a more dynamic environment where the consequences of collisions can be observed more clearly.
For this reason, we have conducted a detailed investigation of three massive subhalos: ID0, ID736368, and ID1018388. Our analysis focused primarily on the stellar mass growth and gas mass evolution throughout cosmic time. In Figure~\ref{Figure_A1} (Appendix), we present a comparison of the star formation rate (SFR), shown in red, and the gas fraction, displayed in green, for each subhalo.
At high redshifts (z > 3), all three subhalos exhibit relatively high gas fractions, reaching values up to 10\%. This is indicative of gas-rich systems in the early Universe, when ample cold gas was available to fuel future star formation. However, during these early epochs, the galaxies show little to no star formation activity, which is why the SFR is not visible in the initial portion of the figure.
Starting around z $\approx$ 3, the star formation rate begins to rise noticeably across the galaxies. This increase coincides with a sharp drop in the gas fraction, suggesting that gas is being rapidly consumed and converted into stars. From z $\approx$ 3 to z $\approx$ 1, the gas fraction remains below its earlier levels, although it shows a gradual increase during this period. This trend may be attributed to a series of gas-rich mergers, which likely replenish the gas supply, as supported by evidence shown in Figures~\ref{Figure_A5}, where the number of mergers is determined by identifying all progenitor galaxies that merge into the same descendant galaxy at a given redshift. 
Similar patterns are also observed in the evolution of the total gas mass, as illustrated in Figures~\ref{Figure_A3}. These figures reveal that the gas mass generally declines as star formation ramps up, but shows periodic increases, again likely due to merging events or accretion from the surrounding environment.
Furthermore, the relationship between star formation rate and stellar mass is analyzed in Figure~\ref{Figure_A2}. Between redshifts z $\approx$ 6 and z $\approx$ 3, both the SFR and stellar mass grow at a similar rate, indicating a close connection between gas consumption and stellar mass buildup. This parallel growth eventually levels off and becomes more stable in the redshift range 1.5 < z < 3. While all three galaxies follow this general pattern of synchronized SFR and stellar mass increase, their evolutionary paths also reflect the influence of other complex events, such as feedback processes or environmental effects, which may modulate their growth trajectories.
Lastly, we calculated the gas depletion time, defined as t$_{depl}$ = M$_{gas}$/ SFR, to better understand how efficiently these galaxies convert gas into stars. The results are shown in Figure~\ref{Figure_A4}. In the early Universe, each subhalo exhibits a gas depletion time of approximately 10 Gyr, although this value shows considerable variability. As the Universe evolves toward lower redshifts, the gas depletion time decreases dramatically, dropping to around 0.1 Gyr. Interestingly, all three galaxies display this same overall trend, highlighting a universal shift toward more efficient star formation in the later stages of cosmic history.

\section{Discussion}

In our study, we investigated whether there is any relationship between the global merger history and the star formation history of the Universe according to the IllustrisTNG simulation.

If we compare the Progenitor and Descendant galaxies, we find that the latter have 1-2 orders of magnitude higher star formation rates than the Progenitors (Figure~\ref{Figure_5}).
The curve differs from the total average of all galaxies SFR in the Universe, with a maximum at around z $\approx$ 2-3  \citet{Hopkins_2006} since the peak of the mergers SFR is at higher redshift, z $\approx$ 4, and the NextProgenitor galaxies have their peak at lower redshift. Using the same simulation \citet{2020MNRAS.494.4969P} investigated merger galaxies' star formation rate and found that massive galaxies' star formation rate density is increasing by 14.5$\%$. \citet{2020MNRAS.493.3716H} investigated galaxy mergers and found that the SFR of the sample examined at 0 $\le$ z $\le$ 1 increased by a factor of 2, and they could not find an evolution in time at this interval.

\begin{figure}[h!]
    \centering
    \includegraphics[width=\columnwidth]{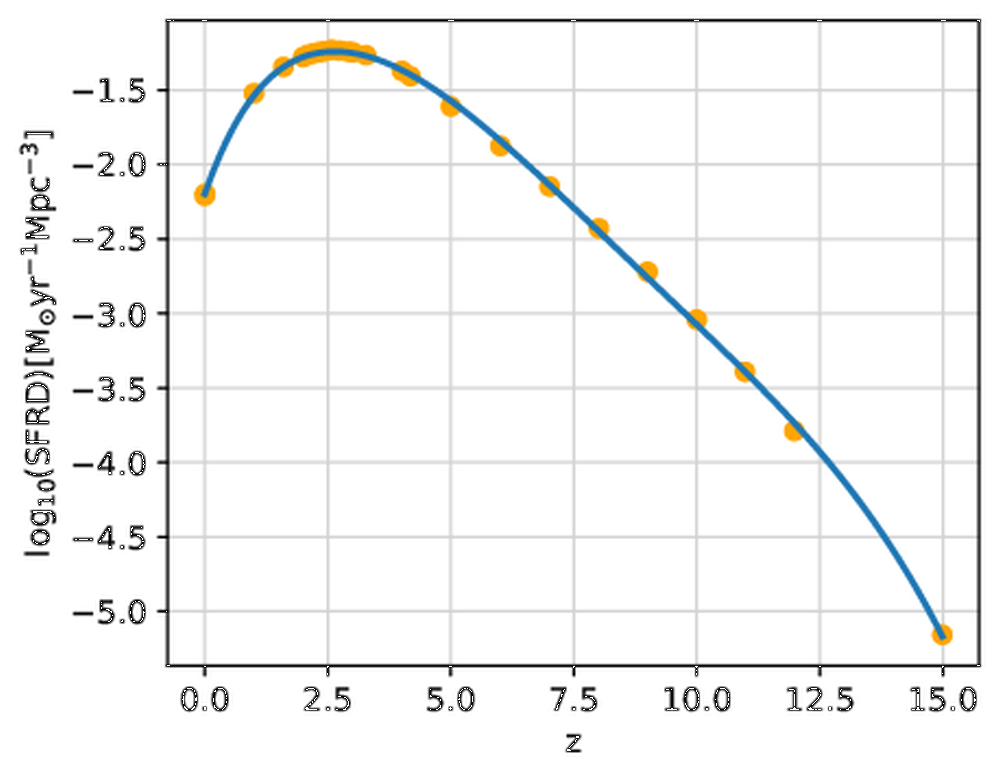}
    \caption{TNG300 cosmic star formation history for all galaxies in the simulation volume. We calculated SFRD values for each redshift z close to a whole number, with a few extra data points added close to the maximum. These values are shown as orange dots. We then fitted these data points with a quadratic B-spline with a smoothing parameter of 0.01, interpolated to 100 bins between $\mathrm{z_{\text{min}}}$ and $\mathrm{z_{\text{max}}}$. We found the bin with the maximum $\mathrm{log_{10}(SFRD)}$ value of $\approx -1.24$ at redshift z=2.57.}
    \label{fig:tng300-sfrd}
\end{figure}

Using the TNG300-1 simulation, we calculated the average star formation rate density for all galaxies at different redshifts (Figure~\ref{fig:tng300-sfrd}). We found that the SFRD peak is at z = 2.57. In our previous work, we examined the SFRD in the TNG100-1 simulation \citet{2023IAUS..373..318J}. If we compare the different Illustris simulations, we find that there is a slight difference in the SFRD history. The peak in the TNG100-1 is at z = 2.73, with a little higher value than in the TNG300-1 result, and in the Illustris simulation is at a bit lower redshift \citet{2018MNRAS.473.4077P}.

From the comparison of the galaxy merger star formation history, we can conclude that on average, Descendant galaxies have 1-2 orders of magnitude higher star formation rates than the Progenitors. The evolution of the specific star formation rate (sSFR) in merging galaxies reveals a notable decline in star formation efficiency over recent cosmic timescales, particularly within the redshift range of 0 < z < 2, which shows a good agreement with the recent COSMOS and GOODS surveys \citet{2015A&A...579A...2I}. This decline is primarily driven by a significant reduction in the star formation rate itself. Over a broader period, from z $\approx$ 8 to z $\approx$ 1, galaxies experienced a steady depletion of their gas content, largely due to intense star formation activity that consumed available gas. However, in the more recent epoch (0 < z < 2), interactions and mergers with gas-rich dwarf galaxies contributed to a resurgence in the gas supply and led to a marked increase in the gas fraction of these systems. Despite this influx of gas, the efficiency with which galaxies convert gas into stars has continued to drop, suggesting a decoupling between gas availability and star formation activity in the later stages of galactic evolution. An interesting next step would be to investigate non-merger galaxies that reside in environments similar to those of merging systems. Comparing their properties, such as gas fraction, SFR, and mass evolution, could offer valuable insights into the role of mergers in galaxy evolution. For instance, non-merger galaxies might retain higher gas fractions, exhibit steadier SFR, or follow different stellar mass growth trends compared to their merging counterparts.
In our study, we also analysed the evolution of galaxy environments by measuring the local density and the 5th nearest neighbor distance (r$_5$) for all galaxies, mergers, and cluster galaxies in the IllustrisTNG simulation across different redshifts. We found that, from z $\approx$10 to z$\approx$6, comoving distances between galaxies decrease, indicating the early stages of clustering. Physical distances, however, remain nearly constant during this period, around 50 kpc for all galaxies and 30 kpc for merging systems. After z < 6, comoving distances for all galaxies remain constant, while those for merging galaxies continue to decline beyond z $\approx$3. In physical coordinates, galaxy separations begin to grow after z < 5, reaching up to $\sim$300 kpc for the total galaxy population and $\sim$130 kpc for mergers. Moreover, galaxy mergers consistently occur in denser regions throughout the entire time interval.

\subsection{Comparison with JWST observations}

New JWST observations make it possible to study galaxies at earlier times than ever before. This also gives the opportunity to compare simulational results to observations at high redshifts which was not possible before the JWST era.

\begin{figure}[h!]
    \centering
    \includegraphics[width=\columnwidth]{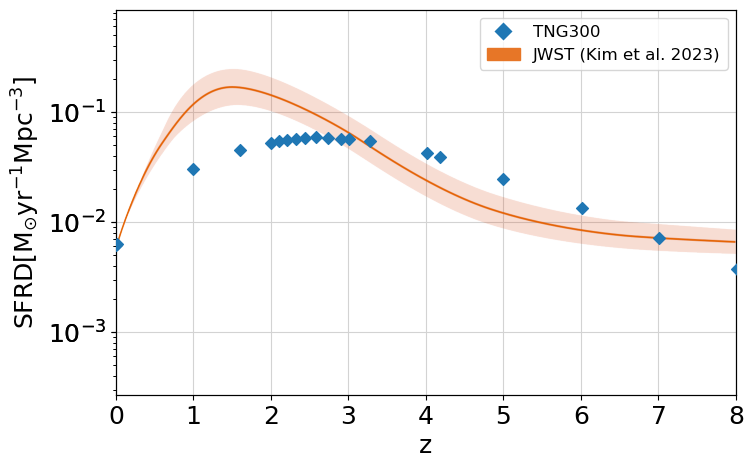}
    \caption{Comparison between the Cosmic Star Formation History (CSFH) derived by \protect\citet{KimCSFH2023} from recent JWST observations and our calculated SFRD data points for TNG300.
    }
    \label{fig:tng300-cfsh}
\end{figure}

\citet{KimCSFH2023} derived the cosmic star formation history (CSFH) using mid-IR source counts of JWST/MIRI observations (\citet{Ling2022_JWST_sourcecounts}; \citet{Wu2023_JWST_sourcecounts}) combined with previous source counts (\citet{Oliver1997_sourcecounts}; \citet{Serjeant2000_sourcecounts}; \citet{Pearson2010_sourcecounts}; \citet{Takagi2011_sourcecounts}; \citet{Pearson2014_sourcecounts}; \cite{Davidge2017_sourcecounts}). Their results show good alignment with previous works up until z $\approx$ 3, all results agreeing on the star formation rate density having a maximum between z = 1 and z = 2. We show a comparison between the CSFH of \citet{KimCSFH2023} and our calculated SFRD data points for TNG300 in Figure~\ref{fig:tng300-cfsh}.
We found a shallower peak of $10^{-1.24}$ at a higher redshift z = 2.57. 
This difference motivates further investigation into whether it may arise from the older CSFH census used to calibrate the TNG300 models or other factors.

\citet{Suess_minormerger_2023} examined massive ($\mathrm{log (M_{*} / M_{\odot}) \geq 10.5}$), quiescent galaxies in the JWST GOODS-S field at redshifts between z = 0.5 and z = 3, and found, that these galaxies have five companions within 35 kpc on average. This is significantly lower than what we calculated for the first merger tree of the TNG300 volume, which we show in Figure~\ref{fig:r5-tng300}. One source of this difference could be a high amount of very faint subhalos in the simulation above the detection limit of JWST, however, it is an interesting question for a possible future investigation.

Amongst the most spectacular findings of JWST is how early in the Universe we see already evolved galaxies. \citet{Atek2022_JWST_z16} revealed three galaxy candidates with masses $\mathrm{log (M_{*} / M_{\odot}) > 9}$ at redshift z between 8.88 and 9.78 in the JWST deep field towards the massive lensing cluster SMACS0723. \citet{Labbe2023_JWST_z7-9} found six galaxies with masses $\mathrm{log (M_{*} / M_{\odot}) > 10}$ in the redshift range from z = 7.4 to z = 9.1. Looking for these massive galaxies in the TNG300 volume, we found, that 1.36 \% of all the galaxies at redshift z = 9 have masses $\mathrm{log (M_{*} / M_{\odot}) > 10}$, with the first ones appearing as early as in the first snapshot at z = 20. This indicates that the TNG300 model allows for the early formation of massive galaxies.



\authorcontributions{Conceptualization, all; methodology, all; validation, all; formal analysis, B.K., A.P.J. and L.V.T.; investigation, all; resources, B.K. and A.P.J. ; data curation, all; writing---original draft preparation, B.K. and A.P.J.; writing---review and editing, all; visualization, B.K. and A.P.J ; supervision, L.V.T and A.B.; project administration, I.H. and L.V.T.}

\funding{This work was partially supported by the Project No. TKP2021-NVA-16 implemented with funding provided by the Ministry of Culture and Innovation of Hungary from the National Research, Development, and Innovation Fund.}

\dataavailability{The IllustrisTNG simulations, including TNG100-1 \& TNG300-1, are publicly available at \url{www.tng-project.org/data} (\cite{nelson2021illustristngsimulationspublicdata}). }

\acknowledgments{The authors would like to express their gratitude to Seong Jin Kim and colleagues for providing the data used in their results.

We gratefully acknowledge the travel support provided by the Physics Doctoral School and the Talent Support Council of Eötvös Loránd University, which made it possible to present this research and gain the attention of a broad scientific audience at the IAU General Assembly 2024, receiving valuable feedback and inspiration for the publication.

We are deeply grateful to Bruce Elmergreen for his insightful comments and suggestions.

We are also grateful to the High Energy Astronomy Research Team (HEART - \url{https://physics.elte.hu/KRFT_heart}) for their support and thoughtful feedback.}

\conflictsofinterest{The authors declare no conflicts of interest.} 


\appendixtitles{yes} 
\appendixstart
\appendix
\begin{appendix} \label{appendix}

\onecolumn
\begin{figure}
\section{Additional figures to the three investigated massive galaxies}
\centering
\includegraphics[width=7.5cm]{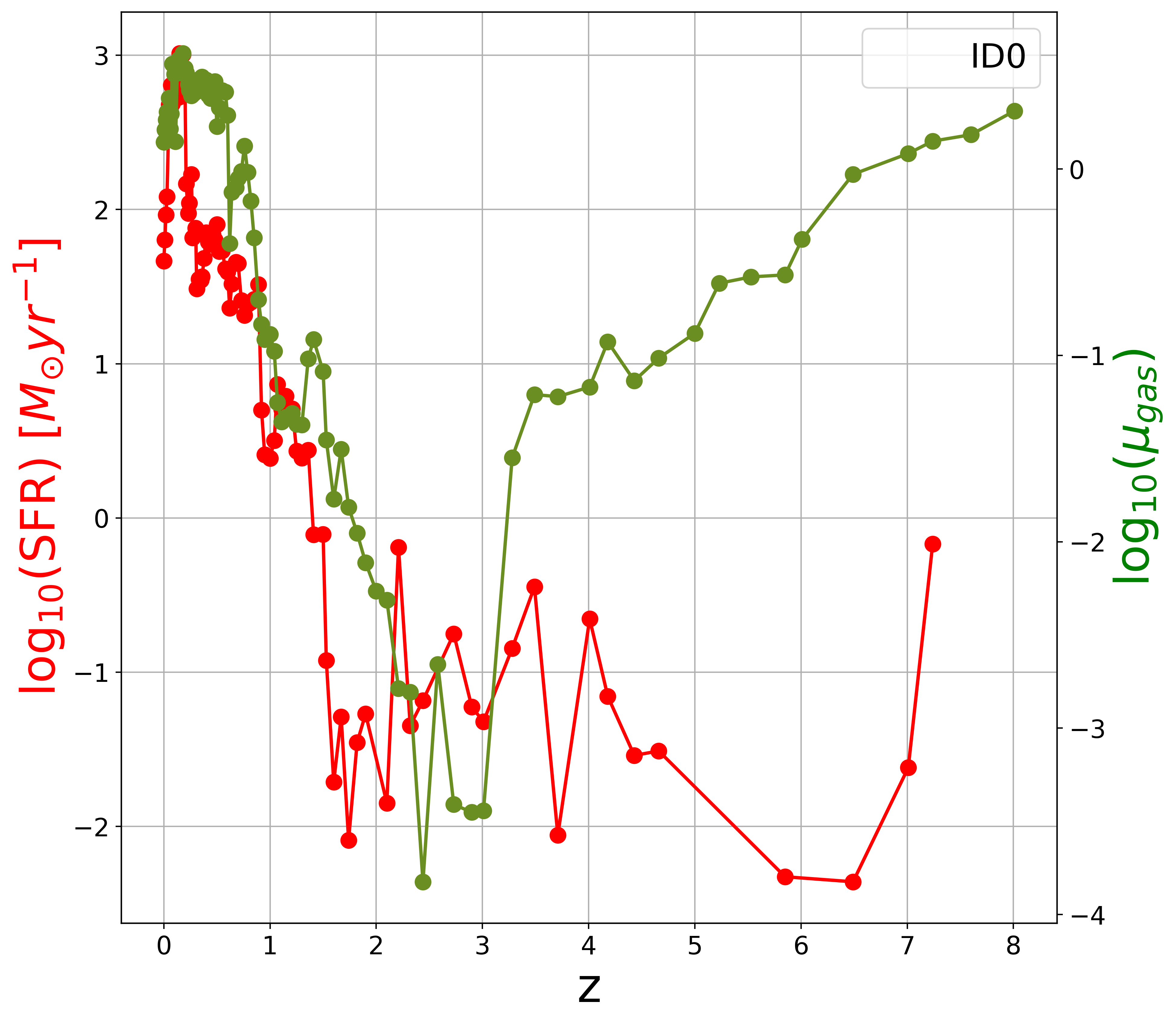} 
\\[3mm]
\includegraphics[width=7.5cm]{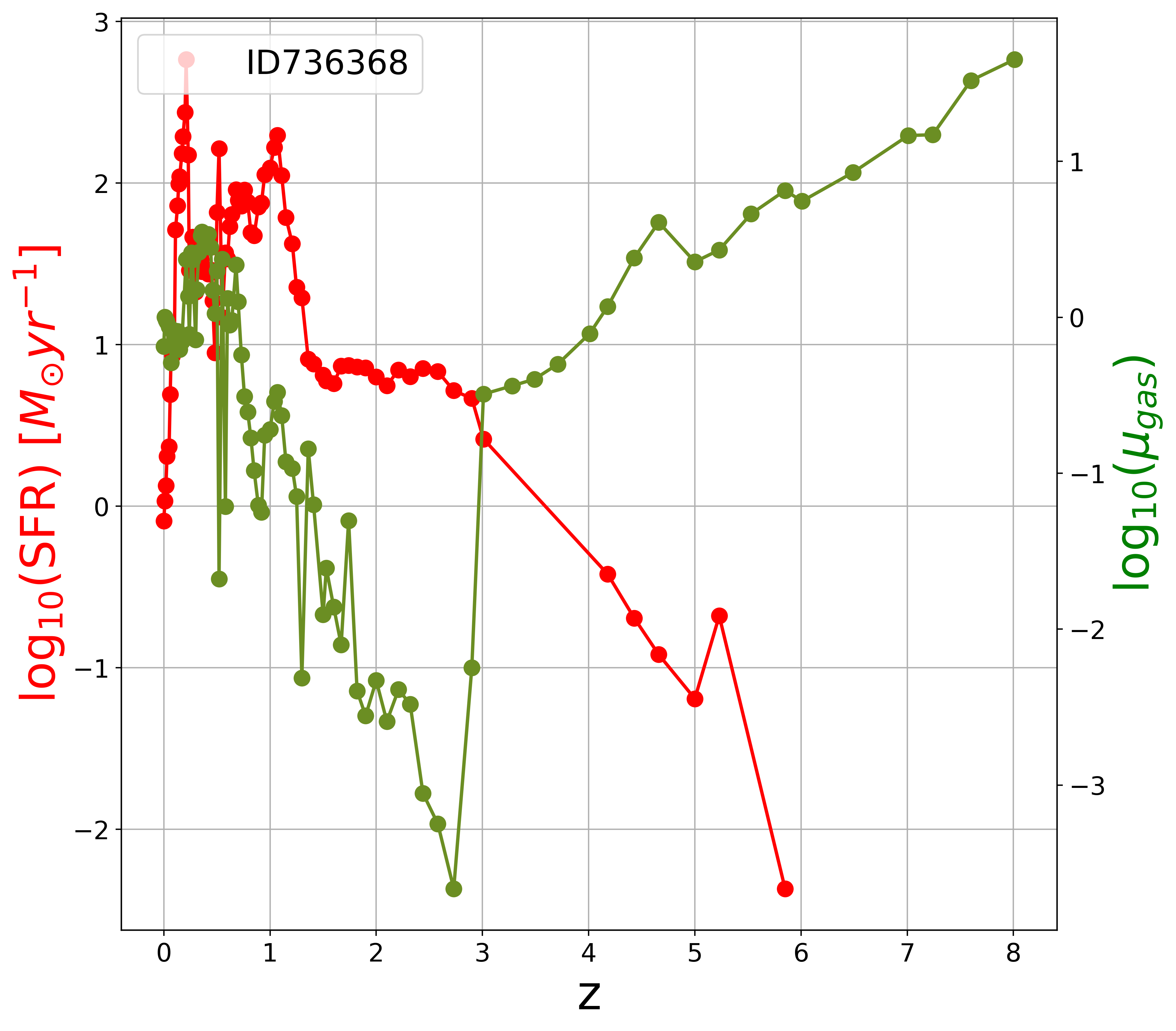}
\\[3mm]
\includegraphics[width=7.5cm]{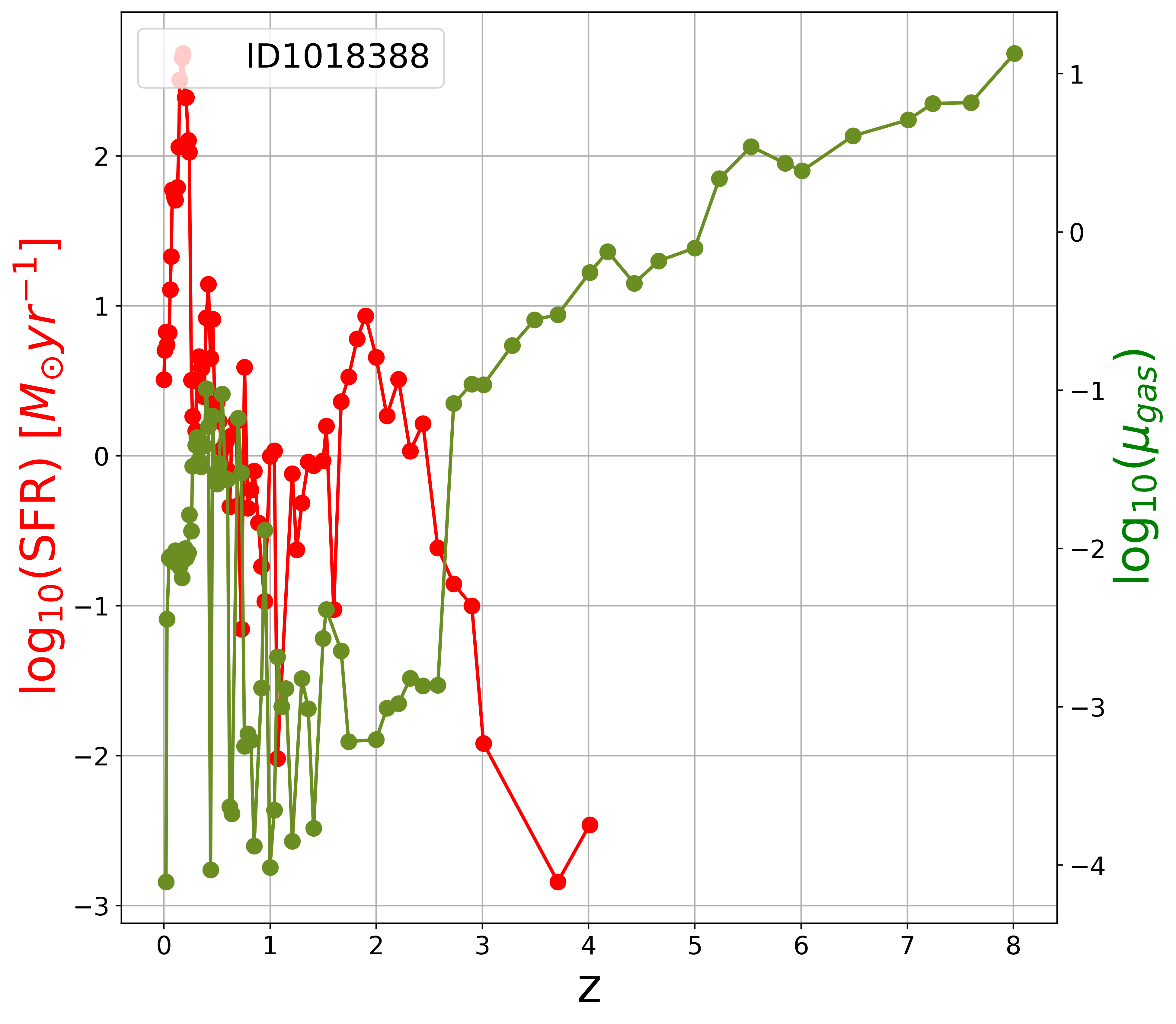} 
\caption{Star formation rate history (red line) and the gas fraction of galaxies (green dots) versus the redshift at 0 < z < 8. On the three panels, different subhalos are shown: ID 0 (top), ID 736368 (middle), and ID 1018388 (bottom).}
\label{Figure_A1}
\end{figure}

\begin{figure}
\centering
\includegraphics[width=8.5cm]{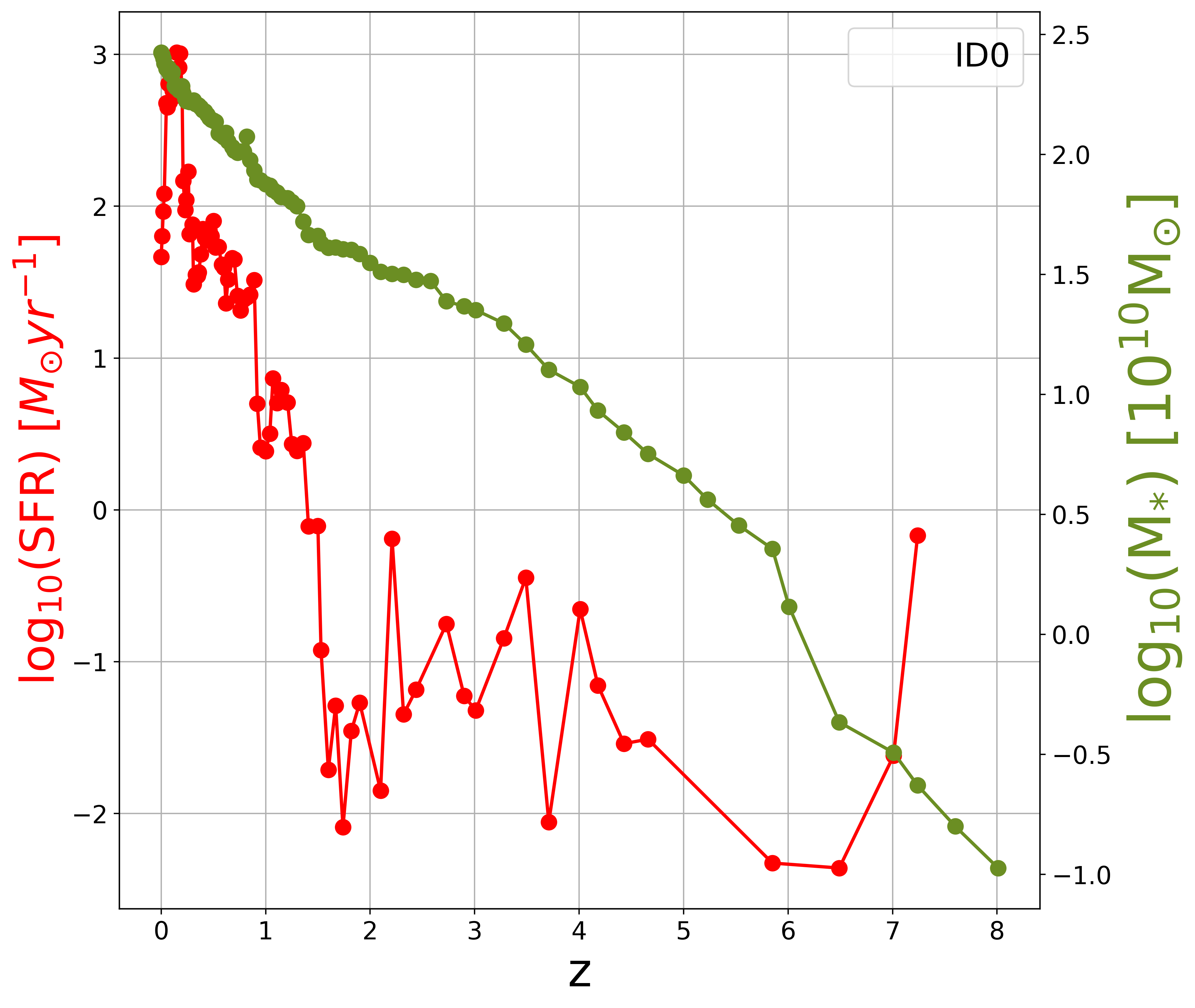} 
\\[3mm]
\includegraphics[width=8.5cm]{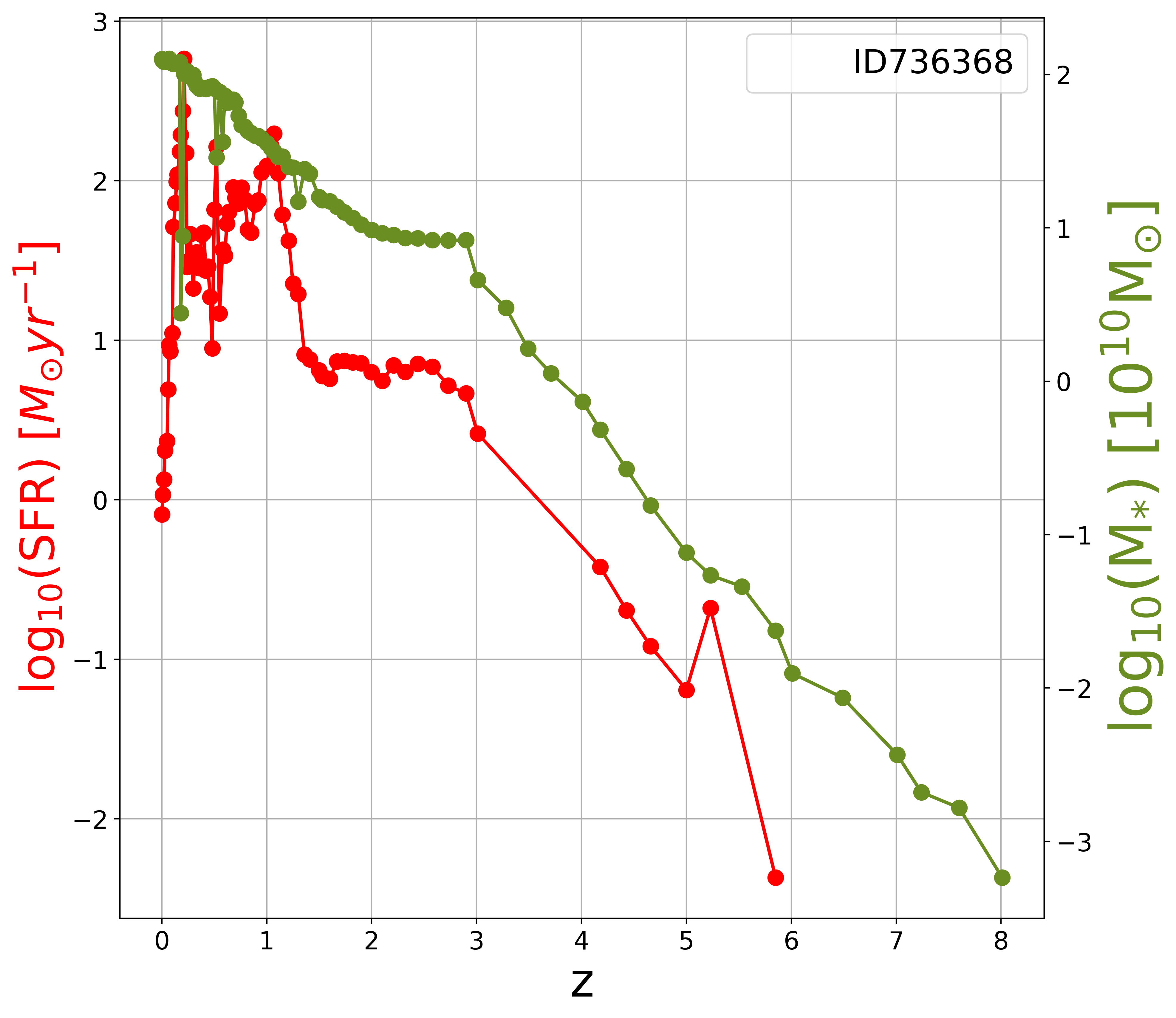}
\\[3mm]
\includegraphics[width=8.5cm]{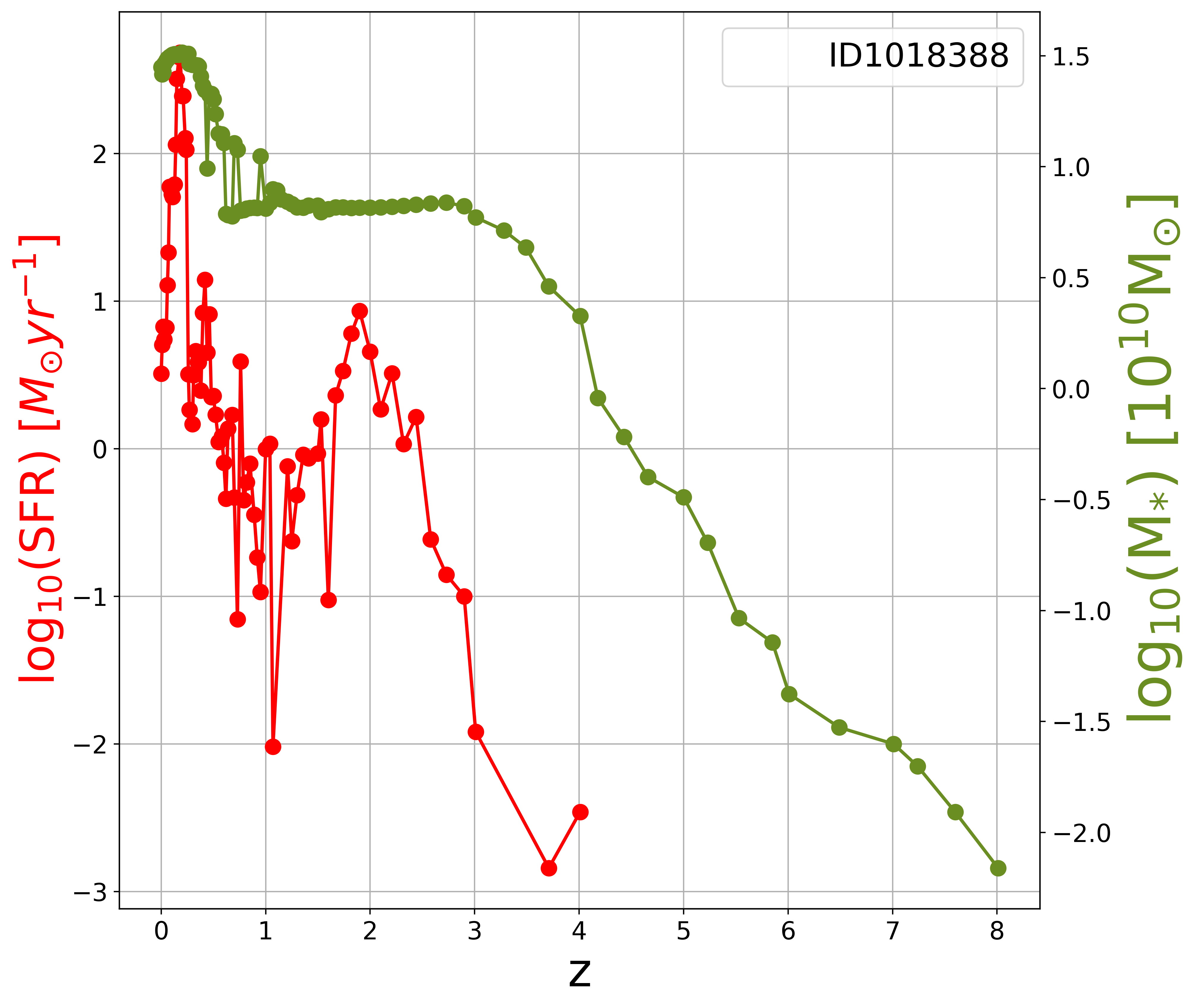} 
\caption{Star formation rate history (red line) and the stellar mass of galaxies (green dots) versus the redshift at 0 < z < 8. On the three panels, different subhalos are shown: ID 0 (top), ID 736368 (middle), and ID 1018388 (bottom).}
\label{Figure_A2}
\end{figure}

\begin{figure}
\centering
\includegraphics[width=8.5cm]{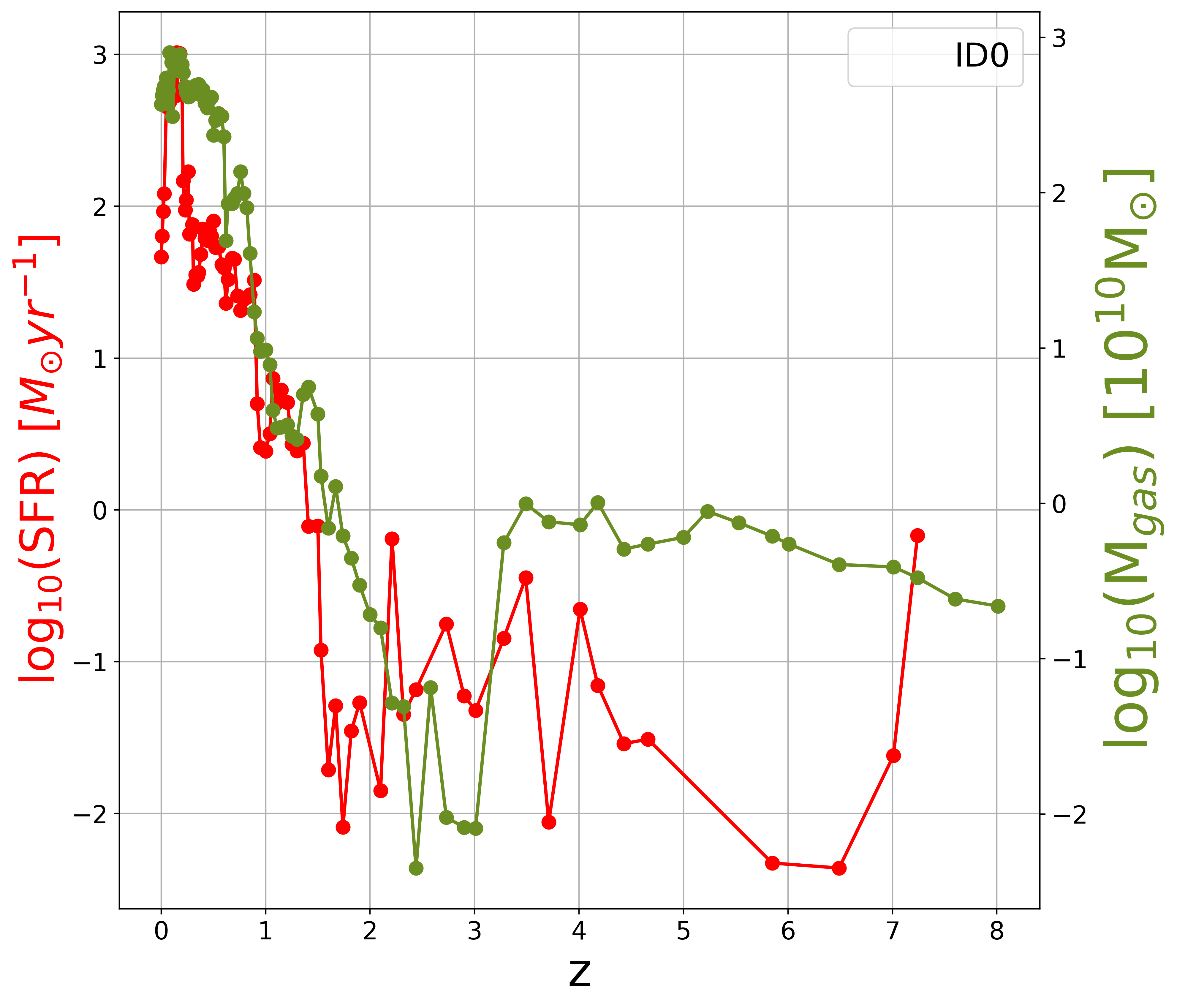} 
\\[3mm]
\includegraphics[width=8.5cm]{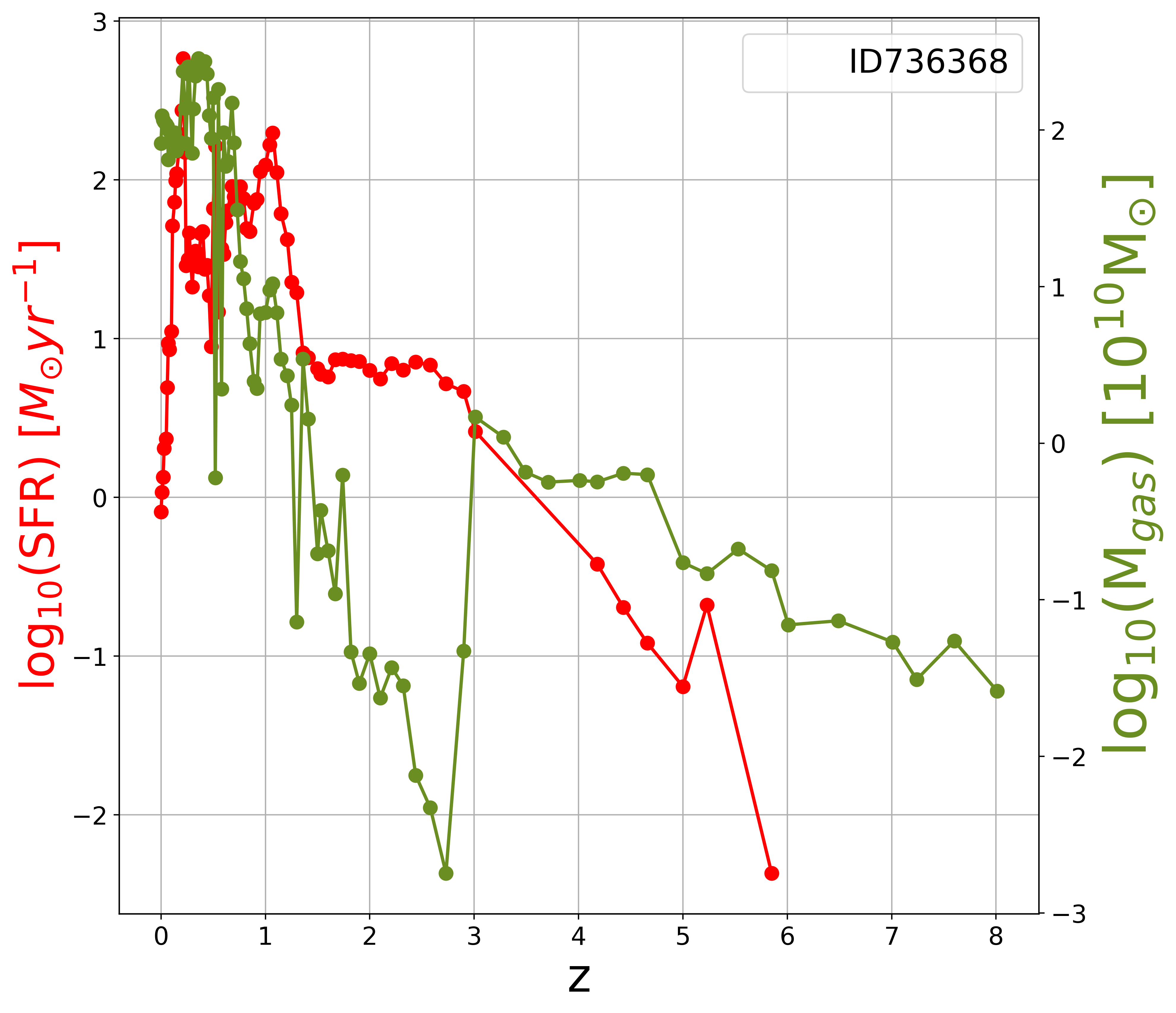}
\\[3mm]
\includegraphics[width=8.5cm]{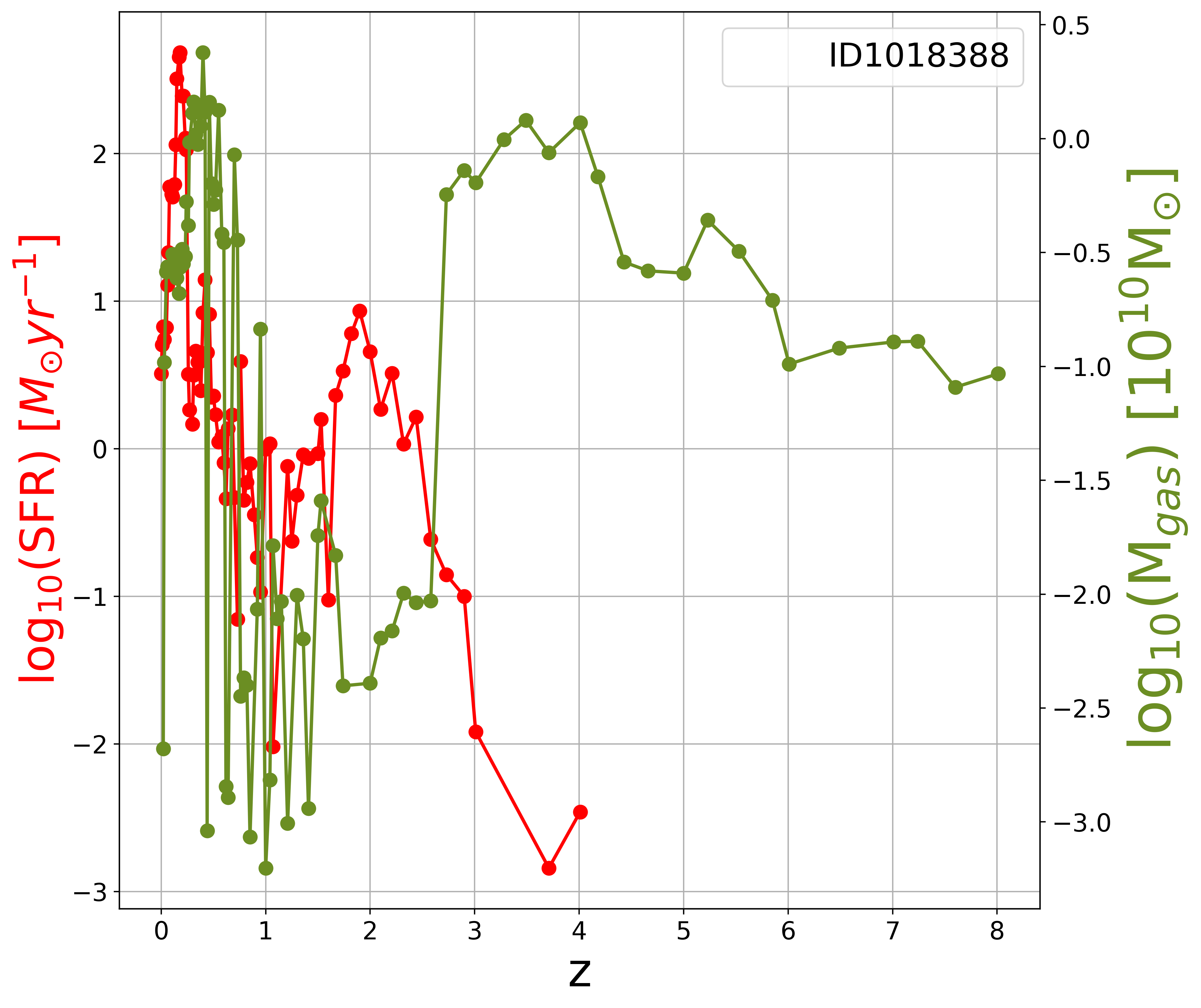} 
\caption{Star formation rate history (red line) and the gas mass of galaxies (green dots) versus the redshift at 0 < z < 8. On the three panels, different subhalos are shown: ID 0 (top), ID 736368 (middle), and ID 1018388 (bottom).}
\label{Figure_A3}
\end{figure}

\begin{figure}
\centering
\includegraphics[width=8.5cm]{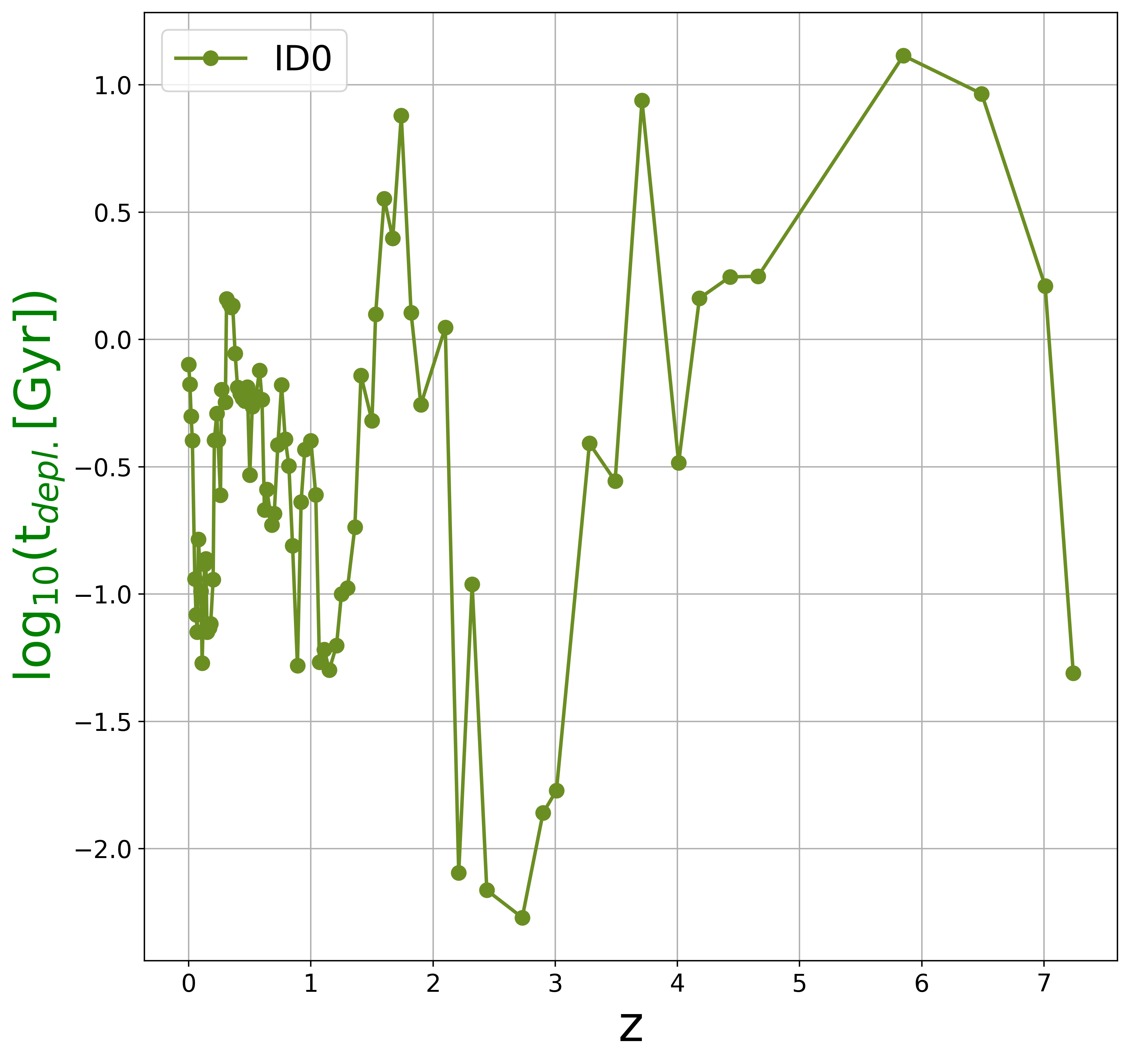} 
\\[3mm]
\includegraphics[width=8.5cm]{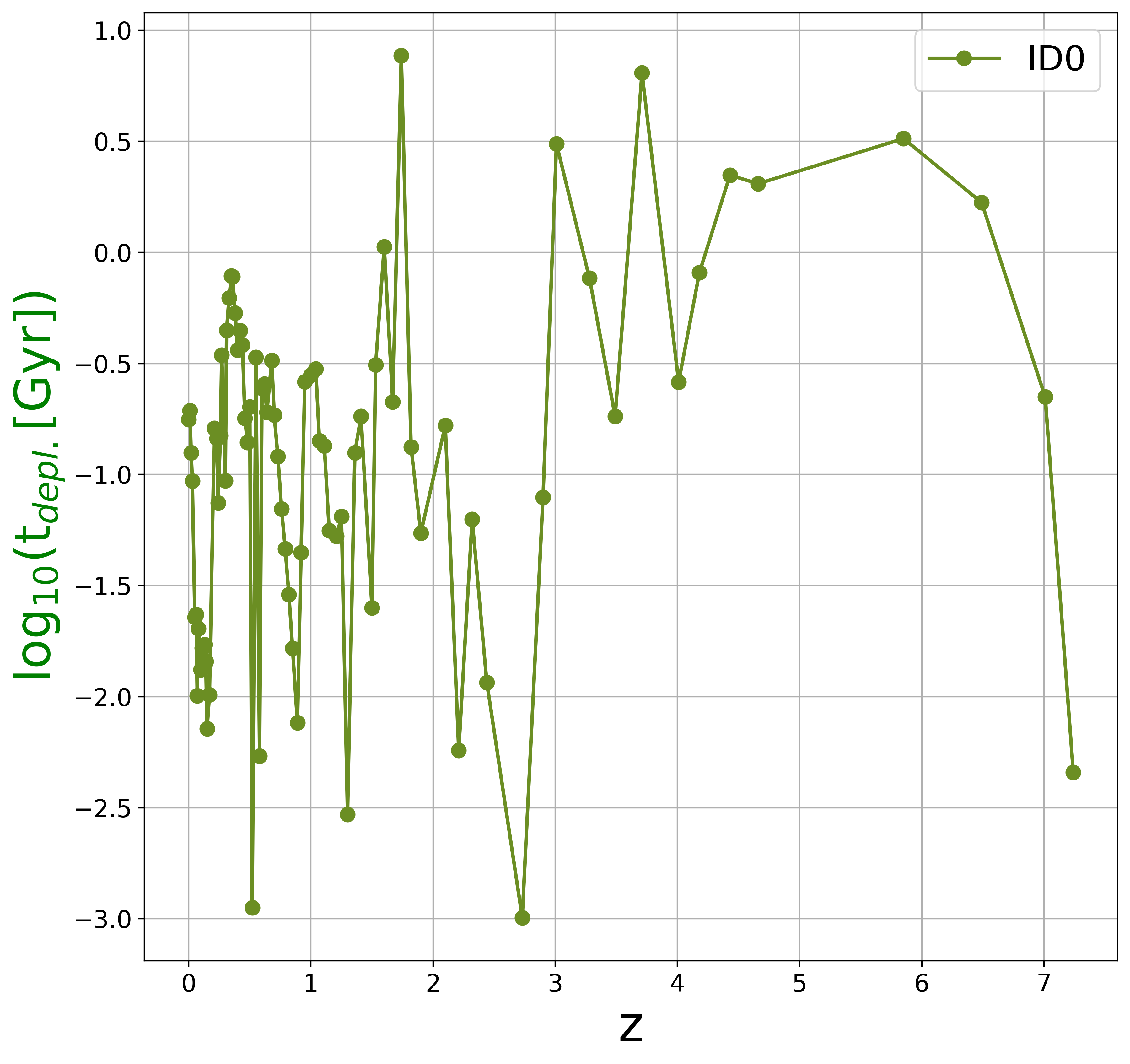}
\\[3mm]
\includegraphics[width=8.5cm]{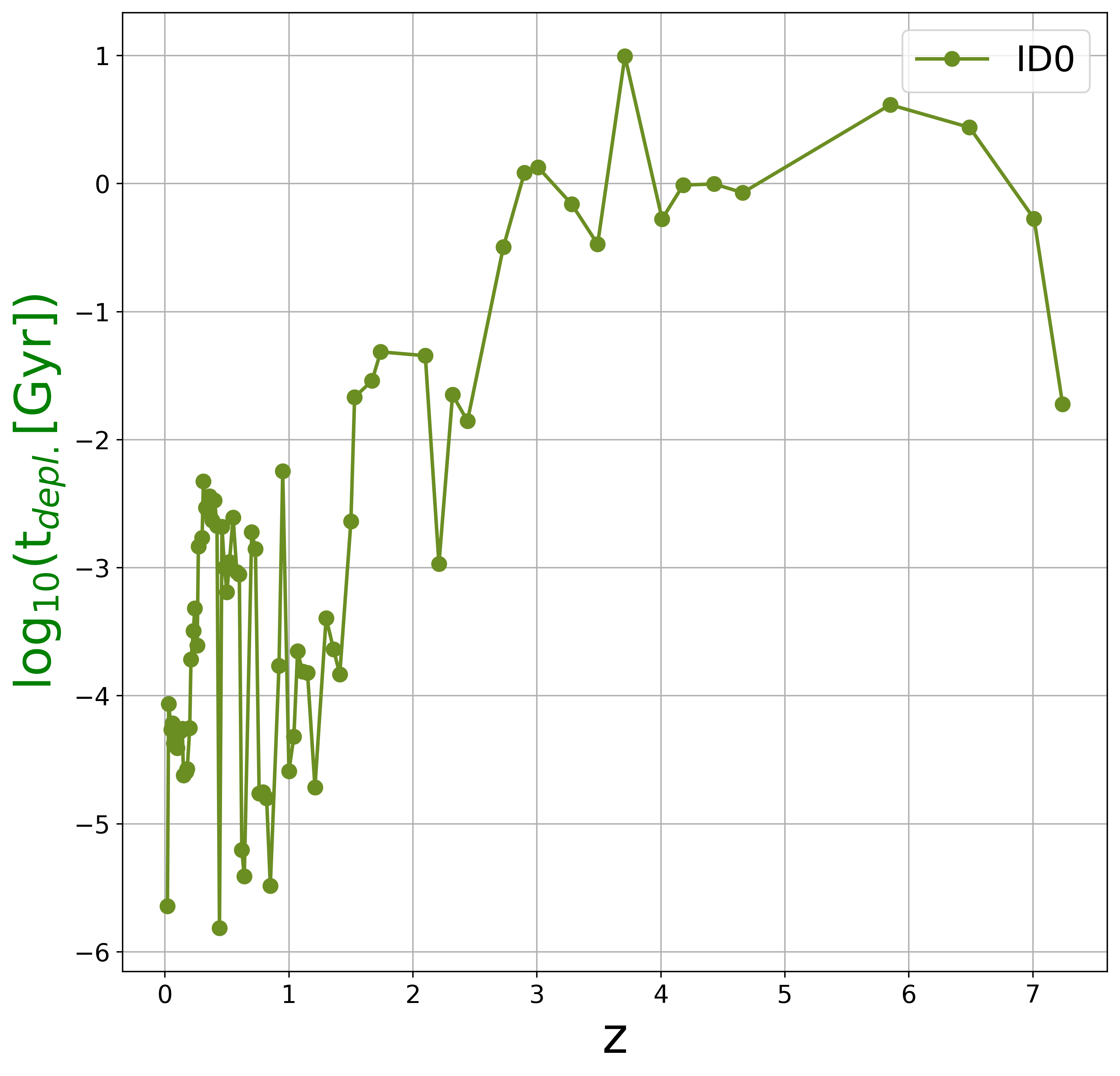} 
\caption{Deplation time of galaxies (green dots) versus the redshift at 0 < z < 8. On the three panels, different subhalos are shown: ID 0 (top), ID 736368 (middle), and ID 1018388 (bottom).}
\label{Figure_A4}
\end{figure}

\begin{figure}
\centering
\includegraphics[width=8.5cm]{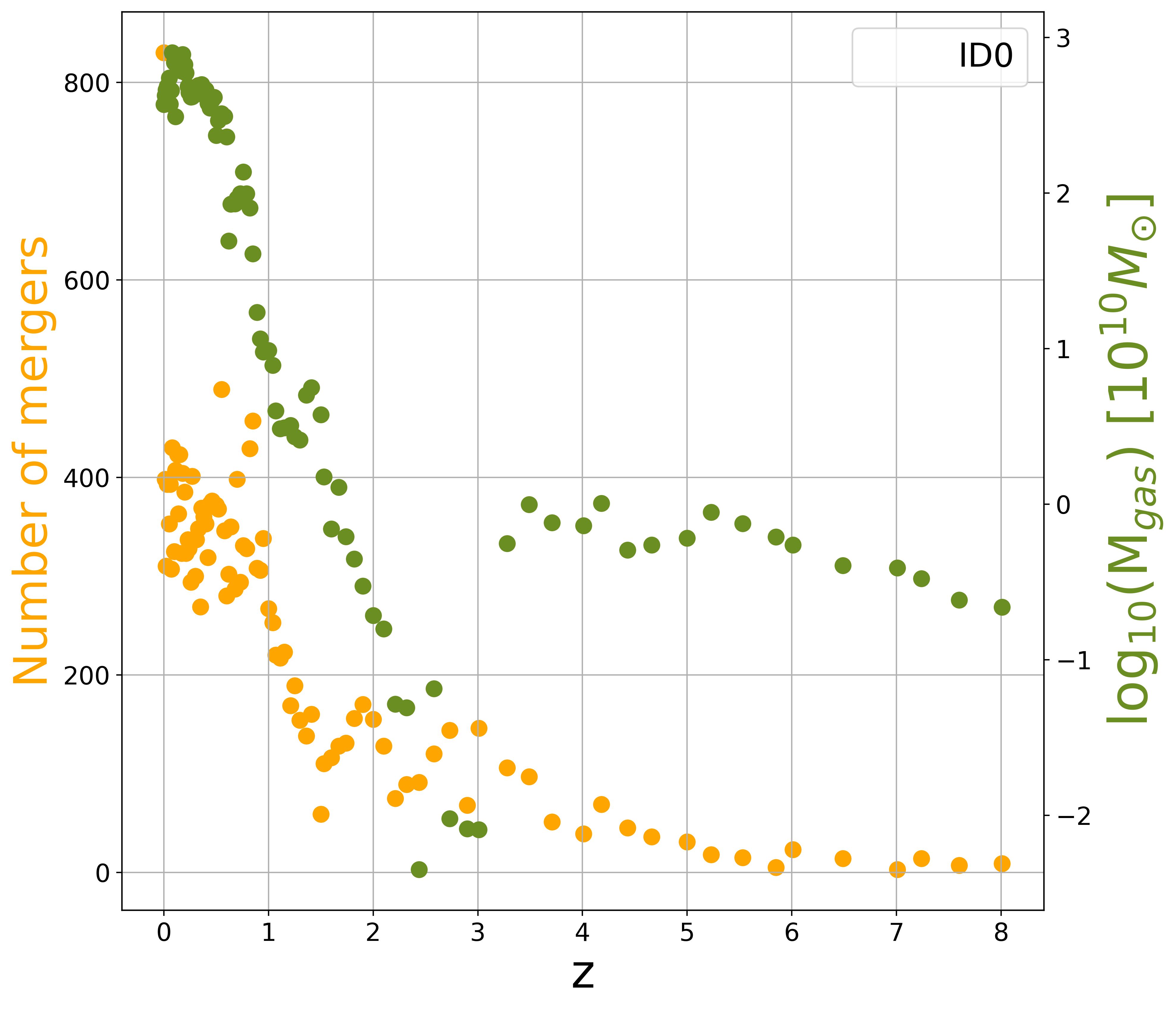} 
\\[3mm]
\includegraphics[width=8.5cm]{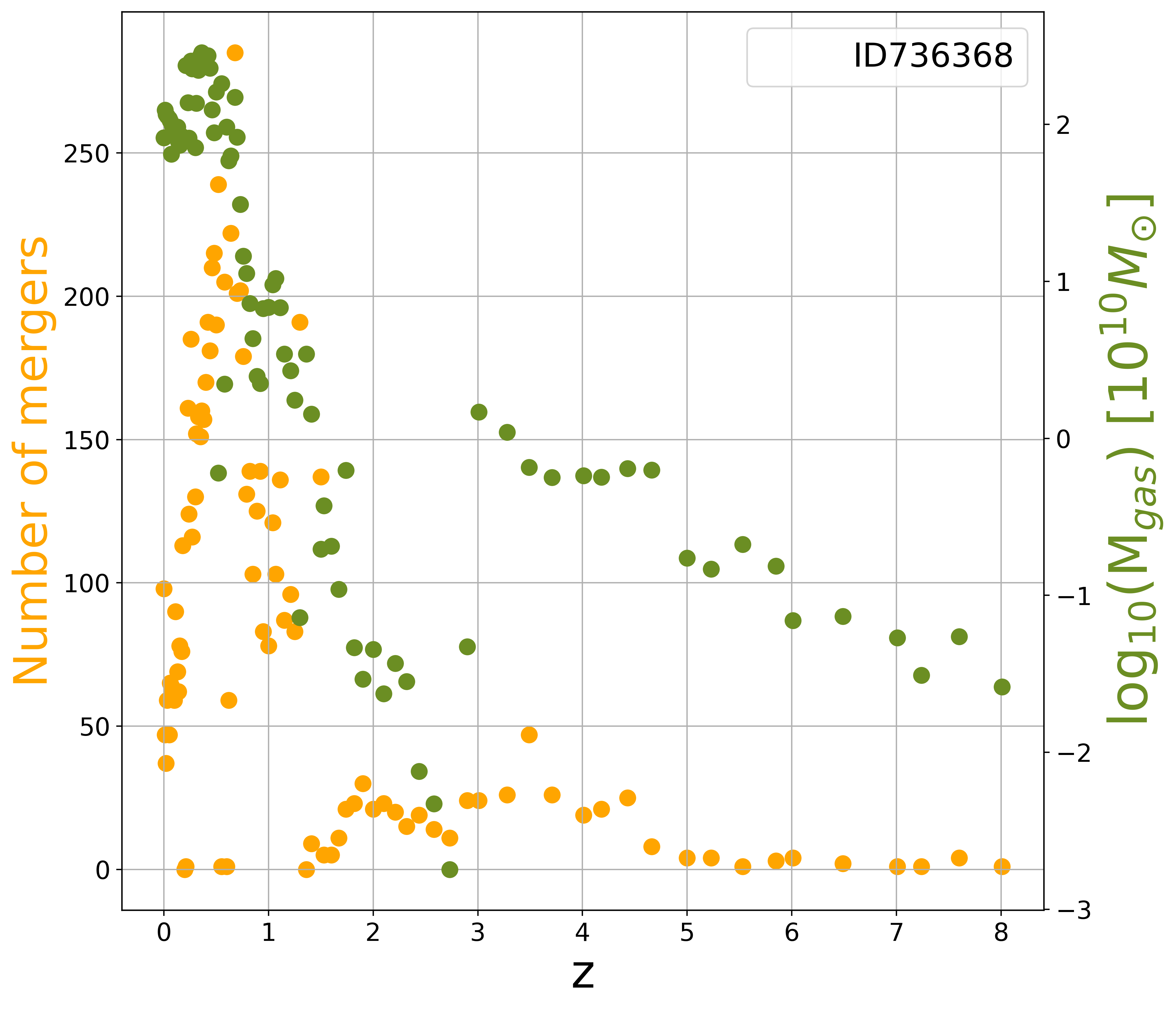}
\\[3mm]
\includegraphics[width=8.5cm]{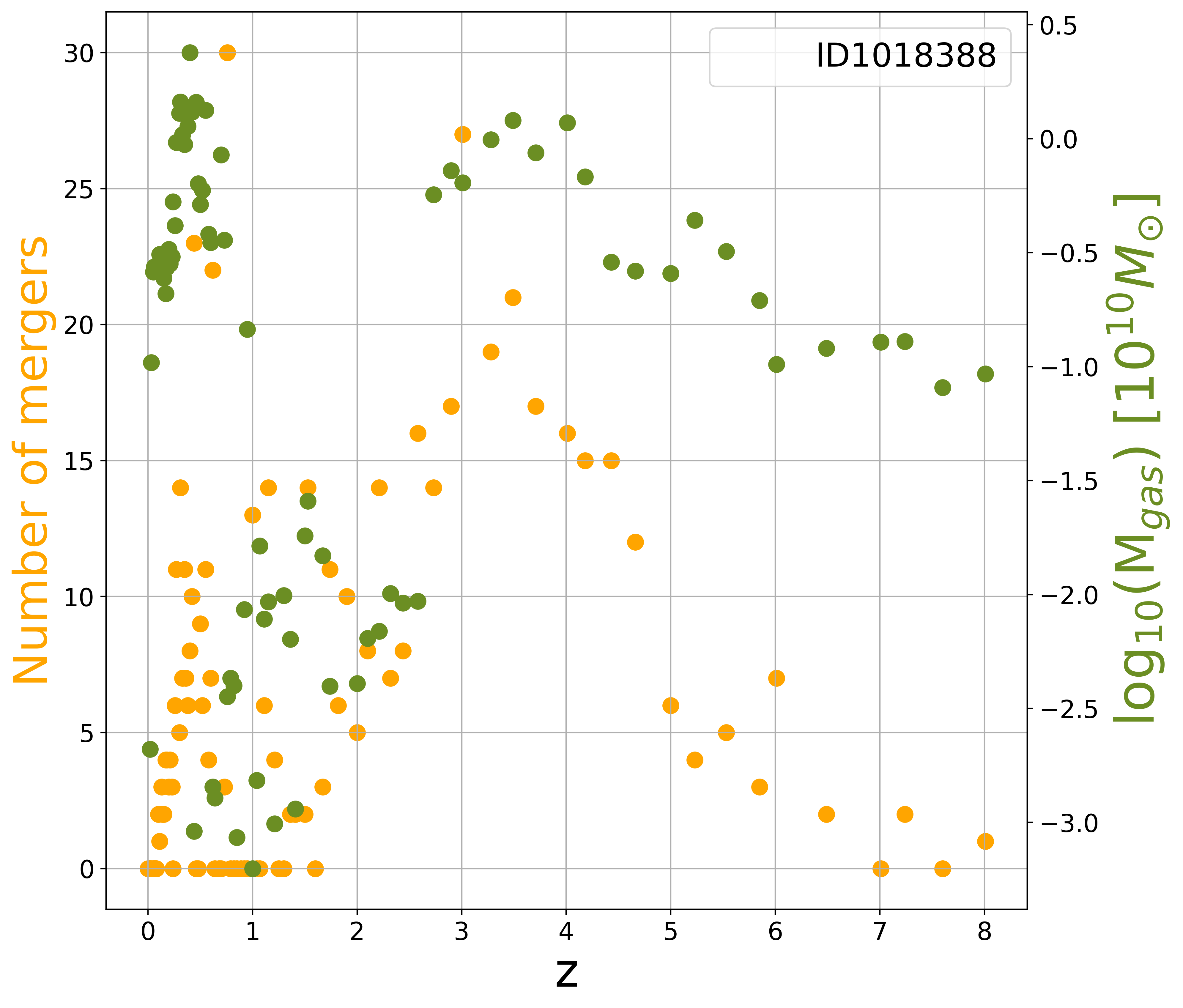} 
\caption{Number of mergers (orange dots) and the gas mass of galaxies (green dots) versus the redshift at 0 < z < 8. On the three panels, different subhalos are shown: ID 0 (top), ID 736368 (middle), and ID 1018388 (bottom).}
\label{Figure_A5}
\end{figure}

\section{Mass cuts for r$_5$ distances} \label{appendix B}

\begin{figure}[h!]
    \centering
    \includegraphics[width=\columnwidth]{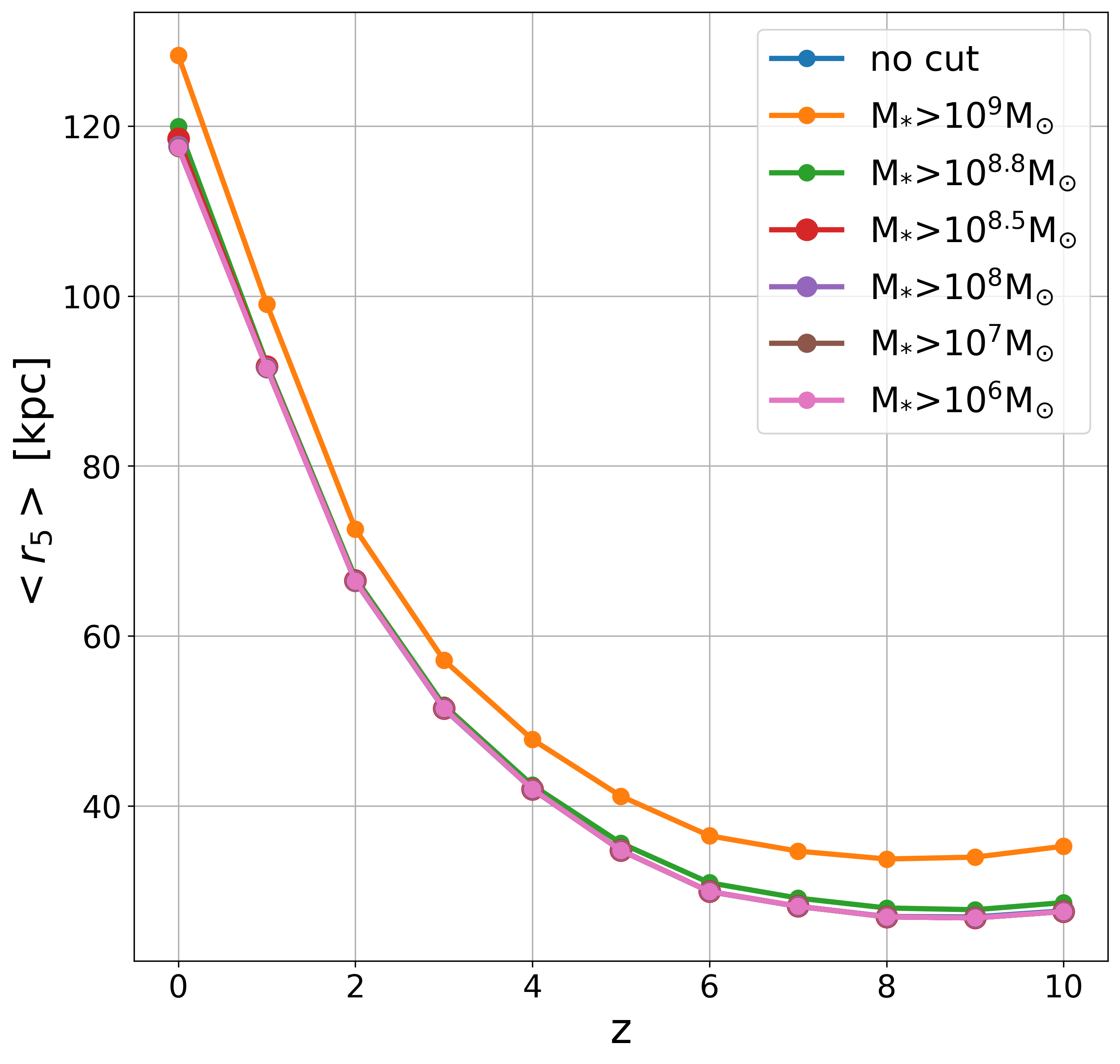}
    \caption{Average physical distance of the 5th closest neighbor (r$_5$) for merger galaxies, with different low-mass cuts at various redshifts. Because of the negligible differences, the figure shows significant overlap among the data points, especially at lower mass cuts.}
    \label{Figure_cut}
\end{figure}

\begin{figure}[h!]
    \centering
    \includegraphics[width=\columnwidth]{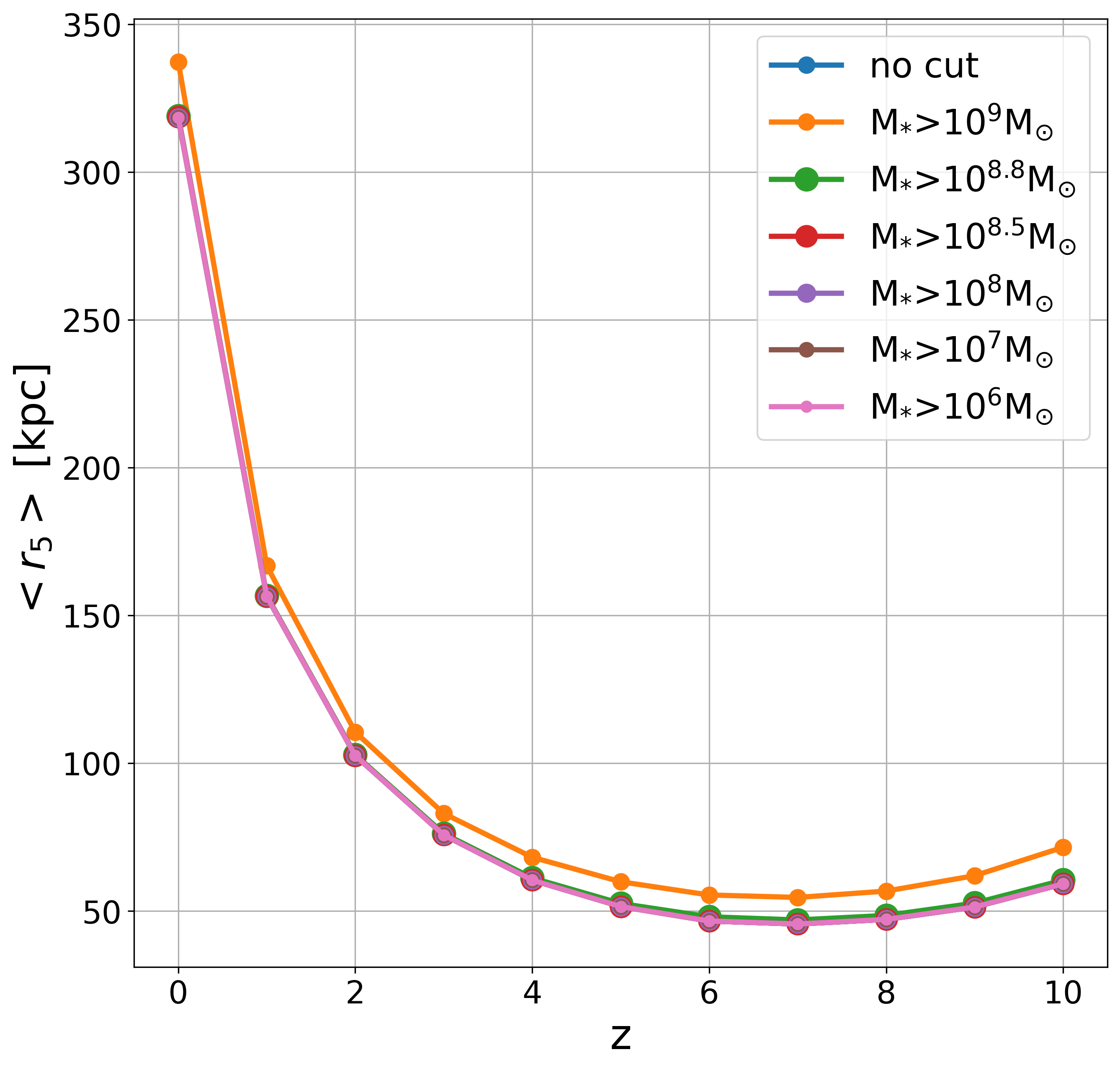}
    \caption{Average physical distance of the 5th closest neighbor (r$_5$) for all galaxies, with different low-mass cuts at various redshifts. Because of the negligible differences, the figure shows significant overlap among the data points, especially at lower mass cuts.}
    \label{Figure_cut_all}
\end{figure}

To demonstrate how the numerical resolution limits of the smallest subhalos impact our results, we plotted the r$_5$ distances using various low-mass thresholds for both all galaxies (Figure~\ref{Figure_cut_all}) and galaxy mergers (Figure~\ref{Figure_cut}).
To calculate the r$_5$ distances for different mass cuts, we first collected all subhalos at the relevant redshifts. We then selected only those galaxies with masses above a given threshold. Finally, we computed the distances to the nearest neighboring galaxies separately for galaxy mergers and the full set of remaining galaxies. For galaxy mergers, all remaining galaxies above the mass threshold were included as potential neighbors. The evolution of r$_5$ is consistent across all mass cuts for both cases. For mass thresholds of 10$^{8.8}$ and lower, the results are nearly identical or show only minor differences. However, at a higher threshold of 10$^9$, the galaxy distances increase by approximately 10 kpc. This is expected, as higher mass cuts exclude more galaxies. When comparing mergers to the full galaxy sample, we observe similar trends, indicating that including all subhalos does not significantly affect the overall results. 

\begin{table}
\section{Parameters of the TNG simulations}
    \captionsetup{justification=centering}
    \caption{Parameters of the three TNG simulation volumes}
    \centering
   
    \label{tab:example_table}
    \begin{tabular}{lcccr} 
        \hline
           &  & TNG50 & TNG100 & TNG300  \\
        \hline
        V & [Mpc$^3$] & 51.7$^3$ & 110.7$^3$ & 302.6$^3$  \\
        L$_{box}$ & [Mpc/h] & 35 & 75 & 205 \\
        N$_{GAS}$ & - & 2160$^3$ & 1820$^3$ & 2500$^3$\\
        N$_{DM}$ & - & 2160$^3$ & 1820$^3$ & 2500$^3$ \\
        N$_{TR}$ & - & 2160$^3$ & 2$\cdot$1820$^3$ & 2500$^3$ \\
        m$_{baryon}$ & [M$_{\odot}$] & 8.5$\cdot$10$^4$ & 1.4$\cdot$10$^6$ & 1.1$\cdot$10$^7$ \\
        m$_{DM}$ & [M$_{\odot}$] & 4.5$\cdot$10$^5$ & 7.5$\cdot$10$^6$ & 5.9$\cdot$10$^7$ \\ 
        \hline
    \end{tabular}
\end{table}

\begin{table}
    \centering
    \captionsetup{justification=centering}
    \caption{Additional parameters for the TNG300 simulation}
    \label{tab:example_table2}
    \begin{tabular}{lr} 
        \hline
          Parameter  &  Value \\
        \hline
        side length of simulation box [ckpc/h] & 2.05$\cdot$10$^5$\\
        average gas cell mass [10$^{10}$ M$_{\odot}$/h] & 7.43$\cdot$10$^{-4}$ \\
        dark matter particle mass [10$^{10}$ M$_{\odot}$/h] &  3.98$\cdot$10$^{-3}$\\
        number of dark matter particles & 1.5625$\cdot$10$^{10}$ \\
        number of gas tracer particles &  1.5625$\cdot$10$^{10}$ \\
        starting redshift & 127.0  \\
        ending redshift & 0.0  \\ 
        \hline
    \end{tabular}
\end{table}

\end{appendix}

\begin{adjustwidth}{-\extralength}{0cm}
\reftitle{References}


\bibliography{ref2,merger,ref,largescale}

\begin{thebibliography}{999}

\bibitem[Hwang et~al.(2019)Hwang, Shin, and Song]{Hwang_2019}
Hwang, H.S.; Shin, J.; Song, H.
\newblock Evolution of star formation rate–density relation over cosmic time in a simulated universe: the observed reversal reproduced.
\newblock {\em Monthly Notices of the Royal Astronomical Society} {\bf 2019}, {\em 489},~339–348.
\newblock {\url{https://doi.org/10.1093/mnras/stz2136}}.

\bibitem[{Mihos} and {Hernquist}(1996)]{1996ApJ...464..641M}
{Mihos}, J.C.; {Hernquist}, L.
\newblock {Gasdynamics and Starbursts in Major Mergers}.
\newblock {\em \apj} {\bf 1996}, {\em 464},~641,  \href{http://arxiv.org/abs/astro-ph/9512099}{{\normalfont [arXiv:astro-ph/astro-ph/9512099]}}.
\newblock {\url{https://doi.org/10.1086/177353}}.

\bibitem[Shah et~al.(2022)Shah, Kartaltepe, Magagnoli, Cox, Wetherell, Vanderhoof, Cooke, Calabro, Chartab, Conselice, Croton, de~la Vega, Hathi, Ilbert, Inami, Kocevski, Koekemoer, Lemaux, Lubin, Mantha, Marchesi, Martig, Moreno, Pampliega, Patton, Salvato, and Treister]{Shah_2022}
Shah, E.A.; Kartaltepe, J.S.; Magagnoli, C.T.; Cox, I.G.; Wetherell, C.T.; Vanderhoof, B.N.; Cooke, K.C.; Calabro, A.; Chartab, N.; Conselice, C.J.;  et~al.
\newblock Investigating the Effect of Galaxy Interactions on Star Formation at 0.5 \&lt; z \&lt; 3.0.
\newblock {\em The Astrophysical Journal} {\bf 2022}, {\em 940},~4.
\newblock {\url{https://doi.org/10.3847/1538-4357/ac96eb}}.

\bibitem[{Nath} and {Chiba}(1995)]{1995ApJ...454..604N}
{Nath}, B.B.; {Chiba}, M.
\newblock {Dwarf Galaxies and the Origin of the Intracluster Medium}.
\newblock {\em \apj} {\bf 1995}, {\em 454},~604,  \href{http://arxiv.org/abs/astro-ph/9505081}{{\normalfont [arXiv:astro-ph/astro-ph/9505081]}}.
\newblock {\url{https://doi.org/10.1086/176514}}.

\bibitem[{Moustakas} et~al.(2013){Moustakas}, {Coil}, {Aird}, {Blanton}, {Cool}, {Eisenstein}, {Mendez}, {Wong}, {Zhu}, and {Arnouts}]{2013ApJ...767...50M}
{Moustakas}, J.; {Coil}, A.L.; {Aird}, J.; {Blanton}, M.R.; {Cool}, R.J.; {Eisenstein}, D.J.; {Mendez}, A.J.; {Wong}, K.C.; {Zhu}, G.; {Arnouts}, S.
\newblock {PRIMUS: Constraints on Star Formation Quenching and Galaxy Merging, and the Evolution of the Stellar Mass Function from z = 0-1}.
\newblock {\em \apj} {\bf 2013}, {\em 767},~50,  \href{http://arxiv.org/abs/1301.1688}{{\normalfont [arXiv:astro-ph.CO/1301.1688]}}.
\newblock {\url{https://doi.org/10.1088/0004-637X/767/1/50}}.

\bibitem[Ellison et~al.(2022)Ellison, Wilkinson, Woo, Leung, Wild, Bickley, Patton, Quai, and Gwyn]{Ellison_2022}
Ellison, S.L.; Wilkinson, S.; Woo, J.; Leung, H.H.; Wild, V.; Bickley, R.W.; Patton, D.R.; Quai, S.; Gwyn, S.
\newblock Galaxy mergers can rapidly shut down star formation.
\newblock {\em Monthly Notices of the Royal Astronomical Society: Letters} {\bf 2022}, {\em 517},~L92–L96.
\newblock {\url{https://doi.org/10.1093/mnrasl/slac109}}.

\bibitem[{Pontzen} et~al.(2017){Pontzen}, {Tremmel}, {Roth}, {Peiris}, {Saintonge}, {Volonteri}, {Quinn}, and {Governato}]{2017MNRAS.465..547P}
{Pontzen}, A.; {Tremmel}, M.; {Roth}, N.; {Peiris}, H.V.; {Saintonge}, A.; {Volonteri}, M.; {Quinn}, T.; {Governato}, F.
\newblock {How to quench a galaxy}.
\newblock {\em \mnras} {\bf 2017}, {\em 465},~547--558,  \href{http://arxiv.org/abs/1607.02507}{{\normalfont [arXiv:astro-ph.GA/1607.02507]}}.
\newblock {\url{https://doi.org/10.1093/mnras/stw2627}}.

\bibitem[{Gabor} et~al.(2010){Gabor}, {Dav{\'e}}, {Finlator}, and {Oppenheimer}]{2010MNRAS.407..749G}
{Gabor}, J.M.; {Dav{\'e}}, R.; {Finlator}, K.; {Oppenheimer}, B.D.
\newblock {How is star formation quenched in massive galaxies?}
\newblock {\em \mnras} {\bf 2010}, {\em 407},~749--771,  \href{http://arxiv.org/abs/1001.1734}{{\normalfont [arXiv:astro-ph.CO/1001.1734]}}.
\newblock {\url{https://doi.org/10.1111/j.1365-2966.2010.16961.x}}.

\bibitem[{Pearson} et~al.(2019){Pearson}, {Wang}, {Alpaslan}, {Baldry}, {Bilicki}, {Brown}, {Grootes}, {Holwerda}, {Kitching}, {Kruk}, and {van der Tak}]{2019A&A...631A..51P}
{Pearson}, W.J.; {Wang}, L.; {Alpaslan}, M.; {Baldry}, I.; {Bilicki}, M.; {Brown}, M.J.I.; {Grootes}, M.W.; {Holwerda}, B.W.; {Kitching}, T.D.; {Kruk}, S.;  et~al.
\newblock {Effect of galaxy mergers on star-formation rates}.
\newblock {\em \aap} {\bf 2019}, {\em 631},~A51,  \href{http://arxiv.org/abs/1908.10115}{{\normalfont [arXiv:astro-ph.GA/1908.10115]}}.
\newblock {\url{https://doi.org/10.1051/0004-6361/201936337}}.

\bibitem[Nelson et~al.(2017)Nelson, Pillepich, Springel, Weinberger, Hernquist, Pakmor, Genel, Torrey, Vogelsberger, Kauffmann, Marinacci, and Naiman]{Nelson_2017}
Nelson, D.; Pillepich, A.; Springel, V.; Weinberger, R.; Hernquist, L.; Pakmor, R.; Genel, S.; Torrey, P.; Vogelsberger, M.; Kauffmann, G.;  et~al.
\newblock First results from the {IllustrisTNG} simulations: the galaxy colour bimodality.
\newblock {\em Monthly Notices of the Royal Astronomical Society} {\bf 2017}, {\em 475},~624--647.
\newblock {\url{https://doi.org/10.1093/mnras/stx3040}}.

\bibitem[Marinacci et~al.(2018)Marinacci, Vogelsberger, Pakmor, Torrey, Springel, Hernquist, Nelson, Weinberger, Pillepich, Naiman, and Genel]{Marinacci_2018}
Marinacci, F.; Vogelsberger, M.; Pakmor, R.; Torrey, P.; Springel, V.; Hernquist, L.; Nelson, D.; Weinberger, R.; Pillepich, A.; Naiman, J.;  et~al.
\newblock First results from the {IllustrisTNG} simulations: radio haloes and magnetic fields.
\newblock {\em Monthly Notices of the Royal Astronomical Society} {\bf 2018}.
\newblock {\url{https://doi.org/10.1093/mnras/sty2206}}.

\bibitem[Pillepich et~al.(2017)Pillepich, Nelson, Hernquist, Springel, Pakmor, Torrey, Weinberger, Genel, Naiman, Marinacci, and Vogelsberger]{Pillepich_2017}
Pillepich, A.; Nelson, D.; Hernquist, L.; Springel, V.; Pakmor, R.; Torrey, P.; Weinberger, R.; Genel, S.; Naiman, J.P.; Marinacci, F.;  et~al.
\newblock First results from the {IllustrisTNG} simulations: the stellar mass content of groups and clusters of galaxies.
\newblock {\em Monthly Notices of the Royal Astronomical Society} {\bf 2017}, {\em 475},~648--675.
\newblock {\url{https://doi.org/10.1093/mnras/stx3112}}.

\bibitem[Springel et~al.(2017)Springel, Pakmor, Pillepich, Weinberger, Nelson, Hernquist, Vogelsberger, Genel, Torrey, Marinacci, and Naiman]{Springel_2017}
Springel, V.; Pakmor, R.; Pillepich, A.; Weinberger, R.; Nelson, D.; Hernquist, L.; Vogelsberger, M.; Genel, S.; Torrey, P.; Marinacci, F.;  et~al.
\newblock First results from the {IllustrisTNG} simulations: matter and galaxy clustering.
\newblock {\em Monthly Notices of the Royal Astronomical Society} {\bf 2017}, {\em 475},~676--698.
\newblock {\url{https://doi.org/10.1093/mnras/stx3304}}.

\bibitem[Naiman et~al.(2018)Naiman, Pillepich, Springel, Ramirez-Ruiz, Torrey, Vogelsberger, Pakmor, Nelson, Marinacci, Hernquist, Weinberger, and Genel]{Naiman_2018}
Naiman, J.P.; Pillepich, A.; Springel, V.; Ramirez-Ruiz, E.; Torrey, P.; Vogelsberger, M.; Pakmor, R.; Nelson, D.; Marinacci, F.; Hernquist, L.;  et~al.
\newblock First results from the {IllustrisTNG} simulations: a tale of two elements {\textendash} chemical evolution of magnesium and europium.
\newblock {\em Monthly Notices of the Royal Astronomical Society} {\bf 2018}, {\em 477},~1206--1224.
\newblock {\url{https://doi.org/10.1093/mnras/sty618}}.

\bibitem[{Cutri} and {McAlary}(1985)]{1985ApJ...296...90C}
{Cutri}, R.M.; {McAlary}, C.W.
\newblock {A statistical study of the relationship between galaxy interactions and nuclear activity.}
\newblock {\em \apj} {\bf 1985}, {\em 296},~90--105.
\newblock {\url{https://doi.org/10.1086/163422}}.

\bibitem[{Ellison} et~al.(2011){Ellison}, {Patton}, {Mendel}, and {Scudder}]{2011MNRAS.418.2043E}
{Ellison}, S.L.; {Patton}, D.R.; {Mendel}, J.T.; {Scudder}, J.M.
\newblock {Galaxy pairs in the Sloan Digital Sky Survey - IV. Interactions trigger active galactic nuclei}.
\newblock {\em \mnras} {\bf 2011}, {\em 418},~2043--2053,  \href{http://arxiv.org/abs/1108.2711}{{\normalfont [arXiv:astro-ph.CO/1108.2711]}}.
\newblock {\url{https://doi.org/10.1111/j.1365-2966.2011.19624.x}}.

\bibitem[{Moreno} et~al.(2019){Moreno}, {Torrey}, {Ellison}, {Patton}, {Hopkins}, {Bueno}, {Hayward}, {Narayanan}, {Kere{\v{s}}}, {Bluck}, and {Hernquist}]{2019MNRAS.485.1320M}
{Moreno}, J.; {Torrey}, P.; {Ellison}, S.L.; {Patton}, D.R.; {Hopkins}, P.F.; {Bueno}, M.; {Hayward}, C.C.; {Narayanan}, D.; {Kere{\v{s}}}, D.; {Bluck}, A.F.L.;  et~al.
\newblock {Interacting galaxies on FIRE-2: the connection between enhanced star formation and interstellar gas content}.
\newblock {\em \mnras} {\bf 2019}, {\em 485},~1320--1338,  \href{http://arxiv.org/abs/1902.02305}{{\normalfont [arXiv:astro-ph.GA/1902.02305]}}.
\newblock {\url{https://doi.org/10.1093/mnras/stz417}}.

\bibitem[{Joseph} and {Wright}(1985)]{1985MNRAS.214...87J}
{Joseph}, R.D.; {Wright}, G.S.
\newblock {Recent star formation in interacting galaxies - II. Super starbursts in merging galaxies.}
\newblock {\em \mnras} {\bf 1985}, {\em 214},~87--95.
\newblock {\url{https://doi.org/10.1093/mnras/214.2.87}}.

\bibitem[{Rupke} et~al.(2005){Rupke}, {Veilleux}, and {Sanders}]{2005ApJS..160..115R}
{Rupke}, D.S.; {Veilleux}, S.; {Sanders}, D.B.
\newblock {Outflows in Infrared-Luminous Starbursts at z < 0.5. II. Analysis and Discussion}.
\newblock {\em \apjs} {\bf 2005}, {\em 160},~115--148,  \href{http://arxiv.org/abs/astro-ph/0506611}{{\normalfont [arXiv:astro-ph/astro-ph/0506611]}}.
\newblock {\url{https://doi.org/10.1086/432889}}.

\bibitem[{Strickland} and {Heckman}(2009)]{2009ApJ...697.2030S}
{Strickland}, D.K.; {Heckman}, T.M.
\newblock {Supernova Feedback Efficiency and Mass Loading in the Starburst and Galactic Superwind Exemplar M82}.
\newblock {\em \apj} {\bf 2009}, {\em 697},~2030--2056,  \href{http://arxiv.org/abs/0903.4175}{{\normalfont [arXiv:astro-ph.CO/0903.4175]}}.
\newblock {\url{https://doi.org/10.1088/0004-637X/697/2/2030}}.

\bibitem[{Rupke} et~al.(2005){Rupke}, {Veilleux}, and {Sanders}]{2005ApJ...632..751R}
{Rupke}, D.S.; {Veilleux}, S.; {Sanders}, D.B.
\newblock {Outflows in Active Galactic Nucleus/Starburst-Composite Ultraluminous Infrared Galaxies1,}.
\newblock {\em \apj} {\bf 2005}, {\em 632},~751--780,  \href{http://arxiv.org/abs/astro-ph/0507037}{{\normalfont [arXiv:astro-ph/astro-ph/0507037]}}.
\newblock {\url{https://doi.org/10.1086/444451}}.

\bibitem[{Woo} et~al.(2017){Woo}, {Son}, and {Bae}]{2017ApJ...839..120W}
{Woo}, J.H.; {Son}, D.; {Bae}, H.J.
\newblock {Delayed or No Feedback? Gas Outflows in Type 2 AGNs. III.}
\newblock {\em \apj} {\bf 2017}, {\em 839},~120,  \href{http://arxiv.org/abs/1702.06681}{{\normalfont [arXiv:astro-ph.GA/1702.06681]}}.
\newblock {\url{https://doi.org/10.3847/1538-4357/aa6894}}.

\bibitem[{Luo} et~al.(2014){Luo}, {Yang}, and {Zhang}]{2014ApJ...789L..16L}
{Luo}, W.; {Yang}, X.; {Zhang}, Y.
\newblock {Connections between Galaxy Mergers and Starburst: Evidence from the Local Universe}.
\newblock {\em \apjl} {\bf 2014}, {\em 789},~L16,  \href{http://arxiv.org/abs/1406.5315}{{\normalfont [arXiv:astro-ph.GA/1406.5315]}}.
\newblock {\url{https://doi.org/10.1088/2041-8205/789/1/L16}}.

\bibitem[{Cibinel} et~al.(2019){Cibinel}, {Daddi}, {Sargent}, {Le Floc'h}, {Liu}, {Bournaud}, {Oesch}, {Amram}, {Calabr{\`o}}, {Duc}, {Pannella}, {Puglisi}, {Perret}, {Elbaz}, and {Kokorev}]{2019MNRAS.485.5631C}
{Cibinel}, A.; {Daddi}, E.; {Sargent}, M.T.; {Le Floc'h}, E.; {Liu}, D.; {Bournaud}, F.; {Oesch}, P.A.; {Amram}, P.; {Calabr{\`o}}, A.; {Duc}, P.A.;  et~al.
\newblock {Early- and late-stage mergers among main sequence and starburst galaxies at 0.2 {\ensuremath{\leq}} z {\ensuremath{\leq}} 2}.
\newblock {\em \mnras} {\bf 2019}, {\em 485},~5631--5651,  \href{http://arxiv.org/abs/1809.00715}{{\normalfont [arXiv:astro-ph.GA/1809.00715]}}.
\newblock {\url{https://doi.org/10.1093/mnras/stz690}}.

\bibitem[Ma et~al.(2022)Ma, Liu, Guo, Cui, Jones, Wang, Zhang, and Davé]{Ma_2022}
Ma, W.; Liu, K.; Guo, H.; Cui, W.; Jones, M.G.; Wang, J.; Zhang, L.; Davé, R.
\newblock Effects of Active Galactic Nucleus Feedback on Cold Gas Depletion and Quenching of Central Galaxies.
\newblock {\em The Astrophysical Journal} {\bf 2022}, {\em 941},~205.
\newblock {\url{https://doi.org/10.3847/1538-4357/aca326}}.

\bibitem[{Lequeux} et~al.(1979){Lequeux}, {Peimbert}, {Rayo}, {Serrano}, and {Torres-Peimbert}]{1979A&A....80..155L}
{Lequeux}, J.; {Peimbert}, M.; {Rayo}, J.F.; {Serrano}, A.; {Torres-Peimbert}, S.
\newblock {Chemical Composition and Evolution of Irregular and Blue Compact Galaxies}.
\newblock {\em \aap} {\bf 1979}, {\em 80},~155.

\bibitem[{Kinman} and {Davidson}(1981)]{1981ApJ...243..127K}
{Kinman}, T.D.; {Davidson}, K.
\newblock {Spectroscopic observations of 10 emission-line dwarf galaxies.}
\newblock {\em \apj} {\bf 1981}, {\em 243},~127--139.
\newblock {\url{https://doi.org/10.1086/158575}}.

\bibitem[{Rubin} et~al.(1984){Rubin}, {Ford}, and {Whitmore}]{1984ApJ...281L..21R}
{Rubin}, V.C.; {Ford}, W.~K., J.; {Whitmore}, B.C.
\newblock {Luminosity-dependent line ratios in disks of spiral galaxies.}
\newblock {\em \apjl} {\bf 1984}, {\em 281},~L21--L24.
\newblock {\url{https://doi.org/10.1086/184276}}.

\bibitem[{Kewley} et~al.(2006){Kewley}, {Geller}, and {Barton}]{2006AJ....131.2004K}
{Kewley}, L.J.; {Geller}, M.J.; {Barton}, E.J.
\newblock {Metallicity and Nuclear Star Formation in Nearby Galaxy Pairs: Evidence for Tidally Induced Gas Flows}.
\newblock {\em Astronomical Journal} {\bf 2006}, {\em 131},~2004--2017,  \href{http://arxiv.org/abs/astro-ph/0511119}{{\normalfont [arXiv:astro-ph/astro-ph/0511119]}}.
\newblock {\url{https://doi.org/10.1086/500295}}.

\bibitem[Ellison et~al.(2008)Ellison, Patton, Simard, and McConnachie]{Ellison_2008}
Ellison, S.L.; Patton, D.R.; Simard, L.; McConnachie, A.W.
\newblock GALAXY PAIRS IN THE SLOAN DIGITAL SKY SURVEY. I. STAR FORMATION, ACTIVE GALACTIC NUCLEUS FRACTION, AND THE LUMINOSITY/MASS–METALLICITY RELATION.
\newblock {\em The Astronomical Journal} {\bf 2008}, {\em 135},~1877.
\newblock {\url{https://doi.org/10.1088/0004-6256/135/5/1877}}.

\bibitem[{Chamberlain} et~al.(2024){Chamberlain}, {Patel}, {Besla}, {Torrey}, and {Rodriguez-Gomez}]{2024ApJ...975..104C}
{Chamberlain}, K.; {Patel}, E.; {Besla}, G.; {Torrey}, P.; {Rodriguez-Gomez}, V.
\newblock {A Physically Motivated Framework to Compare the Merger Timescales of Isolated Low- and High-mass Galaxy Pairs Across Cosmic Time}.
\newblock {\em \apj} {\bf 2024}, {\em 975},~104,  \href{http://arxiv.org/abs/2409.02233}{{\normalfont [arXiv:astro-ph.GA/2409.02233]}}.
\newblock {\url{https://doi.org/10.3847/1538-4357/ad7bad}}.

\bibitem[{Bottrell} et~al.(2024){Bottrell}, {Yesuf}, {Popping}, {Omori}, {Tang}, {Ding}, {Pillepich}, {Nelson}, {Eisert}, {Gao}, {Goulding}, {Kalita}, {Luo}, {Greene}, {Shi}, and {Silverman}]{2024MNRAS.527.6506B}
{Bottrell}, C.; {Yesuf}, H.M.; {Popping}, G.; {Omori}, K.C.; {Tang}, S.; {Ding}, X.; {Pillepich}, A.; {Nelson}, D.; {Eisert}, L.; {Gao}, H.;  et~al.
\newblock {IllustrisTNG in the HSC-SSP: image data release and the major role of mini mergers as drivers of asymmetry and star formation}.
\newblock {\em \mnras} {\bf 2024}, {\em 527},~6506--6539,  \href{http://arxiv.org/abs/2308.14793}{{\normalfont [arXiv:astro-ph.GA/2308.14793]}}.
\newblock {\url{https://doi.org/10.1093/mnras/stad2971}}.

\bibitem[{Montenegro-Taborda} et~al.(2023){Montenegro-Taborda}, {Rodriguez-Gomez}, {Pillepich}, {Avila-Reese}, {Sales}, {Rodr{\'\i}guez-Puebla}, and {Hernquist}]{2023MNRAS.521..800M}
{Montenegro-Taborda}, D.; {Rodriguez-Gomez}, V.; {Pillepich}, A.; {Avila-Reese}, V.; {Sales}, L.V.; {Rodr{\'\i}guez-Puebla}, A.; {Hernquist}, L.
\newblock {The growth of brightest cluster galaxies in the TNG300 simulation: dissecting the contributions from mergers and in situ star formation}.
\newblock {\em \mnras} {\bf 2023}, {\em 521},~800--817,  \href{http://arxiv.org/abs/2302.10943}{{\normalfont [arXiv:astro-ph.GA/2302.10943]}}.
\newblock {\url{https://doi.org/10.1093/mnras/stad586}}.

\bibitem[{Ferreira} et~al.(2024){Ferreira}, {Bickley}, {Ellison}, {Patton}, {Byrne-Mamahit}, {Wilkinson}, {Bottrell}, {Fabbro}, {Gwyn}, and {McConnachie}]{2024MNRAS.533.2547F}
{Ferreira}, L.; {Bickley}, R.W.; {Ellison}, S.L.; {Patton}, D.R.; {Byrne-Mamahit}, S.; {Wilkinson}, S.; {Bottrell}, C.; {Fabbro}, S.; {Gwyn}, S.D.J.; {McConnachie}, A.
\newblock {Galaxy mergers in UNIONS - I. A simulation-driven hybrid deep learning ensemble for pure galaxy merger classification}.
\newblock {\em \mnras} {\bf 2024}, {\em 533},~2547--2569,  \href{http://arxiv.org/abs/2407.18396}{{\normalfont [arXiv:astro-ph.GA/2407.18396]}}.
\newblock {\url{https://doi.org/10.1093/mnras/stae1885}}.

\bibitem[{Margalef-Bentabol} et~al.(2024){Margalef-Bentabol}, {Wang}, {La Marca}, {Blanco-Prieto}, {Chudy}, {Dom{\'\i}nguez-S{\'a}nchez}, {Goulding}, {Guzm{\'a}n-Ortega}, {Huertas-Company}, {Martin}, {Pearson}, {Rodriguez-Gomez}, {Walmsley}, {Bickley}, {Bottrell}, {Conselice}, and {O'Ryan}]{2024A&A...687A..24M}
{Margalef-Bentabol}, B.; {Wang}, L.; {La Marca}, A.; {Blanco-Prieto}, C.; {Chudy}, D.; {Dom{\'\i}nguez-S{\'a}nchez}, H.; {Goulding}, A.D.; {Guzm{\'a}n-Ortega}, A.; {Huertas-Company}, M.; {Martin}, G.;  et~al.
\newblock {Galaxy merger challenge: A comparison study between machine learning-based detection methods}.
\newblock {\em \aap} {\bf 2024}, {\em 687},~A24,  \href{http://arxiv.org/abs/2403.15118}{{\normalfont [arXiv:astro-ph.GA/2403.15118]}}.
\newblock {\url{https://doi.org/10.1051/0004-6361/202348239}}.

\bibitem[{Jung} et~al.(2024){Jung}, {Kim}, {Oh}, {Hong}, {Lee}, and {Kim}]{2024ApJ...965..156J}
{Jung}, M.; {Kim}, J.h.; {Oh}, B.K.; {Hong}, S.E.; {Lee}, J.; {Kim}, J.
\newblock {Merger-tree-based Galaxy Matching: A Comparative Study across Different Resolutions}.
\newblock {\em \apj} {\bf 2024}, {\em 965},~156,  \href{http://arxiv.org/abs/2312.02466}{{\normalfont [arXiv:astro-ph.GA/2312.02466]}}.
\newblock {\url{https://doi.org/10.3847/1538-4357/ad34d1}}.

\bibitem[{Omori} et~al.(2023){Omori}, {Bottrell}, {Walmsley}, {Yesuf}, {Goulding}, {Ding}, {Popping}, {Silverman}, {Takeuchi}, and {Toba}]{2023A&A...679A.142O}
{Omori}, K.C.; {Bottrell}, C.; {Walmsley}, M.; {Yesuf}, H.M.; {Goulding}, A.D.; {Ding}, X.; {Popping}, G.; {Silverman}, J.D.; {Takeuchi}, T.T.; {Toba}, Y.
\newblock {Galaxy mergers in Subaru HSC-SSP: A deep representation learning approach for identification, and the role of environment on merger incidence}.
\newblock {\em \aap} {\bf 2023}, {\em 679},~A142,  \href{http://arxiv.org/abs/2309.15539}{{\normalfont [arXiv:astro-ph.GA/2309.15539]}}.
\newblock {\url{https://doi.org/10.1051/0004-6361/202346743}}.

\bibitem[{Patton} et~al.(2020){Patton}, {Wilson}, {Metrow}, {Ellison}, {Torrey}, {Brown}, {Hani}, {McAlpine}, {Moreno}, and {Woo}]{2020MNRAS.494.4969P}
{Patton}, D.R.; {Wilson}, K.D.; {Metrow}, C.J.; {Ellison}, S.L.; {Torrey}, P.; {Brown}, W.; {Hani}, M.H.; {McAlpine}, S.; {Moreno}, J.; {Woo}, J.
\newblock {Interacting galaxies in the IllustrisTNG simulations - I: Triggered star formation in a cosmological context}.
\newblock {\em \mnras} {\bf 2020}, {\em 494},~4969--4985,  \href{http://arxiv.org/abs/2003.00289}{{\normalfont [arXiv:astro-ph.GA/2003.00289]}}.
\newblock {\url{https://doi.org/10.1093/mnras/staa913}}.

\bibitem[{Hani} et~al.(2020){Hani}, {Gosain}, {Ellison}, {Patton}, and {Torrey}]{2020MNRAS.493.3716H}
{Hani}, M.H.; {Gosain}, H.; {Ellison}, S.L.; {Patton}, D.R.; {Torrey}, P.
\newblock {Interacting galaxies in the IllustrisTNG simulations - II: star formation in the post-merger stage}.
\newblock {\em \mnras} {\bf 2020}, {\em 493},~3716--3731,  \href{http://arxiv.org/abs/2001.04472}{{\normalfont [arXiv:astro-ph.GA/2001.04472]}}.
\newblock {\url{https://doi.org/10.1093/mnras/staa459}}.

\bibitem[{Quai} et~al.(2021){Quai}, {Hani}, {Ellison}, {Patton}, and {Woo}]{2021MNRAS.504.1888Q}
{Quai}, S.; {Hani}, M.H.; {Ellison}, S.L.; {Patton}, D.R.; {Woo}, J.
\newblock {Interacting galaxies in the IllustrisTNG simulations - III. (The rarity of) quenching in post-merger galaxies}.
\newblock {\em \mnras} {\bf 2021}, {\em 504},~1888--1901,  \href{http://arxiv.org/abs/2104.03327}{{\normalfont [arXiv:astro-ph.GA/2104.03327]}}.
\newblock {\url{https://doi.org/10.1093/mnras/stab988}}.

\bibitem[{Byrne-Mamahit} et~al.(2023){Byrne-Mamahit}, {Hani}, {Ellison}, {Quai}, and {Patton}]{2023MNRAS.519.4966B}
{Byrne-Mamahit}, S.; {Hani}, M.H.; {Ellison}, S.L.; {Quai}, S.; {Patton}, D.R.
\newblock {Interacting galaxies in the IllustrisTNG simulations - IV: enhanced supermassive black hole accretion rates in post-merger galaxies}.
\newblock {\em \mnras} {\bf 2023}, {\em 519},~4966--4981,  \href{http://arxiv.org/abs/2212.07342}{{\normalfont [arXiv:astro-ph.GA/2212.07342]}}.
\newblock {\url{https://doi.org/10.1093/mnras/stac3674}}.

\bibitem[{Brown} et~al.(2023){Brown}, {Patton}, {Ellison}, and {Faria}]{2023MNRAS.522.5107B}
{Brown}, W.; {Patton}, D.R.; {Ellison}, S.L.; {Faria}, L.
\newblock {Interacting galaxies in the IllustrisTNG simulations - V. Comparing the influence of star-forming versus passive companions}.
\newblock {\em \mnras} {\bf 2023}, {\em 522},~5107--5122,  \href{http://arxiv.org/abs/2304.14566}{{\normalfont [arXiv:astro-ph.GA/2304.14566]}}.
\newblock {\url{https://doi.org/10.1093/mnras/stad1314}}.

\bibitem[Hopkins and Beacom(2006)]{Hopkins_2006}
Hopkins, A.M.; Beacom, J.F.
\newblock On the Normalization of the Cosmic Star Formation History.
\newblock {\em The Astrophysical Journal} {\bf 2006}, {\em 651},~142--154.
\newblock {\url{https://doi.org/10.1086/506610}}.

\bibitem[{Springel}(2010)]{2010MNRAS.401..791S}
{Springel}, V.
\newblock {E pur si muove: Galilean-invariant cosmological hydrodynamical simulations on a moving mesh}.
\newblock {\em \mnras} {\bf 2010}, {\em 401},~791--851,  \href{http://arxiv.org/abs/0901.4107}{{\normalfont [arXiv:astro-ph.CO/0901.4107]}}.
\newblock {\url{https://doi.org/10.1111/j.1365-2966.2009.15715.x}}.

\bibitem[{Pillepich} et~al.(2018){Pillepich}, {Springel}, {Nelson}, {Genel}, {Naiman}, {Pakmor}, {Hernquist}, {Torrey}, {Vogelsberger}, {Weinberger}, and {Marinacci}]{2018MNRAS.473.4077P}
{Pillepich}, A.; {Springel}, V.; {Nelson}, D.; {Genel}, S.; {Naiman}, J.; {Pakmor}, R.; {Hernquist}, L.; {Torrey}, P.; {Vogelsberger}, M.; {Weinberger}, R.;  et~al.
\newblock {Simulating galaxy formation with the IllustrisTNG model}.
\newblock {\em \mnras} {\bf 2018}, {\em 473},~4077--4106,  \href{http://arxiv.org/abs/1703.02970}{{\normalfont [arXiv:astro-ph.GA/1703.02970]}}.
\newblock {\url{https://doi.org/10.1093/mnras/stx2656}}.

\bibitem[{Weinberger} et~al.(2017){Weinberger}, {Springel}, {Hernquist}, {Pillepich}, {Marinacci}, {Pakmor}, {Nelson}, {Genel}, {Vogelsberger}, {Naiman}, and {Torrey}]{2017MNRAS.465.3291W}
{Weinberger}, R.; {Springel}, V.; {Hernquist}, L.; {Pillepich}, A.; {Marinacci}, F.; {Pakmor}, R.; {Nelson}, D.; {Genel}, S.; {Vogelsberger}, M.; {Naiman}, J.;  et~al.
\newblock {Simulating galaxy formation with black hole driven thermal and kinetic feedback}.
\newblock {\em \mnras} {\bf 2017}, {\em 465},~3291--3308,  \href{http://arxiv.org/abs/1607.03486}{{\normalfont [arXiv:astro-ph.GA/1607.03486]}}.
\newblock {\url{https://doi.org/10.1093/mnras/stw2944}}.

\bibitem[{Pillepich} et~al.(2019){Pillepich}, {Nelson}, {Springel}, {Pakmor}, {Torrey}, {Weinberger}, {Vogelsberger}, {Marinacci}, {Genel}, {van der Wel}, and {Hernquist}]{2019MNRAS.490.3196P}
{Pillepich}, A.; {Nelson}, D.; {Springel}, V.; {Pakmor}, R.; {Torrey}, P.; {Weinberger}, R.; {Vogelsberger}, M.; {Marinacci}, F.; {Genel}, S.; {van der Wel}, A.;  et~al.
\newblock {First results from the TNG50 simulation: the evolution of stellar and gaseous discs across cosmic time}.
\newblock {\em \mnras} {\bf 2019}, {\em 490},~3196--3233,  \href{http://arxiv.org/abs/1902.05553}{{\normalfont [arXiv:astro-ph.GA/1902.05553]}}.
\newblock {\url{https://doi.org/10.1093/mnras/stz2338}}.

\bibitem[{Nelson} et~al.(2019){Nelson}, {Pillepich}, {Springel}, {Pakmor}, {Weinberger}, {Genel}, {Torrey}, {Vogelsberger}, {Marinacci}, and {Hernquist}]{2019MNRAS.490.3234N}
{Nelson}, D.; {Pillepich}, A.; {Springel}, V.; {Pakmor}, R.; {Weinberger}, R.; {Genel}, S.; {Torrey}, P.; {Vogelsberger}, M.; {Marinacci}, F.; {Hernquist}, L.
\newblock {First results from the TNG50 simulation: galactic outflows driven by supernovae and black hole feedback}.
\newblock {\em \mnras} {\bf 2019}, {\em 490},~3234--3261,  \href{http://arxiv.org/abs/1902.05554}{{\normalfont [arXiv:astro-ph.GA/1902.05554]}}.
\newblock {\url{https://doi.org/10.1093/mnras/stz2306}}.

\bibitem[{Rodriguez-Gomez} et~al.(2015){Rodriguez-Gomez}, {Genel}, {Vogelsberger}, {Sijacki}, {Pillepich}, {Sales}, {Torrey}, {Snyder}, {Nelson}, {Springel}, {Ma}, and {Hernquist}]{2015MNRAS.449...49R}
{Rodriguez-Gomez}, V.; {Genel}, S.; {Vogelsberger}, M.; {Sijacki}, D.; {Pillepich}, A.; {Sales}, L.V.; {Torrey}, P.; {Snyder}, G.; {Nelson}, D.; {Springel}, V.;  et~al.
\newblock {The merger rate of galaxies in the Illustris simulation: a comparison with observations and semi-empirical models}.
\newblock {\em \mnras} {\bf 2015}, {\em 449},~49--64,  \href{http://arxiv.org/abs/1502.01339}{{\normalfont [arXiv:astro-ph.GA/1502.01339]}}.
\newblock {\url{https://doi.org/10.1093/mnras/stv264}}.

\bibitem[{Springel} et~al.(2005){Springel}, {White}, {Jenkins}, {Frenk}, {Yoshida}, {Gao}, {Navarro}, {Thacker}, {Croton}, {Helly}, {Peacock}, {Cole}, {Thomas}, {Couchman}, {Evrard}, {Colberg}, and {Pearce}]{2005Natur.435..629S}
{Springel}, V.; {White}, S.D.M.; {Jenkins}, A.; {Frenk}, C.S.; {Yoshida}, N.; {Gao}, L.; {Navarro}, J.; {Thacker}, R.; {Croton}, D.; {Helly}, J.;  et~al.
\newblock {Simulations of the formation, evolution and clustering of galaxies and quasars}.
\newblock {\em Nature} {\bf 2005}, {\em 435},~629--636,  \href{http://arxiv.org/abs/astro-ph/0504097}{{\normalfont [arXiv:astro-ph/astro-ph/0504097]}}.
\newblock {\url{https://doi.org/10.1038/nature03597}}.

\bibitem[{Koncz} et~al.(2023){Koncz}, {Jo{\'o}}, and {Pint{\'e}r}]{2023CoSka..53d.153K}
{Koncz}, B.; {Jo{\'o}}, A.P.; {Pint{\'e}r}, S.
\newblock {Investigating star formation in Illustris TNG galaxy mergers}.
\newblock {\em Contributions of the Astronomical Observatory Skalnate Pleso} {\bf 2023}, {\em 53},~153--163.
\newblock {\url{https://doi.org/10.31577/caosp.2023.53.4.153}}.

\bibitem[Das et~al.(2023)Das, Pandey, and Sarkar]{Das_2023}
Das, A.; Pandey, B.; Sarkar, S.
\newblock Galaxy Interactions in Filaments and Sheets: Effects of the Large-scale Structures Versus the Local Density.
\newblock {\em Research in Astronomy and Astrophysics} {\bf 2023}, {\em 23},~025016.
\newblock {\url{https://doi.org/10.1088/1674-4527/acab44}}.

\bibitem[{Casertano} and {Hut}(1985)]{1985ApJ...298...80C}
{Casertano}, S.; {Hut}, P.
\newblock {Core radius and density measurements in N-body experiments Connections with theoretical and observational definitions}.
\newblock {\em \apj} {\bf 1985}, {\em 298},~80--94.
\newblock {\url{https://doi.org/10.1086/163589}}.

\bibitem[{Jo{\'o}} et~al.(2023){Jo{\'o}}, {Koncz}, {Pinter}, and {T{\'o}th}]{2023IAUS..373..318J}
{Jo{\'o}}, A.P.; {Koncz}, B.; {Pinter}, S.; {T{\'o}th}, L.V.
\newblock {Star Formation History in the Illustris TNG Simulation}.
\newblock In Proceedings of the Resolving the Rise and Fall of Star Formation in Galaxies; {Wong}, T.; {Kim}, W.T., Eds.,  1 2023, Vol. 373, {\em IAU Symposium}, pp. 318--321.
\newblock {\url{https://doi.org/10.1017/S1743921323000157}}.

\bibitem[{Ilbert} et~al.(2015){Ilbert}, {Arnouts}, {Le Floc'h}, {Aussel}, {Bethermin}, {Capak}, {Hsieh}, {Kajisawa}, {Karim}, {Le F{\`e}vre}, {Lee}, {Lilly}, {McCracken}, {Michel-Dansac}, {Moutard}, {Renzini}, {Salvato}, {Sanders}, {Scoville}, {Sheth}, {Silverman}, {Smol{\v{c}}i{\'c}}, {Taniguchi}, and {Tresse}]{2015A&A...579A...2I}
{Ilbert}, O.; {Arnouts}, S.; {Le Floc'h}, E.; {Aussel}, H.; {Bethermin}, M.; {Capak}, P.; {Hsieh}, B.C.; {Kajisawa}, M.; {Karim}, A.; {Le F{\`e}vre}, O.;  et~al.
\newblock {Evolution of the specific star formation rate function at z< 1.4 Dissecting the mass-SFR plane in COSMOS and GOODS}.
\newblock {\em \aap} {\bf 2015}, {\em 579},~A2,  \href{http://arxiv.org/abs/1410.4875}{{\normalfont [arXiv:astro-ph.GA/1410.4875]}}.
\newblock {\url{https://doi.org/10.1051/0004-6361/201425176}}.

\bibitem[Kim et~al.(2023)Kim, Goto, Ling, Wu, Hashimoto, Kilerci, Ho, Uno, Wang, and Lin]{KimCSFH2023}
Kim, S.J.; Goto, T.; Ling, C.T.; Wu, C.K.W.; Hashimoto, T.; Kilerci, E.; Ho, S.C.C.; Uno, Y.; Wang, P.Y.; Lin, Y.W.
\newblock {Cosmic star-formation history and black hole accretion history inferred from the JWST mid-infrared source counts}.
\newblock {\em Monthly Notices of the Royal Astronomical Society} {\bf 2023}, {\em 527},~5525--5539,  \href{http://arxiv.org/abs/https://academic.oup.com/mnras/article-pdf/527/3/5525/53971412/stad3499.pdf}{{\normalfont [https://academic.oup.com/mnras/article-pdf/527/3/5525/53971412/stad3499.pdf]}}.
\newblock {\url{https://doi.org/10.1093/mnras/stad3499}}.

\bibitem[Ling et~al.(2022)Ling, Kim, Wu, Goto, Kilerci, Hashimoto, Lin, Wang, Ho, and Hsiao]{Ling2022_JWST_sourcecounts}
Ling, C.T.; Kim, S.J.; Wu, C.K.W.; Goto, T.; Kilerci, E.; Hashimoto, T.; Lin, Y.W.; Wang, P.Y.; Ho, S.C.C.; Hsiao, T.Y.Y.
\newblock Galaxy source counts at 7.7, 10, and 15$\mu$m with the James Webb Space Telescope.
\newblock {\em Monthly Notices of the Royal Astronomical Society} {\bf 2022}, {\em 517},~853–857.
\newblock {\url{https://doi.org/10.1093/mnras/stac2716}}.

\bibitem[Wu et~al.(2023)Wu, Ling, Goto, Kim, Hashimoto, Kilerci, Lin, Wang, Uno, Ho, and Hsiao]{Wu2023_JWST_sourcecounts}
Wu, C.K.W.; Ling, C.T.; Goto, T.; Kim, S.J.; Hashimoto, T.; Kilerci, E.; Lin, Y.W.; Wang, P.Y.; Uno, Y.; Ho, S.C.C.;  et~al.
\newblock Source counts at 7.7–21$\mu$m in CEERS field with JWST.
\newblock {\em Monthly Notices of the Royal Astronomical Society} {\bf 2023}, {\em 523},~5187–5197.
\newblock {\url{https://doi.org/10.1093/mnras/stad1769}}.

\bibitem[Oliver et~al.(1997)Oliver, Goldschmidt, Franceschini, Serjeant, Efstathiou, Verma, Gruppioni, Eaton, Mann, Mobasher, Pearson, Rowan-Robinson, Sumner, Danese, Elbaz, Egami, Kontizas, Lawrence, McMahon, Norgaard-Nielsen, Perez-Fournon, and Gonzalez-Serrano]{Oliver1997_sourcecounts}
Oliver, S.J.; Goldschmidt, P.; Franceschini, A.; Serjeant, S.B.G.; Efstathiou, A.; Verma, A.; Gruppioni, C.; Eaton, N.; Mann, R.G.; Mobasher, B.;  et~al.
\newblock Observations of the Hubble Deep Field with the Infrared Space Observatory - III. Source counts and P(D) analysis.
\newblock {\em Monthly Notices of the Royal Astronomical Society} {\bf 1997}, {\em 289},~471–481.
\newblock {\url{https://doi.org/10.1093/mnras/289.2.471}}.

\bibitem[Serjeant et~al.(2000)Serjeant, Oliver, Rowan-Robinson, Crockett, Missoulis, Sumner, Gruppioni, Mann, Eaton, Elbaz, Clements, Baker, Efstathiou, Cesarsky, Danese, Franceschini, Genzel, Lawrence, Lemke, McMahon, Miley, Puget, and Rocca-Volmerange]{Serjeant2000_sourcecounts}
Serjeant, S.; Oliver, S.; Rowan-Robinson, M.; Crockett, H.; Missoulis, V.; Sumner, T.; Gruppioni, C.; Mann, R.G.; Eaton, N.; Elbaz, D.;  et~al.
\newblock The European Large Area ISO Survey -- II. Mid-infrared extragalactic source counts.
\newblock {\em Monthly Notices of the Royal Astronomical Society} {\bf 2000}, {\em 316},~768–778.
\newblock {\url{https://doi.org/10.1046/j.1365-8711.2000.03551.x}}.

\bibitem[Pearson et~al.(2010)Pearson, Oyabu, Wada, Matsuhara, Lee, Kim, Takagi, Goto, Im, Serjeant, Lee, Ko, White, and Ohyama]{Pearson2010_sourcecounts}
Pearson, C.P.; Oyabu, S.; Wada, T.; Matsuhara, H.; Lee, H.M.; Kim, S.J.; Takagi, T.; Goto, T.; Im, M.S.; Serjeant, S.;  et~al.
\newblock Source counts at 15 microns from the AKARI NEP survey.
\newblock {\em Astronomy and Astrophysics} {\bf 2010}, {\em 514},~A8.
\newblock {\url{https://doi.org/10.1051/0004-6361/200913382}}.

\bibitem[Takagi et~al.(2011)Takagi, Matsuhara, Goto, Hanami, Im, Imai, Ishigaki, Lee, Lee, Malkan, Ohyama, Oyabu, Pearson, Serjeant, Wada, and White]{Takagi2011_sourcecounts}
Takagi, T.; Matsuhara, H.; Goto, T.; Hanami, H.; Im, M.; Imai, K.; Ishigaki, T.; Lee, H.M.; Lee, M.G.; Malkan, M.;  et~al.
\newblock The AKARI NEP-Deep survey: a mid-infrared source catalogue.
\newblock {\em Astronomy \&amp; Astrophysics} {\bf 2011}, {\em 537},~A24.
\newblock {\url{https://doi.org/10.1051/0004-6361/201117759}}.

\bibitem[Pearson et~al.(2014)Pearson, Serjeant, Oyabu, Matsuhara, Wada, Goto, Takagi, Lee, Im, Ohyama, Kim, and Murata]{Pearson2014_sourcecounts}
Pearson, C.P.; Serjeant, S.; Oyabu, S.; Matsuhara, H.; Wada, T.; Goto, T.; Takagi, T.; Lee, H.M.; Im, M.; Ohyama, Y.;  et~al.
\newblock The first source counts at 18$\mu$m from the AKARI NEP Survey.
\newblock {\em Monthly Notices of the Royal Astronomical Society} {\bf 2014}, {\em 444},~846–859.
\newblock {\url{https://doi.org/10.1093/mnras/stu1472}}.

\bibitem[Davidge et~al.(2017)Davidge, Serjeant, Pearson, Matsuhara, Wada, Dryer, and Barrufet]{Davidge2017_sourcecounts}
Davidge, H.; Serjeant, S.; Pearson, C.; Matsuhara, H.; Wada, T.; Dryer, B.; Barrufet, L.
\newblock AKARI/IRC source catalogues and source counts for the IRAC Dark Field, ELAIS North and the AKARI Deep Field South.
\newblock {\em Monthly Notices of the Royal Astronomical Society} {\bf 2017}, {\em 472},~4259–4286.
\newblock {\url{https://doi.org/10.1093/mnras/stx1935}}.

\bibitem[Suess et~al.(2023)Suess, Williams, Robertson, Ji, Johnson, Nelson, Alberts, Hainline, D’Eugenio, Übler, Rieke, Rieke, Bunker, Carniani, Charlot, Eisenstein, Maiolino, Stark, Tacchella, and Willott]{Suess_minormerger_2023}
Suess, K.A.; Williams, C.C.; Robertson, B.; Ji, Z.; Johnson, B.D.; Nelson, E.; Alberts, S.; Hainline, K.; D’Eugenio, F.; Übler, H.;  et~al.
\newblock Minor Merger Growth in Action: JWST Detects Faint Blue Companions around Massive Quiescent Galaxies at 0.5 $\leq$ z $\leq$ 3.0.
\newblock {\em The Astrophysical Journal Letters} {\bf 2023}, {\em 956},~L42.
\newblock {\url{https://doi.org/10.3847/2041-8213/acf5e6}}.

\bibitem[Atek et~al.(2022)Atek, Shuntov, Furtak, Richard, Kneib, Mahler, Zitrin, McCracken, Charlot, Chevallard, and Chemerynska]{Atek2022_JWST_z16}
Atek, H.; Shuntov, M.; Furtak, L.J.; Richard, J.; Kneib, J.P.; Mahler, G.; Zitrin, A.; McCracken, H.J.; Charlot, S.; Chevallard, J.;  et~al.
\newblock Revealing galaxy candidates out to z $\sim$ 16 with JWST observations of the lensing cluster SMACS0723.
\newblock {\em Monthly Notices of the Royal Astronomical Society} {\bf 2022}, {\em 519},~1201--1220,  \href{http://arxiv.org/abs/https://academic.oup.com/mnras/article-pdf/519/1/1201/48377308/stac3144.pdf}{{\normalfont [https://academic.oup.com/mnras/article-pdf/519/1/1201/48377308/stac3144.pdf]}}.
\newblock {\url{https://doi.org/10.1093/mnras/stac3144}}.

\bibitem[Labbé et~al.(2023)Labbé, van Dokkum, Nelson, Bezanson, Suess, Leja, Brammer, Whitaker, Mathews, Stefanon, and Wang]{Labbe2023_JWST_z7-9}
Labbé, I.; van Dokkum, P.; Nelson, E.; Bezanson, R.; Suess, K.A.; Leja, J.; Brammer, G.; Whitaker, K.; Mathews, E.; Stefanon, M.;  et~al.
\newblock A population of red candidate massive galaxies ~600 Myr after the Big Bang.
\newblock {\em Nature} {\bf 2023}, {\em 616},~266–269.
\newblock {\url{https://doi.org/10.1038/s41586-023-05786-2}}.

\bibitem[Nelson et~al.(2021)Nelson, Springel, Pillepich, Rodriguez-Gomez, Torrey, Genel, Vogelsberger, Pakmor, Marinacci, Weinberger, Kelley, Lovell, Diemer, and Hernquist]{nelson2021illustristngsimulationspublicdata}
Nelson, D.; Springel, V.; Pillepich, A.; Rodriguez-Gomez, V.; Torrey, P.; Genel, S.; Vogelsberger, M.; Pakmor, R.; Marinacci, F.; Weinberger, R.;  et~al.
\newblock The IllustrisTNG Simulations: Public Data Release,  2021,  \href{http://arxiv.org/abs/1812.05609}{{\normalfont [arXiv:astro-ph.GA/1812.05609]}}.

\end{thebibliography}


%


\PublishersNote{}
\isPreprints{}
\end{adjustwidth}
\end{document}